\documentclass[prd,aps,onecolumn,preprintnumbers,amsmath,amssymb,superscriptaddress]{revtex4}
\pdfoutput=1

\usepackage{amsmath}
\usepackage{graphicx}
\usepackage{dcolumn}
\usepackage{bm}
\usepackage{amssymb}
\usepackage{latexsym}
\usepackage{epstopdf}
\usepackage{threeparttable}
\usepackage{booktabs}
\usepackage{tabularx}
\usepackage{ulem}
\usepackage{float}
\usepackage{mathrsfs}
\usepackage{hhline}
\usepackage{multirow}
\usepackage{color}
\usepackage{subfigure}
\usepackage[colorlinks,linkcolor=magenta,anchorcolor=blue,citecolor=blue]{hyperref}

\makeatletter

\newcommand{\Rmnum}[1]{\expandafter\@slowromancap\romannumeral #1@}
\makeatother

\def\be{\begin{equation}}
	\def\ee{\end{equation}}
\def\bea{\begin{eqnarray}}
	\def\eea{\end{eqnarray}}

\begin{document}
\title{On Primordial Black Holes and secondary gravitational waves generated from inflation with solo/multi-bumpy potential}
	
\author{Ruifeng Zheng}
\email{zrf2021@mails.ccnu.edu.cn}
\affiliation{Institute of Astrophysics, Central China Normal University, Wuhan 430079, China}
	
\author{Jiaming Shi}
\email[Co-first author: ]{2016jimshi@mails.ccnu.edu.cn}
\affiliation{Institute of Astrophysics, Central China Normal University, Wuhan 430079, China}
	
\author{Taotao Qiu}
\email[Corresponding author: ]{qiutt@hust.edu.cn}
\affiliation{School of Physics, Huazhong University of Science and Technology, Wuhan, 430074, China}

\begin{abstract}
It is well known that a primordial black hole (PBH) can be generated in the inflation process of the early universe, especially when the inflation field has a number of non-trivial features that could break the slow-roll condition. In this study, we investigate a toy model of inflation with bumpy potential, which has one or several bumps. We determined that the potential with multi-bump can generate power spectra with multi-peaks in small-scale region, which can in turn predict the generation of primordial black holes in various mass ranges. We also consider the two possibilities of PBH formation by spherical and elliptical collapses. Finally, we discuss the scalar-induced gravitational waves(SIGWs)generated by linear scalar perturbations at second-order.
\end{abstract}
	
\maketitle

\section{introduction}
Since the beginning of the 21st century, the science of astronomy and astrophysics have been developing rapidly, and researchers have been paying increasing attention to a type of evenly mysterious objects: black holes. From 2015 to date, there have been dozens of gravitational wave events, where gravitational wave signals are released from either binary black holes or other compact objects like white dwarfs and neutron stars \cite{Abbott:2016blz, TheLIGOScientific:2017qsa}. In 2019, the working group of Event Horizon Telescope (EHT) claimed to obtain the first images for the shadow of the supermassive black hole in the center of M87 galaxy \cite{Akiyama:2019cqa}. Moreover, the recent (2020) Nobel Prize in Physics was also awarded to studies on the black hole. Along with the development of modern science and technology, it is expected that more information about this object will be discovered with time.

Black holes are classified into several categories, based on how they are formed. One of these black-hole acategories is of particular interest to astronomers and cosmologists and are called primordial black holes (PBHs) \cite{Hawking:1971ei, Carr:1974nx, Khlopov_2010}. Unlike astrophysical black holes, PBHs are not formed from the collapse of stars. Instead, they are formed in the primordial universe, which exists significantly earlier than the star formation, owing to the gravitational collapse by the overdensity of cosmic spacetime in local patches. For this reason, the formed PBHs can have broad mass ranges, which are not constrained by Chandrasekhar and Oppenheimer limits. 

To form PBHs, small-scale inhomogeneities are required, and these can be provided by the mechanism of primordial perturbation production during the inflation process in the early universe. In the inflation scenario \cite{Guth:1980zm, Linde:1981mu, Starobinsky:1980te}, where the universe expands to very large extensions within a short time, the quantum fluctuations of the inflation field's vacuum will be stretched out of the horizon and become classical perturbations (see, e.g., \cite{Riotto:2002yw} for a comprehensive review). If the inflation has some feature that makes these perturbations exceed certain threshold value at small scales, then when the perturbations re-enters the horizon, it will cause large inhomogeneities of the universe, and PBHs will be formed. There have been long discussions on the formation of PBHs from inflation models \cite{Cai:2019bmk, GarciaBellido:1996qt, Drees:2011yz, Garcia-Bellido:2017mdw, Domcke:2017fix, Ballesteros:2017fsr, Kannike:2017bxn, Carr:2017edp, Germani:2017bcs, Motohashi:2017kbs, Gong:2017qlj, Pi:2017gih, Ozsoy:2018flq, Biagetti:2018pjj, Cai:2018tuh, Gao:2018pvq, Cai:2018dig, Ballesteros:2018wlw, Byrnes:2018txb, Dalianis:2018ymb, Pi:2019ihn, Vallejo-Pena:2019lfo, Dalianis:2019asr, Bhaumik:2019tvl, Fu:2019ttf, Xu:2019bdp, Liu:2019lul, Chen:2019zza, Arya:2019wck, Mahbub:2019uhl, Mishra:2019pzq, Lin:2020goi, Fu:2020lob, Ballesteros:2020sre, Ozsoy:2020kat, Anguelova:2020nzl, Solbi:2021wbo, Choi:2021yxz, Gao:2021vxb, Inomata:2021uqj, Biagetti:2021eep, Drees:2019xpp, Carr:2018poi}. Among these models, the simplest is a single field inflation in the framework of General Relativity; however, as has been pointed out in \cite{Germani:2017bcs, Motohashi:2017kbs}, the slow-roll condition has to be violated to form PBHs that consists all the dark matter. Consequently, slow-roll violating models become an interesting alternative, including ultra-slow-roll inflation \cite{Gong:2017qlj}, inflation with inflection points or bumps \cite{Germani:2017bcs, Bhaumik:2019tvl, Gao:2018pvq, Xu:2019bdp, Mishra:2019pzq, Ozsoy:2020kat, Atal:2019cdz}, etc. 

However, although it is well-known that PBHs can function as dark matter, they have been constrained by several experiments, such as  Subaru Hyper Suprime-Cam (Subaru-HSC) \cite{Niikura:2017zjd}, Experience de Recherche d’Objets Sombres (EROS) \cite{Tisserand:2006zx}, Optical Gravitational Lensing Experiment (OGLE) \cite{Niikura:2019kqi}, Cosmic Microwave Background (CMB) \cite{Serpico:2020ehh}, Femtolensing of Gamma-ray Bursts (FL) \cite{Barnacka:2012bm}, White Dwarf Explosions (WD) \cite{Graham:2015apa} (see also controversies in \cite{Katz:2018zrn, Montero-Camacho:2019jte}), 511 keV gamma-ray line \cite{Laha:2019ssq, Dasgupta:2019cae, Laha:2020ivk, Cai:2020fnq}, BH evaporation \cite{Carr:2009jm}, Neutron Stars (NS) \cite{Capela:2013yf}, NANOGrav \cite{Chen:2019xse, Wong:2020yig}, LIGO \cite{Kavanagh:2018ggo}, Leo-I dwarf galaxy \cite{Lu:2019ktw}, Gravitational-Wave Lensing (GW-Lensing) \cite{Jung:2017flg} and so on \footnote{The information of the constraints, along with the references, are obtained from the publicly available {\bf Python} code \href{https://github.com/bradkav/PBHbounds}{\bf PBHbounds} \cite{bradley_j_kavanagh_2019_3538999}. }. Owing to Hawking radiation, PBHs smaller than $ 10^{15} $ g can now be evaporated; however, there are still a few observational constraints, such as big bang nucleosynthesis (BBN) \cite{Carr:2009jm}, Extragalactic $ \gamma $-rays \cite{Carr:2009jm, Ray:2021mxu}, CMB spectral distortions, and anisotropies (CMB sda) \cite{Acharya:2020jbv}. For the constraints of the small-mass PBHs, refer to \cite{Carr:2020gox, Allahverdi:2020bys}. Currently, the upper limits on the PBH fraction of dark matter $f_{PBH}$ provided by these experiments have covered almost all the mass ranges of PBHs; however, most of the limits can only reach up to about $10^{-5}$. Therefore there are still large blanks for the PBHs in other mass ranges to have larger fractions. In this study, we investigate the generalized bumpy inflation model, whose potential contains multiple bumps. For the multi-bumpy potential, it is expected that the PBHs can be formed in various mass ranges; hence, by comparing it to a single mass range, the detectability of the PBHs can get improved. If the accuracies of these experiments get further improved and the upper limits are put even lower in the future, we can expect that the PBHs generated in our model are more likely to be detected. For example, recently, researchers have studied the electromagnetic signals from the bubbles of a single PBH star and the possibility of their detection via high-energy gamma-ray experiments, especially for small-mass primordial black holes \cite{Cai:2021zxo}. Note that similar multi-range PBH formations can be realized in \cite{Tada:2019amh, Braglia:2020eai}, using a multi-field inflation model. 
\par
However, the enhancement of the primordial curvature perturbation will also induce the generation of secondary gravitational waves \cite{Ananda:2006af, Baumann:2007zm}. Therefore, the observation of SIGWs will further constrain the enhancement of the amplitude of the primordial curvature perturbation during the inflation period.  
\par

The remainder of this paper is organized as follows: In Sec. \ref{sec2}, we present the general formulation of the inflation process at both background and perturbation levels. For analyticity, we impose the slow-roll and ultra-slow-roll conditions. In Sec. \ref{sec3}, we introduce our inflation models, where the inflaton potential has one or multiple bumps (we take the example of 3 bumps), and then calculate the slow-roll parameter power spectrum numerically. In Sec. \ref{sec4}, we calculate the fraction of generated PBHs from our models in dark matter, using the Press-Schechter formalism. We also consider the formation of PBHs in the case of spherical collapse and ellipsoidal collapses. In Sec. \ref{sec5}, we discuss the generation of SIGWs, which is constrained by the latest observations. In Sec. \ref{sec6}, we present our conclusions and discussions. Moreover, in this study, we work in mostly-plus signatures for the metric, (-,+,+,+), and the unit where $c=\hbar=1$ and $M_{pl}=1/\sqrt{8\pi G}$ in this paper.

\section{Inflation: basic formulations}
\label{sec2}
First, we start with the inflation scenario driven by a single scalar field with an arbitrary potential. The action of the inflation model is expressed as
	\begin{equation}
		S=\int d^4x\sqrt{-g}\left[\frac{M_{pl}^2R}{2}-\frac{1}{2}\partial_\mu\phi\partial^\mu\phi-V(\phi)\right]~.
	\end{equation}
From the above action, we can obtain the Friedmann equation:
	\begin{equation}
		H^2=\frac{1}{3M_{pl}^2}\left( \frac{1}{2}\dot{\phi}^2+V\left( \phi \right) \right)~,
	\end{equation}
where $H=\dot a/a$ represents the Hubble parameter; $a$ is the scale factor and dot denotes a derivative with respect to cosmic time $t$. We can also obtain the equation of motion for the inflaton field $\phi$:
	\begin{equation}
	\label{eom}
		\ddot{\phi}+3H\dot{\phi}+V'(\phi)=0~,
	\end{equation}
where $V'(\phi) =dV(\phi)/d\phi$.
To maintain the inflation process, it is usually required that the slow-roll condition should be satisfied. We define the slow-roll parameters as:
	\begin{eqnarray}
		\epsilon _H&=&-\frac{\dot{H}}{H^2}~,\\
		\eta _H&=&-\frac{\ddot{H}}{2\dot{H}H}~,\\
		\xi _H&=&\frac{\dddot{H}}{2H^2\dot{H}}-2\eta _{H}^{2}~,
	\end{eqnarray}
where the suscript $H$ denotes that these parameters are defined for the Hubble parameter, or in other words, the ``whole universe". The slow-roll condition states that during inflation, all these parameters have values that are significantly less than $1$, namely
	\begin{equation}
		|\epsilon_H|~,~|\eta_H|~,~|\xi_H|\ll 1~.
		\label{srcondition}
	\end{equation}
However, another set of slow roll parameters can also be defined using the inflation potential itself:
	\begin{eqnarray}
	 	\epsilon_V&=&\frac{M_{pl}^{2}}{2}\left(\frac{V'(\phi)}{V(\phi)}\right)^2~,\\
		\label{equation3}
		\eta_V&=&M_{pl}^{2}\frac{V''(\phi)}{V(\phi)}~,
	\end{eqnarray}
and when the slow-roll condition (\ref{srcondition}) holds, $\epsilon_H\simeq \epsilon_V$, $\eta_H\simeq\eta_V-\epsilon_V$ is obtained. Moreover, the total number of e-foldings is usually adopted to describe the duration of inflation:
	\begin{equation}
		N\equiv N_{i}-N_{e}=\int_{t_i}^{t_{e}}{H\left( t \right) dt}=\ln\left(\frac{a_{e}}{a_i}\right)~.
	\end{equation}
where the subscript $i$ and $e$ denotes the parameters at the beginning and ending time of the inflation, respectively. To solve the notorious Big-Bang problems, it is usually required that $N\geq 60$ \cite{Baumann:2009ds}.

Now we turn to the perturbations generated during inflation. The perturbed FRW metric in its 3+1 decomposed form is \cite{Arnowitt:1962hi}:
	\begin{equation}
		ds^2=a^2(\tau)[-(1+2\alpha)d\tau^2+2\partial_i\beta d\tau dx^i+ e^{h_{ij}+2\zeta \delta_{ij}}dx^idx^j]~,
	\end{equation}
where $\alpha$, $\beta$, $\zeta$ denote the scalar-type perturbations, and $h_{ij}$ represents the tensor-type perturbation. The comoving time $\tau$ is defined by the relationship $d\tau\equiv a^{-1}(t)dt$. In uniform-$\phi$ gauge, $\alpha$ and $\beta$ become constraint degrees of freedom only, and the only dynamical scalar perturbation is $\zeta$ \cite{Maldacena:2002vr, Chen:2006nt}. To obtain an analytical solution of $\zeta$, we define the well-known Mukhanov-Sasaki (MS) variable $u$ as:
	\begin{equation}
		u\equiv z\zeta~,~~~z\equiv a\frac{\dot{\phi}}{H}~,
	\end{equation}
and this variable satisfies Mukhanov-Sasaki equation in the Flourier space as \cite{Sasaki:1986hm, Mukhanov:1988jd}:
	\begin{equation}
	\label{MSequation}
		u''_{k}+\left( k^2-\frac{z''}{z} \right) u_k=0~,
	\end{equation}\\
where prime denotes the derivative with respect to comoving time $\tau$, and the effective potential term is given by the following expression:
	\begin{equation}
		\frac{z''}{z}=2a^2H^2\left( 1+\epsilon _H-\frac{3}{2}\eta _H+\epsilon _{H}^{2}+\frac{1}{2}\eta _{H}^{2}-2\epsilon _H\eta _H+\frac{1}{2}\xi _H \right)~.
	\end{equation}
	
For a given mode $k$, and for a sub-Hubble region with $k\gg aH$ at sufficiently early times, we can assume $u_{k}$ to be quantum fluctuations in the Bunch-Davies vacuum \cite{Bunch:1978yq} satisfying
	\begin{equation}
		u_k(\tau) \rightarrow \frac{1}{\sqrt{2k}}e^{-ik\tau}~.
	\end{equation}
The above solution can be viewed as the initial condition of the MS variable $u_k$. However, because the inflation process will lead to the decline of comoving Hubble radius, the $k$-mode will exit the horizon and enter the super-Hubble region with $k\ll aH$, while the quantum fluctuations will be decoherent and become classical perturbations. By solving the MS equation and adopting the relationship between the MS variable and its dimensionless primordial power-spectrum \cite{Ballesteros:2017fsr}
	\begin{equation}
		P_S=\left. \frac{k^3}{2\pi ^2}\frac{\left| u_k \right|^2}{z^2} \right|_{k\ll aH}~,
	\end{equation}
we obtain the power spectrum expression under the slow roll approximation:
	\begin{equation}
		P_S=\frac{1}{8\pi ^2\epsilon _H}\left( \frac{H}{M_{pl}} \right) ^2~.
		\label{equation12}
	\end{equation}
	
On the large cosmological scale accessible to CMB observations, the power spectrum usually takes the power-law form: 
	\begin{equation}
		P_R(k) =A_S\left( \frac{k}{k_*} \right) ^{n_s-1}
	\end{equation}
where $A_S=P_S(k_*)$ is the amplitude of the scalar power spectrum at the pivot scale. In the slow-roll approximation, the
scalar spectral tilt $n_{s}$ are given by \cite{Baumann:2009ds}
	\begin{equation}
		n_s=2\eta _H-4\epsilon _H+1~.
	\end{equation}

Recent CMB observations suggest $P_S \simeq 2.1\times 10^{-9}$ at $ k_*=0.05\text{Mpc}^{-1} $ \cite{Akrami:2018odb}; however, to increase the abundance of PBH to an appropriate order of magnitude, $P_S$ must reach the amount of $O(10^{-2})$. Note that when the slow-roll condition holds, $P_S$ will be conserved in time and will not deviate significantly from its pivot value; hence, it is impossible to generate sufficient PBHs, and we must break the condition. This is also pointed out in recent papers such as \cite{Germani:2017bcs, Motohashi:2017kbs}. A possible approach is to have $\epsilon_H$ decrease rapidly at small scales, while results at large scales are kept unchanged. One such model is the ultra-slow-roll (USR) inflation model \cite{Kinney:2005vj, Namjoo:2012aa, Martin:2012pe}. This model has an absolutely flat potential; therefore,
	\begin{equation}
		\ddot{\phi}+3H\dot{\phi} =-V'\left( \phi \right) =0~,~~~\epsilon_H\propto a^{-6}~.
		\label{equation20}
	\end{equation}
Therefore, $\epsilon_H$ decreases with the explosion of $a$, triggering an abrupt increase in $P_{S}$ according to Eq.(\ref{equation12}) \cite{Gong:2017qlj}. Moreover, in this case, the slow-roll condition gets violated because $\eta_H=3+\epsilon_H>1$, and the above relationship between two sets of the slow-roll parameters does not hold any longer. Another example is the featured potiential like the one with a bump \cite{Mishra:2019pzq, Ozsoy:2020kat}. When the inflation passes through the bump, it will enter a state similar to USR, which will achieve a local enhancement of the power spectrum.

\section{Our models}
\label{sec3}
\subsection{solo-bumpy potential}
\begin{figure}[htbp]
	\centering
	\subfigure{
		\includegraphics[height=4.0cm,width=6.5cm]{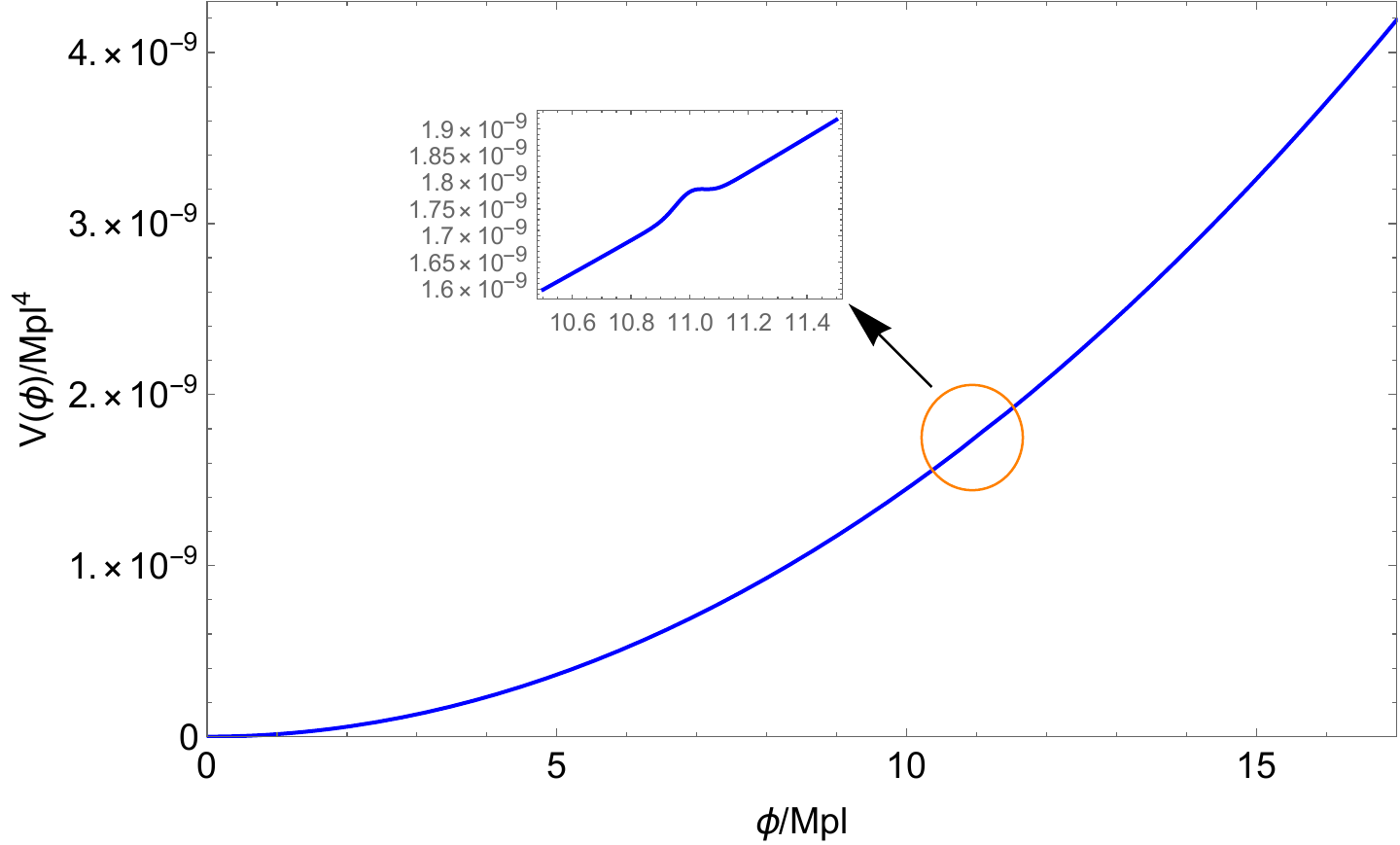}
	}
	\subfigure{
		\includegraphics[height=4.0cm,width=6.5cm]{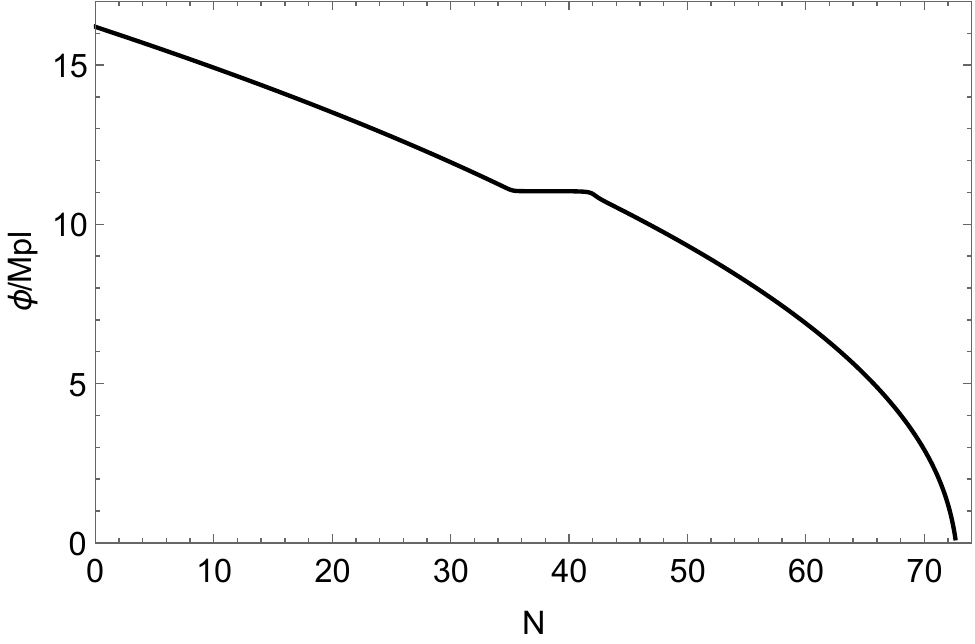}
	}
	\subfigure{
	\includegraphics[height=4.0cm,width=6.5cm]{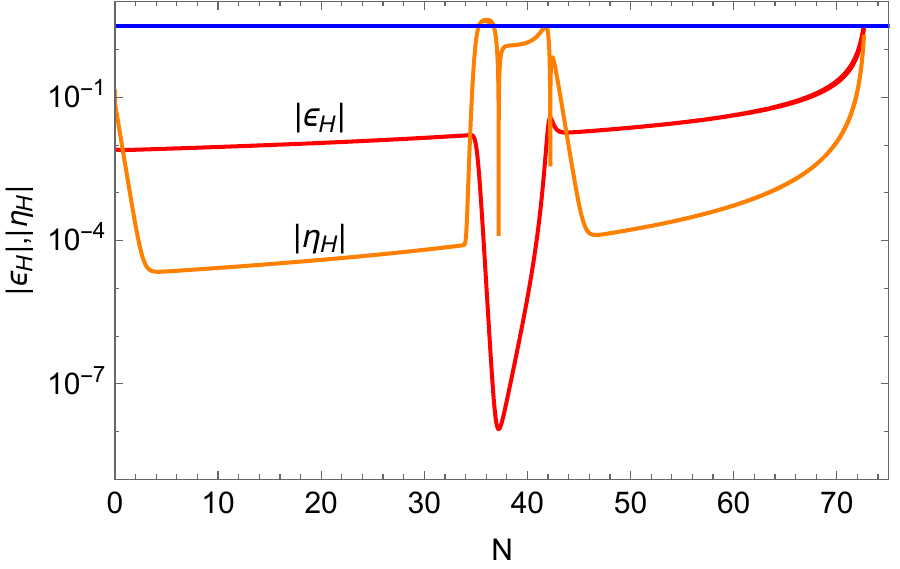}
}
	\subfigure{
	\includegraphics[height=4.0cm,width=6.5cm]{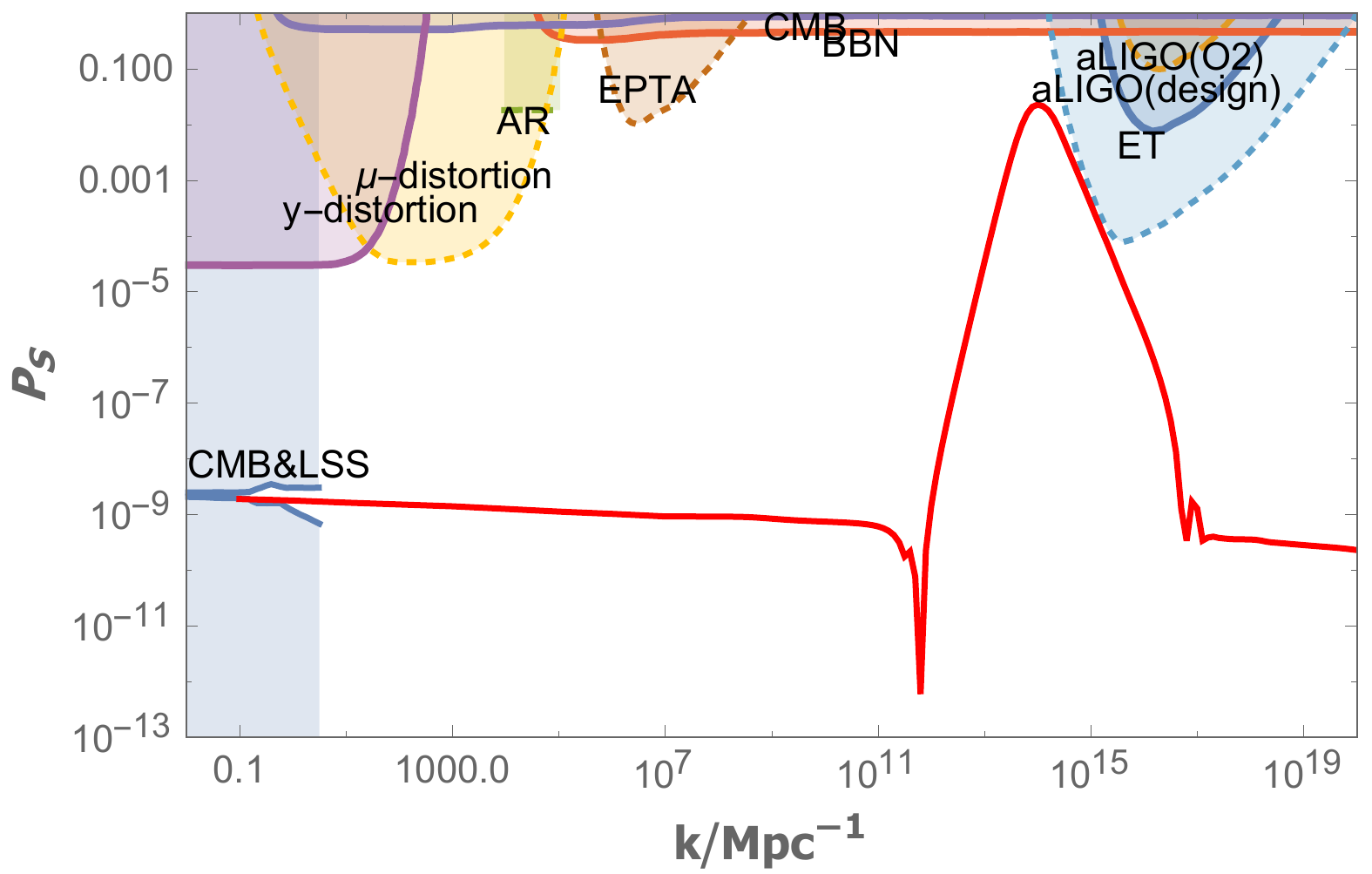}
}
	\caption{{\it Upper left figure:} the inflation potential of $p=2$, in which the bump is located at $\phi=11M_{pl}$. {\it Upper right figure:} the evolution of $ \phi $ with the parameter $ N $, which enters the USR-like stage near $N=40$. {\it Lower left figure:} the evolution of the slow roll parameters $|\epsilon_{H}|$ and $|\eta_{H}|$, in which the blue line corresponds to the value of $3$ on the vertical axis.  {\it Lower right figure:} the relationship between power spectrum $P_{S}$ (in Logarithmic scale) and the wave-number $k$. Our results are consistent with the observational constraints.}
	\label{FIG1}
\end{figure}

In this subsection, we construct inflation field with one bump in its potential. We consider the potential to be in the following form:
	\begin{equation}
		V(\phi)=V_{0}(\frac{\phi}{M_{pl}})^p\left( 1+be^{-\frac{\left( \phi -c \right) ^2}{d}} \right)~.
	\label{solopotential}
	\end{equation}
Here the basic potential is of the power-law form while there is Gaussian function-like bump on the potential, with the parameters $b$, $c$ and $d$ denoting the height, central value in $\phi$, and width of the bump. In general, the power spectrum of the slow-roll inflation potential does not contribute to the peak in the power spectrum, which is required for the formation of the primordial black holes. In this case, it is common to add a local Gaussian bump in the inflation potential, which has been studied in \cite{Mishra:2019pzq, Ragavendra:2021qdu, Atal:2019cdz}. In the following part, we will see that by choosing the parameters practically, the addition of local Gaussian bumps can effectively slow down the inflation field and enhance the scalar power spectrum associated with the formation of the primordial black hole. From this potential, the following can be obtained:
	\begin{equation}
 \epsilon _V=\frac{M_{pl}^{2}}{2}\frac{\left( pde^{\frac{\left( c-\phi \right) ^2}{d}}+b\left( pd+2\left( c-\phi \right) \phi \right) \right) ^2}{d^2\left( b+e^{\frac{\left( c-\phi \right) ^2}{d}} \right) ^2\phi ^2}~,
	\end{equation}
	\begin{equation}
\eta _V=M_{pl}^{2}(
\frac{p\left( p-1 \right)}{\phi ^2}-\frac{2b\phi \left( \phi d+2p\phi d+4c\phi ^2-2\phi ^3-2cpd-2c^2\phi \right)}{\phi ^2\left( b+e^{\frac{\left( c-\phi \right) ^2}{d}} \right) d^2})~.
	\end{equation}
In Fig. \ref{FIG1}, we plot various relationships between variables involved in this case. In the numerical calculation, we select the parameters as follows:
	\begin{equation} 
	\label{eqn2}
		V_{0}=1.45\times 10^{-11}M_{pl}^{4}~,~~~p=2~,~~~b=1.5718\times 10^{-2}~,~~~c=11M_{pl}~,~~~d=5\times 10^{-3}M_{pl}^{2}~.
	\end{equation}
From the plot, it can be easily observed that there is a bump in the potential at $\phi=11M_{pl}$. Owing to the bump, the inflation will stay around $\phi=11M_{pl}$ for a short period. However, if the initial velocity of $\phi$ is insufficient, inflation may stop at the bump forever and will not continue to roll, leading to an eternal inflation. Conversely, if the initial velocity of the inflation is too large, the inflation will pass through the bump very quickly, which will make the peak value of the power spectrum too low to have sufficient abundance of PBHs. Therefore, both the initial velocity and the parameters of the bump need to be carefully settled. In the numerical calculation, we set the initial value of $\phi$ to be $\phi_i=16.2M_{pl}$, from which $\phi$ rolls down, and the initial velocity of $\phi$ to be $\dot\phi_i=4.5\times 10^{-6}$, to satisfy the observational constraint from CMB.

We solve the equation of motion (\ref{eom}) for the scalar field $\phi$ to obtain its evolution with respect to the e-folding number $N$. From the plot, it can be observed that the inflation meets the bump near $N=40$. We also plot the evolution of $\epsilon_H$ and $\eta_H$ with respect to $N$. To increase the power spectrum amplitude to $10^{-2}$, the inflation will stay at the bump for about 7 e-folding numbers. Because of the very slow motion of the inflation, the slow roll parameter $\epsilon_H$ gets a negligible value, while $\eta_H$ becomes significantly large ($\sim O(1)$), just like the case of USR inflation models. Moreover, the inflation can stop when $\epsilon=1$ at approximately $N=70$.

We also solve the MS equation (\ref{MSequation}) numerically and obtain the variation of the power spectrum $P_{S}$ with respect to the wave number $k$. It can be observed that the power spectrum fits the constraints from CMB on large scales with a peak value of $O(10^{-2})$ (the numerical value is $0.02284$) at $k=1\times 10^{14}$ Mpc$ ^{-1} $.

\begin{figure}[htbp]
	\centering
	\subfigure{
		\includegraphics[height=4.0cm,width=6.5cm]{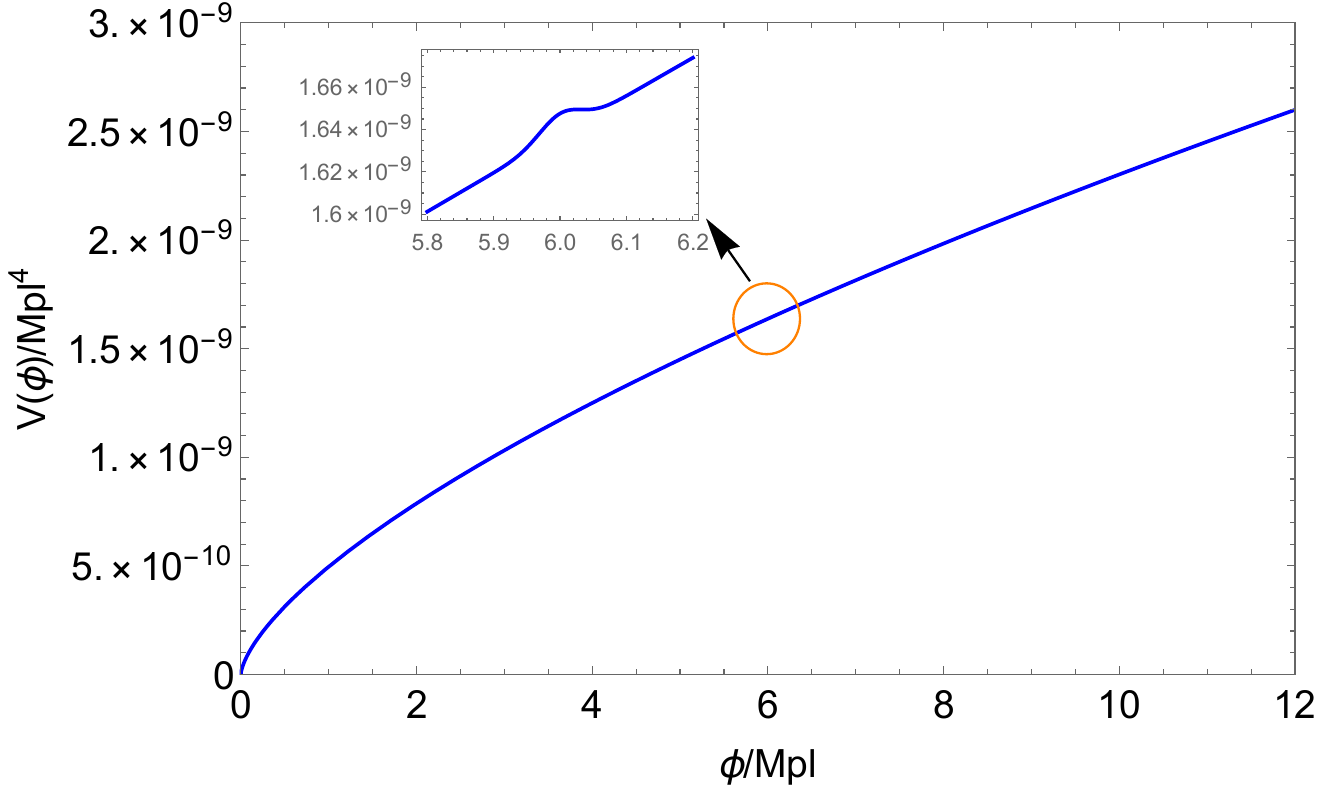}
	}
	\subfigure{
		\includegraphics[height=4.0cm,width=6.5cm]{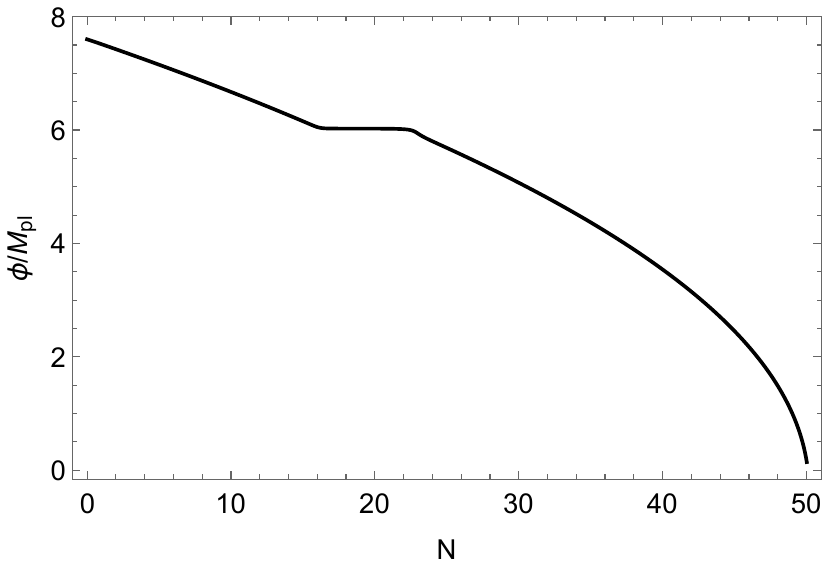}
	}
	\subfigure{
	\includegraphics[height=4.0cm,width=6.5cm]{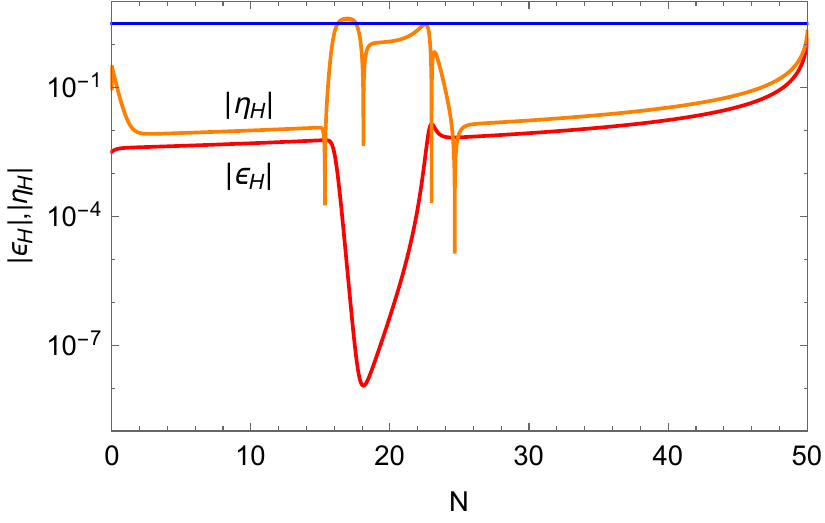}
}
	\subfigure{
	\includegraphics[height=4.0cm,width=6.5cm]{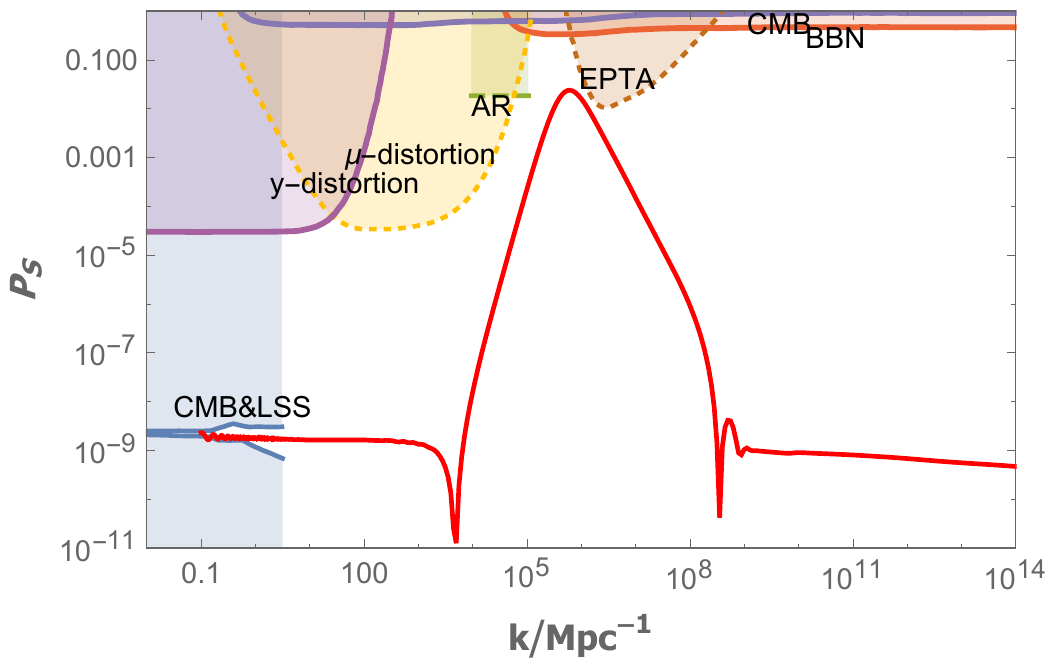}
}
	\caption{{\it Upper left figure:} the inflation potential of $ p=2/3 $, where the bump is located at  $\phi=6M_{pl}$. {\it Upper right figure:} the evolution of $ \phi $ with the parameter $ N $, which enters the USR-like stage near $N=20$. {\it Lower left figure:} the evolution of the slow roll parameters $|\epsilon_{H}|$ and $|\eta_{H}|$, where the blue line corresponds to the value of $3$ on the vertical axis.  {\it Lower right figure:} the relationship between power spectrum $P_{S}$ (in Logarithmic scale) and the wave-number $k$. Our results are consistent with the observational constraints.}
	\label{FIG2}
\end{figure}

For a further verification, we choose a second set of parameters as 
	\begin{equation} 
		V_{0}=4.96\times 10^{-10}M_{pl}^{4}~,~~~p=2/3~,~~~b=5.9921\times 10^{-3}~,~~~c=6M_{pl}~,~~~d=2\times 10^{-3}M_{pl}^{2}~.
	\end{equation}
and plot the same figures in Fig. \ref{FIG2}. In this case, it can be
observed that the decrease in $\epsilon_H$ occurs around $N=20$ during inflation, and a peak appears in the power spectrum at around $k=1\times 10^{6}\text{Mpc}^{-1}$. Therefore, we find that the Gaussian-function-like bump on the potential does lead to the peak on the spectrum at small scales, provided we correctly choose the parameters. Such a feature in the spectrum can in principle generate PBHs, which we will discuss in the next section.

\subsection{multi-bumpy potential}
\begin{figure}[htbp]
	\centering
	\subfigure{
		\includegraphics[height=4.0cm,width=6.5cm]{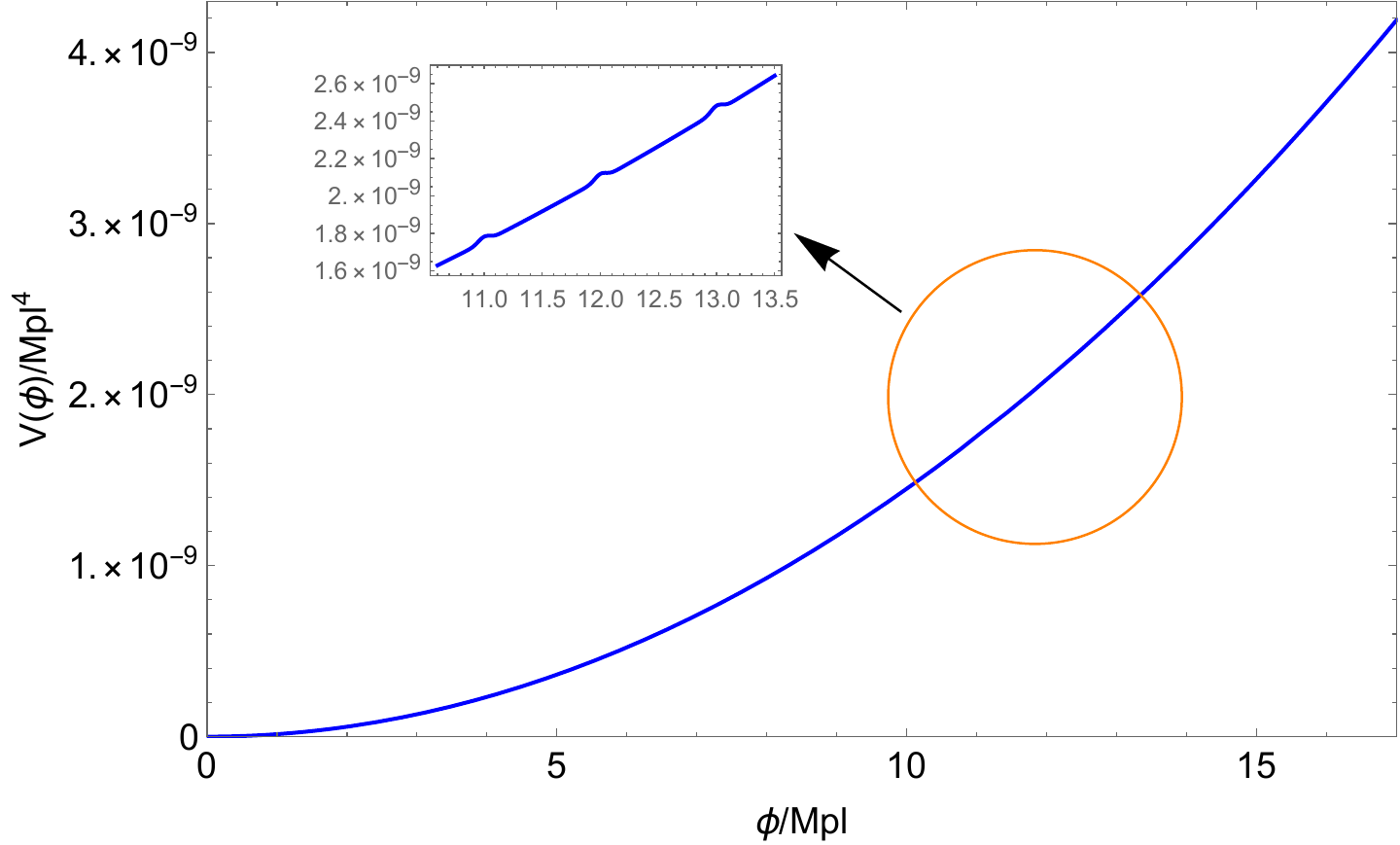}
	}
	\subfigure{
		\includegraphics[height=4.0cm,width=6.5cm]{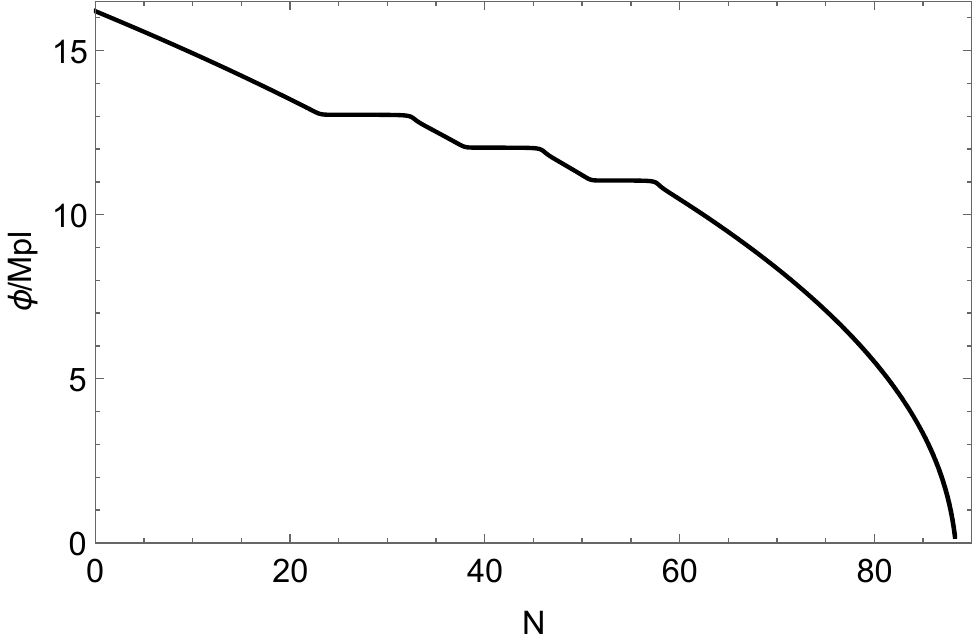}
	}
	\subfigure{
	\includegraphics[height=4.0cm,width=6.5cm]{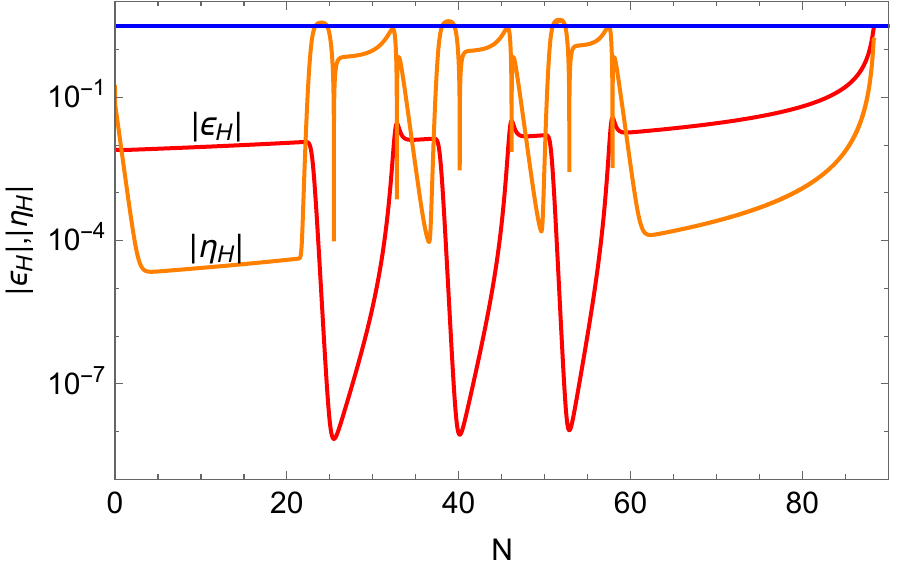}
}
	\subfigure{
	\includegraphics[height=4.0cm,width=6.5cm]{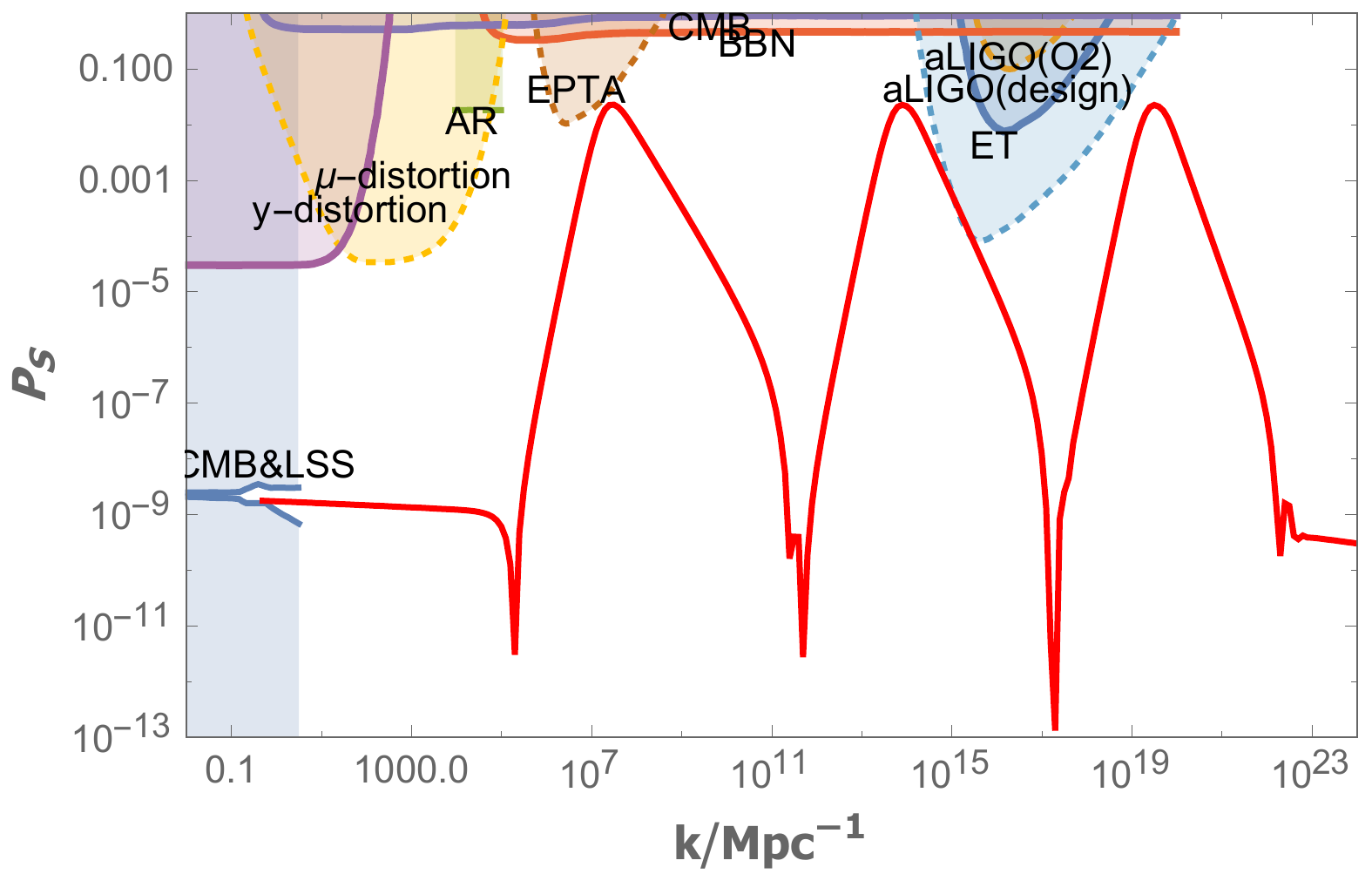}
}
	\caption{ {\it Upper left figure:} the inflation potential with three bumps locating at $\phi=11M_{pl}$, $\phi=12M_{pl}$ and $\phi=13M_{pl}$. {\it Upper right figure:} the evolution of $ \phi $ with the parameter $ N $, which enters the USR-like stage three times near $N=20$, $N=35$ and $N=50$. {\it Lower left figure:} the evolution of the slow roll parameters $|\epsilon_{H}|$ and $|\eta_{H}|$, where the blue line corresponds to the value of $3$ on the vertical axis. {\it Lower right figure:} the relationship between power spectrum $P_{S}$ (in Logarithmic scale) and the wave-number $k$. Our results are consistent with the observational constraints.}
	\label{FIG3}
\end{figure}

In this subsection, we extend the model discussed above to the case where the potential have more than one bump. In this case, there will be multiple peaks in the power spectrum, leading to formation of PBHs in multiple mass ranges. It is interesting in the sense that, in the future, as the accuracy of the experiments improve continuously, it is expected that PBHs can be caught in various mass ranges, which can thus be explained by our model. In the following, we take the potential to be of the following form:
	\begin{equation}
		V(\phi) =V_{0}\left( \frac{\phi}{M_{pl}} \right) ^2\left( 1+b_{1}e^{-\frac{\left( \phi -c_{1} \right) ^2}{d_{1}}}+b_{2}e^{-\frac{\left( \phi -c_{2} \right) ^2}{d_{2}}}+b_{3}e^{-\frac{\left( \phi -c_{3} \right) ^2}{d_{3}}}+\cdot \cdot \cdot \right)~.
	\label{multipotential}
	\end{equation}
In principle, there could be arbitrary numbers of bumps in the potential; however, in this paper, we take an example of three Gaussian function-like bumps for simplicity and analycity. The specific parameters of Eq.(\ref{multipotential}) are given as
	\begin{eqnarray}
	 	&&b_{1}=1.571806\times 10^{-2}~,~c_{1}=11M_{pl}~,~d_{1}=5\times 10^{-3}M_{pl}^{2}~, \nonumber\\
		&&b_{2}=1.41793\times 10^{-2}~,~c_{2}=12M_{pl}~,~d_{2}=5\times 10^{-3}M_{pl}^{2}~, \nonumber\\
		&&b_{3}=1.29364\times 10^{-2}~,~c_{3}=13M_{pl}~,~d_{3}=5\times 10^{-3}M_{pl}^{2}~. 
		\label{parameter-multi}
	\end{eqnarray}
In this case, the inflation rolls down from the value of $\phi=16.2M_{pl}$ and satisfies the observation constraint of CMB from Fig. \ref{FIG3}. We set $ V_{0}=1.45\times 10^{-11}M_{pl}^{4} $. 

 It is easy to observe that there are three USR stages in Fig. \ref{FIG3}, which means that there will be three peaks in the power spectrum. For initial conditions, we still set $\phi_i=16.2M_{pl}$ and $\dot\phi_i=4.5\times 10^{-6}$, which is the same as for the solo-bumpy model. However, it differs from the solo-bumpy ones, where we have to take care of not only the shape of each bump (height, width, etc), as well as the relative distance of the bumps. This is important because, when the inflation field passes through one bump, it will lose kinetic energy, and if the bumps are far from each other, it may not have enough energy to pass through the next ones. Therefore, in this study, we set the bumps close to each other, namely at $\phi=11M_{pl}$, $\phi=12M_{pl}$ and $\phi=13M_{pl}$. It can be observed from the plot that we will have three times $\eta_H$, which breaks the slow-roll condition at approximately $N=20$, $N=35$ and $N=50$, leading to three peaks on the power spectrum in small scales, while the inflation finally ends at  approximately $N=87$, and both $\epsilon_H$ and $\eta_H$ exceeds unity. As will be demonstrated in the next section, this will form PBHs in three different mass ranges.  

\section{PBH formation and abundance}
\label{sec4}
First, We briefly review how the PBHs are formed from inflation and how to relate the mass of PBHs to the inflation power spectrum. Generally, if the fluctuations re-entering the horizon become sufficiently large, the primordial black holes with extensive mass will form due to gravitational collapse. 
The mass of PBHs formed at a certain epoch in the radiation dominated era with fluctuation mode $k_{PBH}$ is given by the Hubble mass at that epoch up to an efficiency factor $\gamma$ \cite{Sasaki:2018dmp, Green:1997sz}:
	\begin{equation}
		M_{PBH}=\gamma M_H=\gamma \frac{4\pi}{3}H_{form}^{-3}\rho _{form}~,
	\label{equation23}
	\end{equation}
such that when the PBHs are formed, the Hubble radius and the energy density of the universe are denoted by $H_{form}^{-1}$ and $\rho_{form}$ respectively, and $M_H$ is the total mass inside the Hubble radius. Moreover, in the radiation dominated era, we obtain $\gamma\simeq0.2$. With the Friedmann equation: $3H_{form}^{2}=8\pi G\rho _{form}$, Eq. (\ref{equation23}) can be written in another form:
	\begin{equation}
		M_{PBH}=\gamma \sqrt{\frac{2\pi}{3G}}\rho _{form}^{1/2}H_{form}^{-2}~.
	\end{equation}

Note that in the radiation dominated era, $ \rho =\frac{\pi ^2}{30}g_*T^4 $, and in the adiabatic environment of the early universe, entropy density: $ s\sim g_*T^3\sim a^{-3}$, therefore $T\sim g_{*}^{-1/3}a^{-1}$ \cite{Kolb:1990vq}. Then we can obtain \cite{Inomata:2017okj}
	\begin{align}
		M_{PBH} & =\gamma \sqrt{\frac{2\pi}{3G}}\left( \frac{g_*}{g_{*eq}} \right) ^{1/2}\left( \frac{T_{form}}{T_{eq}} \right) ^2\left( \frac{H_{form}}{H_{eq}} \right) ^{-2}\rho _{eq}^{1/2}H_{eq}^{-2}\nonumber\\
 		& =\gamma \sqrt{\frac{2\pi}{3G}}\left( \frac{g_*}{g_{*eq}} \right) ^{1/2}\left( \frac{g_*}{g_{*eq}} \right) ^{-2/3}\left( \frac{a_{form}}{a_{eq}} \right) ^{-2}\left( \frac{H_{form}}{H_{eq}} \right) ^{-2}\rho _{eq}^{1/2}H_{eq}^{-2}\nonumber\\
		& =\gamma \frac{1}{2G}H_{eq}^{-1}\left( \frac{g_*}{g_{*eq}} \right) ^{-1/6}\left( \frac{k_{form}}{k_{eq}} \right) ^{-2}\nonumber\\
		& =2\times 10^{48}g\times \left( \frac{\gamma}{0.2} \right) \left( \frac{g_*}{106.75} \right) ^{-1/6}\left( \frac{k_{form}}{0.07~\rm Mpc   ^{-1}} \right) ^{-2}~,
	\label{M_PBH}
	\end{align}
where $g_*$ is the total effective degree of freedom of the universe in the radiation dominated era. The subscript ``form" denotes variables that are evaluated when PBHs are formed, and the subscript ``eq" denotes variables that are evaluated when radiation and matter are equal to each other. Note that in the radiation/matter equality, $g_{*eq}\simeq3.38 $ and $k_{eq}\simeq0.07\varOmega_{m0}h^2\text{Mpc}^{-1} $. Hence, it can be observed that the mass of PBHs are determined by the scale of their formation denoted by $k_{form}$. For a given $k_{form}$, we can obtain the corresponding $M_{PBH}$.

We are interested in whether the PBHs formed can act as dark matter, or more precisely, how much fraction of dark matter can be contributed from the PBHs. To evaluate this, we need to define the fraction of PBHs in dark matter, namely
	\begin{equation}
		f_{PBH}=\left. \frac{\rho _{PBH}}{\rho _{DM}} \right|_{form}~,
	\end{equation}
and this can be connected to the fraction of PBHs in the entire universe 
	\begin{equation}
		\beta(M_{PBH})\equiv\left. \frac{\rho _{PBH}}{\rho _{tot}} \right|_{form}~,
	\end{equation}
via the equation:
	\begin{align}
		f_{PBH}&=\left. \beta(M_{PBH})\left(\frac{\rho_{tot}}{\rho _{DM}}\right)\right|_{form}~\nonumber\\
		&=\beta(M_{PBH})\Omega_{DM0}^{-1}\left(\frac{a_0}{a_{form}}\right)^{-3}\left(\frac{H_{form}}{H_0}\right)^{2}~,
	\label{f_PBH}
	\end{align}
where subscript ``0" denotes variables evaluated today, and $\Omega_{DM0}$ is the today's density fraction of dark matter that can be obtained from the observations. Combining Eqs. (\ref{M_PBH}) and (\ref{f_PBH}) and doing some manipulations (the details of which could be found in, e.g. \cite{Carr:2009jm}), we can obtain the relationship between $ \beta(M) $ and $ f_{PBH} $:
	\begin{equation}
		f_{PBH}\left( M_{PBH} \right) =1.68\times 10^8\left( \frac{\gamma}{0.2} \right) ^{1/2}\left( \frac{g_*}{106.75} \right) ^{-1/4}\left( \frac{M_{PBH}}{M_{\odot}} \right) ^{-1/2}\beta \left( M_{PBH} \right)~.
	\label{equation31}
	\end{equation}	

The standard treatment of $\beta(M)$ is based on the Press-Schechter formalism \cite{Press:1973iz} of the gravitational collapse that is widely adopted in large-scale structure studies \cite{Sureda:2020vgi}. In this formalism, $\beta(M)$ is given by the probability that the fractional overdensity $\delta\equiv\delta \rho/\rho$ is above a certain threshold $\delta_{c}$ for PBH formation \cite{Wu:2020ilx, Mahbub:2020row}. For Gaussian primordial fluctuations, $ \beta(M)$ is given by \cite{Inomata:2017uaw, Carr:1975qj}:
	\begin{align}\label{EQ35}
		\beta \left( M\left( k \right) \right) & =2\int_{\delta _c}^{\infty}{\exp \left( -\frac{\delta ^2}{2\sigma ^2\left( M\left( k \right) \right)} \right)}\frac{d\delta}{\sqrt{2\pi}\sigma \left( M\left( k \right) \right)}\nonumber \\ 
		& =\sqrt{\frac{2}{\pi}}\frac{\sigma \left( M\left( k \right) \right)}{\delta _c}\exp \left( -\frac{\delta _{c}^{2}}{2\sigma ^2\left( M\left( k \right) \right)} \right)~,
	\end{align}
As can be deduced from the above formula, the value of $\beta(M)$ is uniquely determined by the variance $\sigma^2\left(M\left(k\right)\right) $, which is assumed to be coarse-grained variance smoothed on a scale of $R=k^{-1}$. During the radiation dominated era, it is given by the following expression \cite{Blais:2002gw}:
	\begin{equation}\label{EQ36}
		\sigma^2\left( M\left( k \right) \right) =\frac{16}{81}\int_0^{\infty}{d\ln q}\left( \frac{q}{k} \right) ^4W\left( \frac{q}{k} \right) ^2P _{S}\left( q \right)~,
	\end{equation}
where $P_{S}(q)$ is the power spectrum of curvature perturbation and $ W(x) $ is smoothing window function \cite{Tokeshi:2020tjq, Young:2019osy}. In this study, we use the Gaussian form: $ W(x)=\exp(-x^2/2)$. Moreover, In general, in the radiation dominated period, the threshold density $ \delta _c=c_{s}^{2}=1/3\approx 0.33 $. However, some studies have shown that the threshold density $ \delta _c $ can be between 0.33 and 0.66 \cite{Musco:2020jjb, Sato-Polito:2019hws}.

\begin{figure}[htbp]
	\centering
	\includegraphics[height=6.0cm,width=9.75cm]{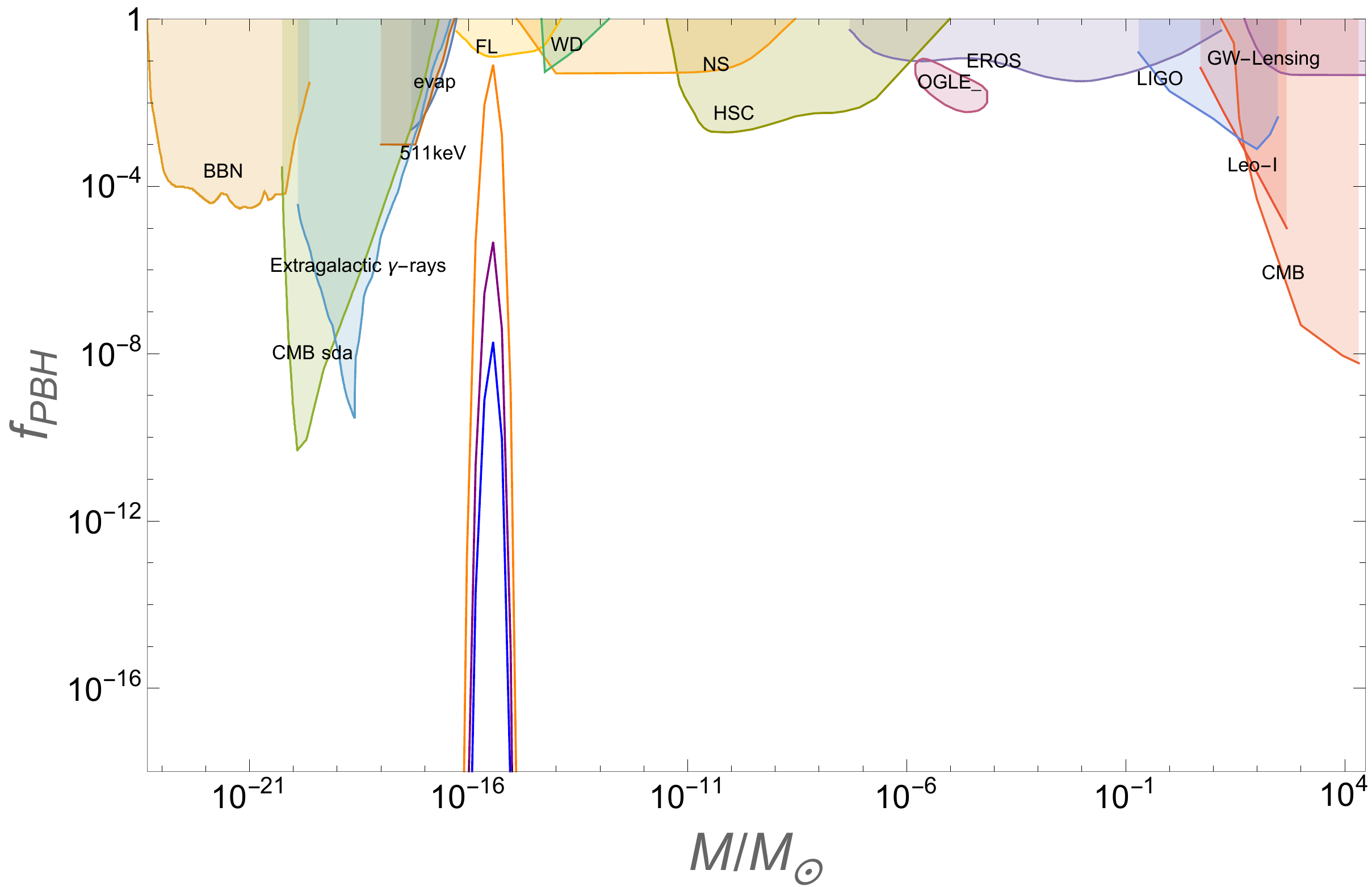}
	\caption{We plot $f_{PBH}$ for the potential (\ref{solopotential}) with $p=2$ using different threshold densities $\delta _c$, where the yellow line corresponds to $\delta _c=0.41$, the purple line corresponds to $\delta _c=0.46$, and the blue line corresponds to $\delta _c=0.486$. Our results are consistent with the constraints from current observations \cite{Siraj:2021arp}. The specific parameters are shown in Table \ref{Table1}. Constraints are obtained from the publicly available {\bf Python} code \href{https://github.com/bradkav/PBHbounds}{\bf PBHbounds} \cite{bradley_j_kavanagh_2019_3538999}.}
	\label{FIG4}
\end{figure}
\begin{table}
	\centering
	\begin{tabular}{|c|c|c|c|}
		\hline\rule{0pt}{10pt}
	$ k_{form}/Mpc^{-1} $ & $ M_{PBH} $	& $ \delta _c $  &  $ f_{PBH} $  \\
		\hline\rule{0pt}{10pt}
	$ 1.1539\times 10^{14} $ &	$ 3.68\times 10^{-16}M_{\odot} $	& 0.41 & $ 8.0455\times 10^{-2} $  \\
		\hline\rule{0pt}{10pt}
	$ 1.1539\times 10^{14} $ &	$ 3.68\times 10^{-16}M_{\odot} $	& 0.46 & $ 4.6573\times 10^{-6} $  \\
		\hline\rule{0pt}{10pt}
	$ 1.1539\times 10^{14} $ &	$ 3.68\times 10^{-16}M_{\odot} $	& 0.486 & $ 1.8905\times 10^{-8} $  \\
		\hline
	\end{tabular}\\
	\caption{$f_{PBH}$ for potential (\ref{solopotential}) with $p=2$ for different threshold densities $\delta _c$.}
\label{Table1}
\end{table}

In Fig. \ref{FIG4}, we numerically plot the abundances of $f_{PBH}$ with different threshold energy $\delta_c$, taking a solo-bumpy potential with $p=2$ as an example. Values of $\delta_c$ are as presented in Table \ref{Table1}. We find that the larger the $\delta_c$ value, the smaller the abundance, and this is easy to understand: the larger the threshold energy, the more difficult it is to form black holes. For a small value of $\delta_c=0.41$, the abundance of PBHs can reach approximately $10\%$ of dark matter. The mass range of the PBHs formed is approximately $10^{-15}M_\odot$, namely the asteroid mass.

Moreover, we also considered the case where the PBHs formed in the early universe were in an ellipsoidal collapse instead of spherical collapse \cite{Arbey:2020yzj, Mirbabayi:2019uph}. The difference between these two collapse models is that the threshold density for forming PBHs is different. compared with the spherical collapse, the PBHs formed by the ellipsoidal collapse will increase the ellipticity of the formed PBHs, which will lead to the correction of the threshold density. In the following, we uniformly use PBHs/e-PBHs to represent the primordial black hole formed by spherical/ellipsoidal collapse and $\delta _c$/$\delta_{ec}$ to describe the threshold density of PBHs/e-PBHs, respectively. According to \cite{Kuhnel:2016exn, Sheth:1999su} we have the following relationship:
	\begin{equation}
		\delta _{ec}=\delta _c\left[ 1+\kappa \left( \frac{\sigma ^2}{\delta _{c}^{2}} \right) ^{\nu} \right]~,
	\end{equation}
where $\kappa=9/\sqrt{10\pi}$ and $\nu=1/2$. From Eq. (\ref{equation31}), we can easily obtain the abundance of PBHs and e-PBHs with a given $ \delta _c $. The calculation shows that the abundance of e-PBHs is lower than that of PBHs, owing to the difference in their threshold densities.
\par
\begin{figure}[h]
	\subfigure{
	\includegraphics[height=4.5cm,width=7cm]{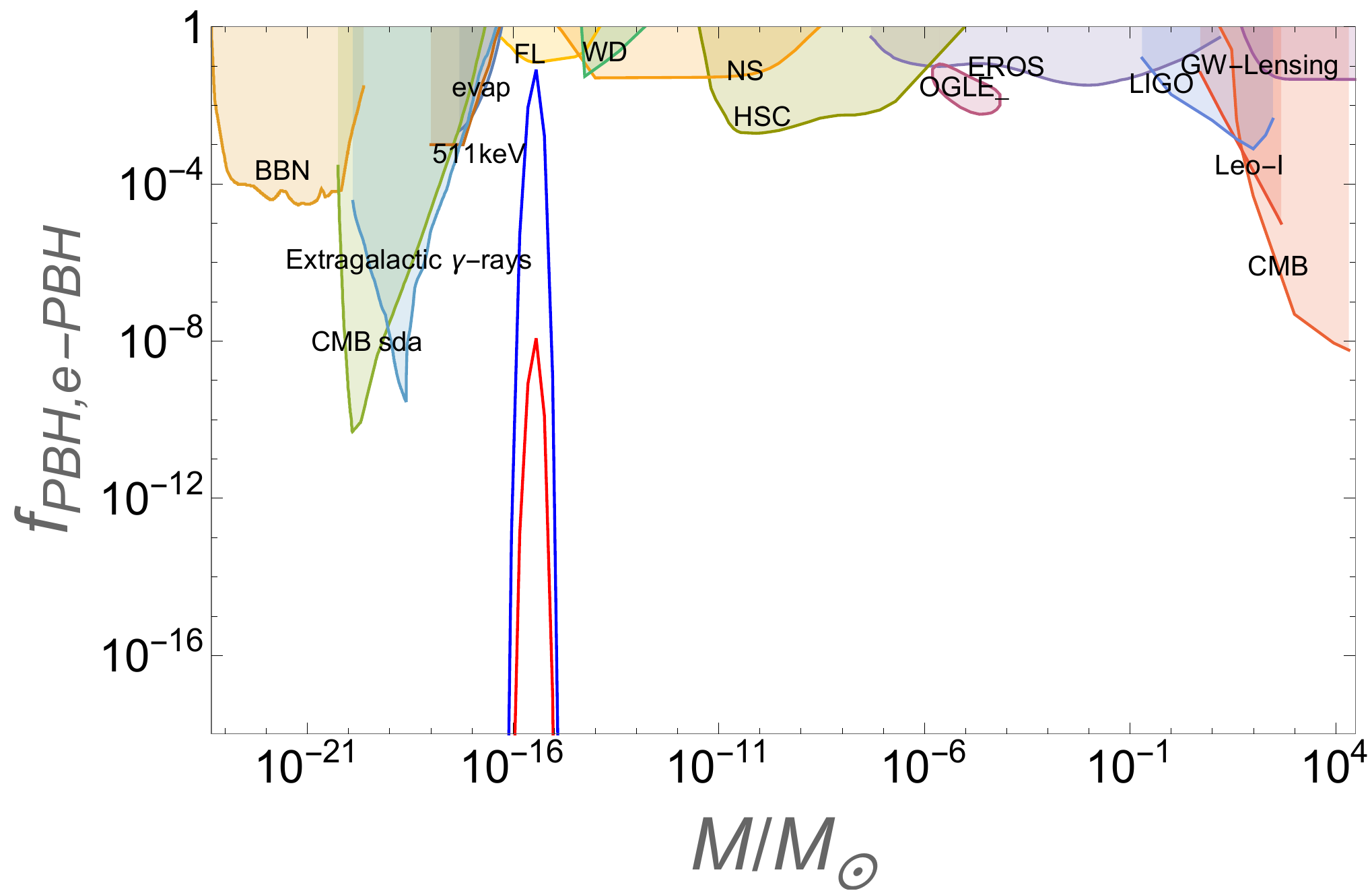}
	}
	\subfigure{
	\includegraphics[height=4.5cm,width=7cm]{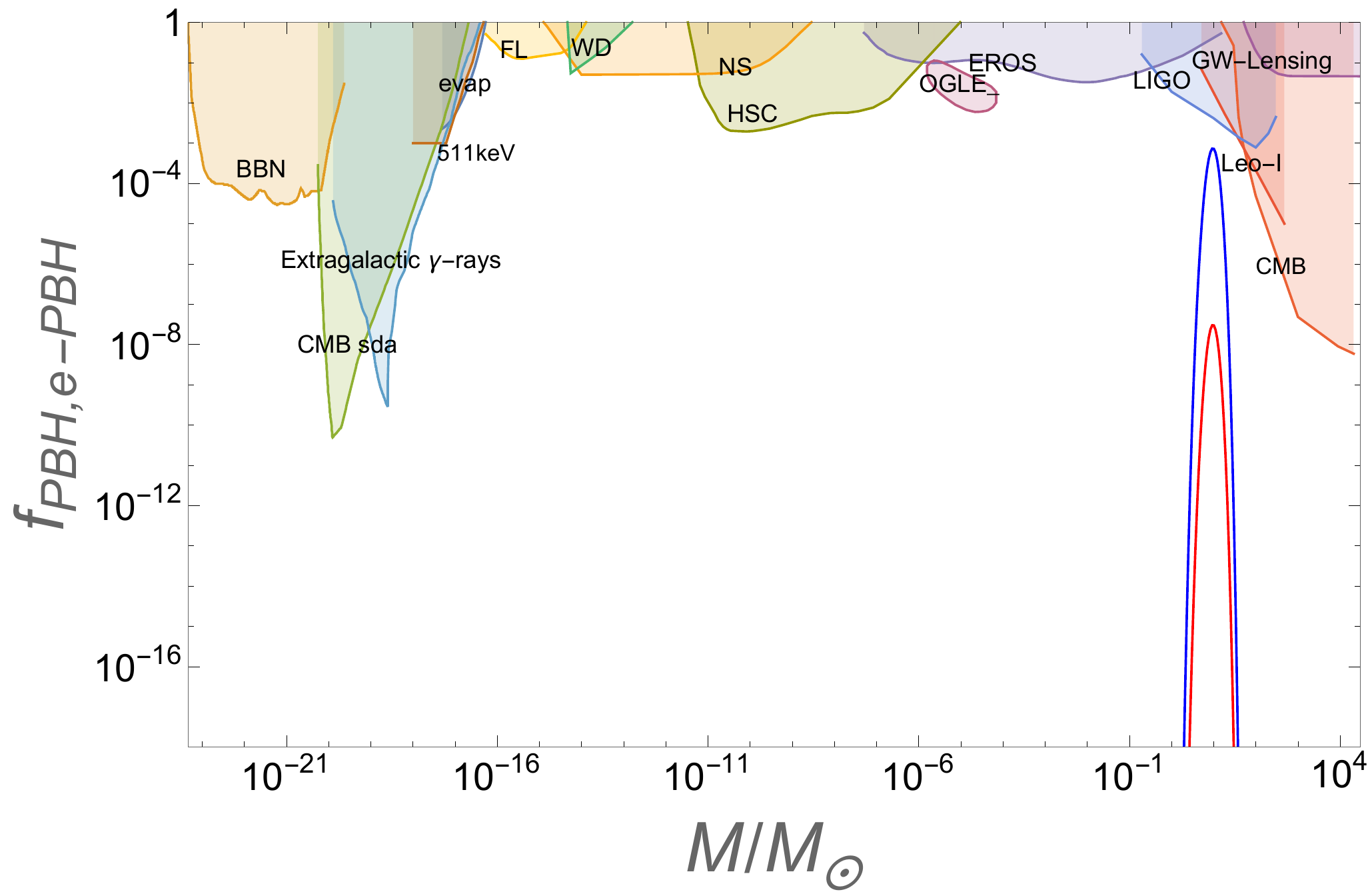}
	}
	\caption{The figures above show the constraints on primordial black holes acting as dark matter, in which the colored region is excluded by various observations. The blue lines correspond to $f_{PBH}$, and the red lines correspond to $f_{e-PBH}$. The left figure is for potential (\ref{solopotential}) with $p=2$ and $\delta _c=0.41$, while the right figure is for potential (\ref{solopotential}) with $p=2/3$ and $\delta _c=0.33$. Constraints are obtained from the publicly available {\bf Python} code \href{https://github.com/bradkav/PBHbounds}{\bf PBHbounds} \cite{bradley_j_kavanagh_2019_3538999}.}
	\label{FIG5}
\end{figure}

\begin{table}
	\centering
	\begin{tabular}{|c|c|c|c|c|}
		\hline\rule{0pt}{10pt}
	$ k_{form}/Mpc^{-1} $ & $ M_{PBH} $	& $ \delta _c $  &  $ f_{PBH} $ & $ f_{e-PBH} $ \\
		\hline\rule{0pt}{10pt}
	$ 1.1539\times 10^{14} $ &	$ 3.68\times 10^{-16}M_{\odot} $	& 0.41 & $ 8.0455\times 10^{-2} $ &  $ 1.1781\times 10^{-8} $  \\
		\hline
	\end{tabular}\\
	\caption{$f_{PBH}$ and $f_{e-PBH}$ for potential (\ref{solopotential}) with $p=2$ and $\delta _c=0.41$.}
\label{Table2}
\end{table}
\par
\begin{table}
	\centering
	\begin{tabular}{|c|c|c|c|c|}
		\hline\rule{0pt}{10pt}
	$ k_{form}/Mpc^{-1} $ & $ M_{PBH} $	& $ \delta _c $  &  $ f_{PBH} $ & $ f_{e-PBH} $ \\
		\hline\rule{0pt}{10pt}
		$ 6.0256\times 10^{5} $ & $ 9.2437M_{\odot} $	& 0.33 & $ 7.2256\times 10^{-4} $ &  $ 2.9602\times 10^{-8} $  \\
		\hline
	\end{tabular}\\
	\caption{$f_{PBH}$ and $f_{e-PBH}$ for potential (\ref{solopotential}) with $p=2/3$ and $\delta _c=0.33$.}
	\label{Table3}
\end{table}
In Fig. \ref{FIG5}, We plot the abundances of $f_{PBH}$ and $f_{e-PBH}$ from solo-bumpy potential (\ref{solopotential}), with parameter sets \{$p=2$, $\delta _c=0.41$\} and \{$p=2/3$, $\delta _c=0.33$\} respectively, as well as the constraints from various observations. For $p=2$ case, the fraction of PBHs could reach $O(10\%)$ at the mass range of $\sim 10^{-15}M_\odot$ (asteroid mass), while for $p=2/3$ case, the fraction is also nearly $O(0.1\%)$ at the mass range of $\sim 1M_\odot$ (solar mass). The specific parameters are presented in Table. \ref{Table2} and Table. \ref{Table3}. 
 
\begin{figure}[htbp]
	\subfigure{
	\includegraphics[height=4.5cm,width=7cm]{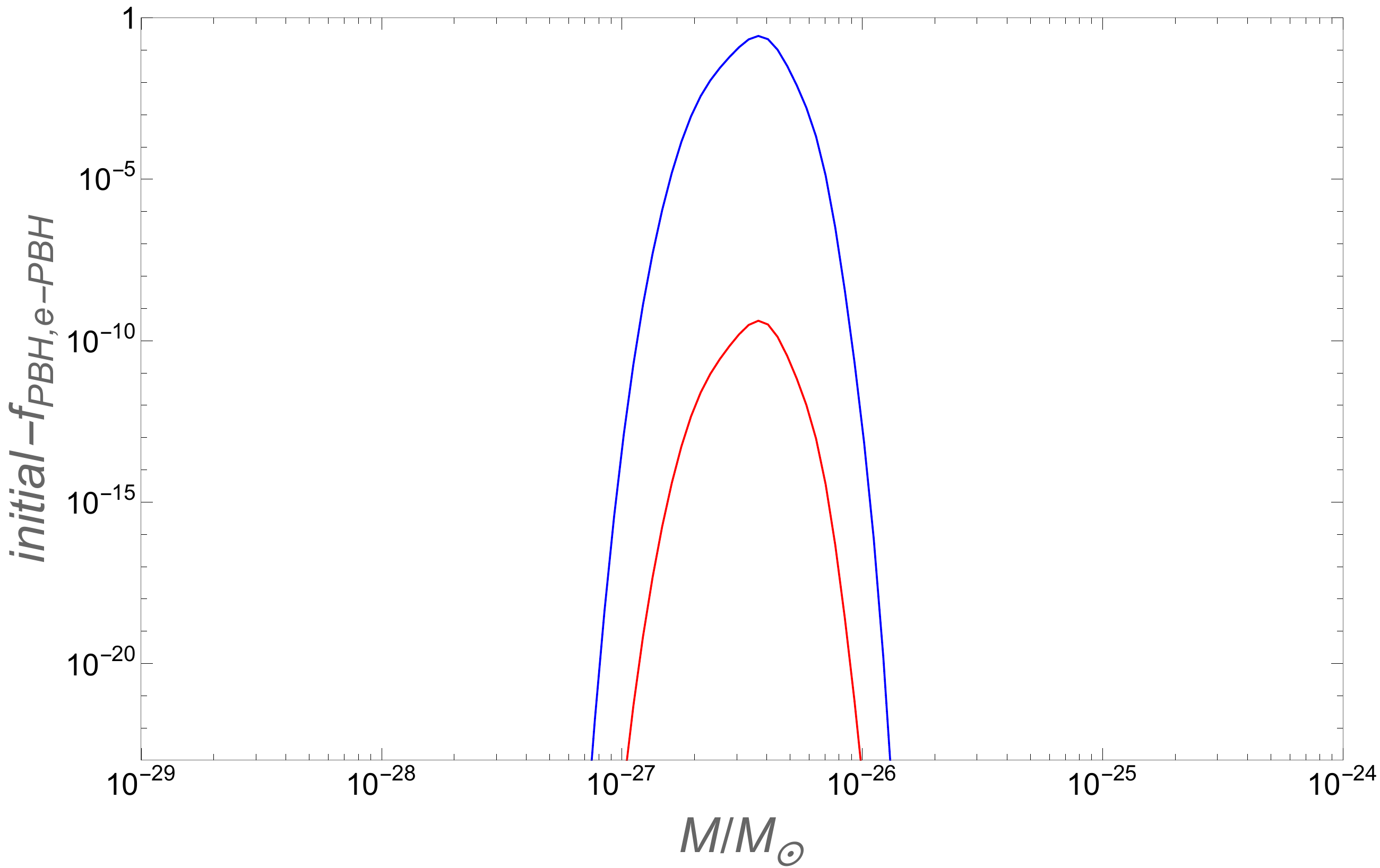}
}
\subfigure{
	\includegraphics[height=4.5cm,width=7cm]{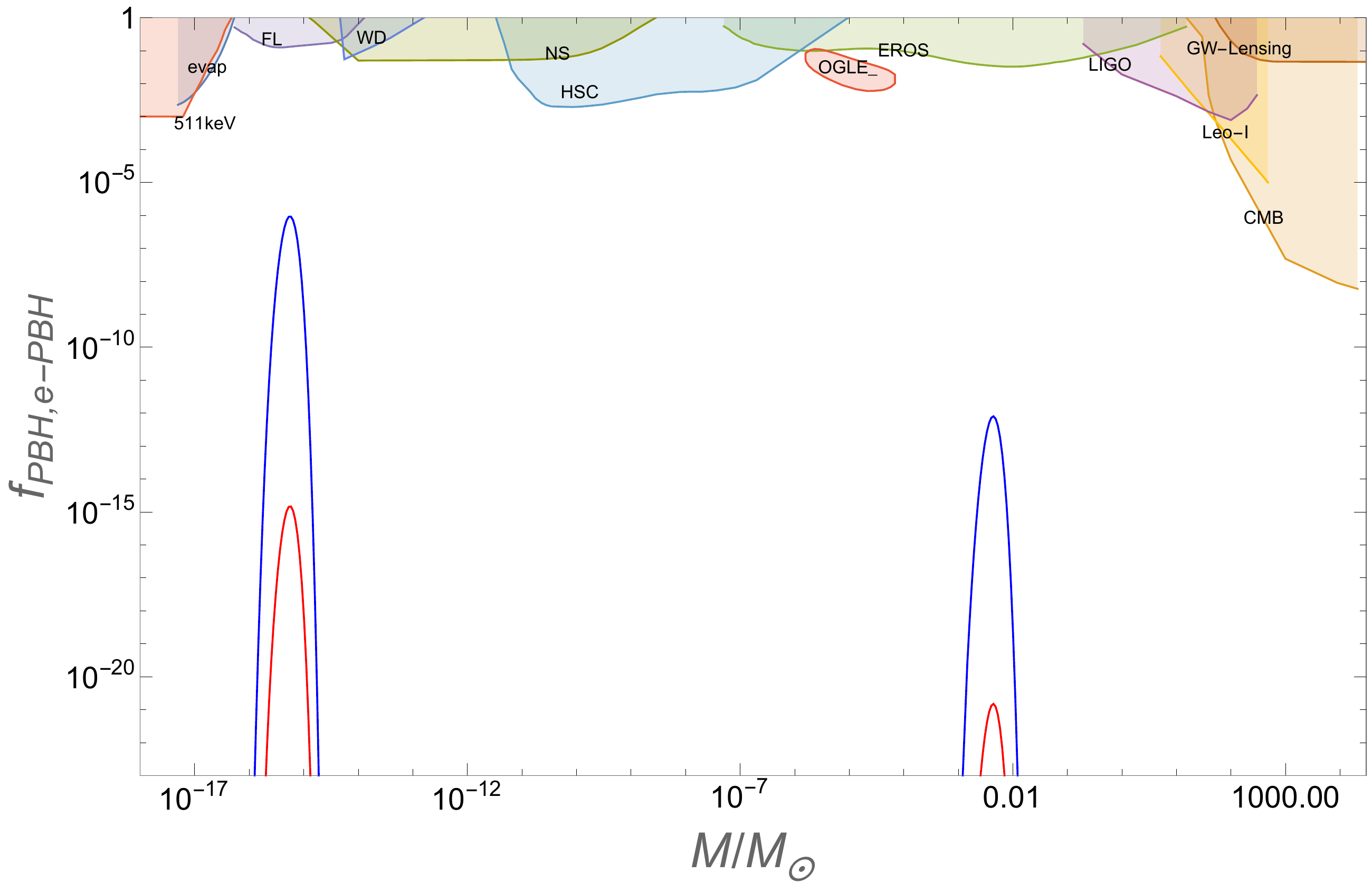}
}
	\caption{The figures above present the constraints on primordial black holes acting as dark matter, in which the colored region is excluded by various observations. The plot is for potential (\ref{multipotential}) and $\delta _c=0.465$. Potential (\ref{multipotential}) can form three mass ranges of PBHs. Because the PBHs in one of the mass ranges is negligible, it cannot be constrained by existing observations; hence, we plot it separately and use $ initial-f_{PBH} $ to represent it (left figure). The other two mass ranges of PBHs can still be constrained by existing observations (right figure). The blue line correspond to $f_{PBH}$ while the red lines correspond to $f_{e-PBH}$. From left to right, the masses of PBHs are $ 3.6975\times 10^{-27}M_{\odot} $, $ 5.8601\times 10^{-16}M_{\odot} $ and $ 3.6975\times 10^{-3}M_{\odot} $ respectively. Constraints are obtained from the publicly available {\bf Python} code \href{https://github.com/bradkav/PBHbounds}{\bf PBHbounds} \cite{bradley_j_kavanagh_2019_3538999}.}
	\label{FIG6}
\end{figure}

Similarly, we plot the abundances of $f_{PBH}$ and $f_{e-PBH}$ from the multi-bumpy potential (\ref{multipotential}) (with parameter sets (\ref{parameter-multi})) in Fig. \ref{FIG6}.
As can be deduced from the figure, PBHs can be formed in three mass ranges. The leftmost peak in Fig. \ref{FIG6} can have a large amount of PBHs, the abundance of which can reach $ 27\% $ of dark matter; however, because the mass range of PBHs is negligible (around $3.6975\times 10^{-27}M_{\odot}$), if one considers effects of Hawking radiation, such PBHs may not have chance to live till today. Nonetheless, it may have some non-trivial effects in the evolution of the Universe, which we will mention in the next section. However, the other two peaks refer to PBHs of $10^{-15}M_\odot$ (asteroid mass) and $10^{-3}M_\odot$ (planet mass), which can live till today and may act as dark matter. The specific values are presented in Table. \ref{Table4}. 
\begin{table}
	\centering
	\par
	\begin{tabular}{|c|c|c|c|c|}
		\hline\rule{0pt}{10pt}
		$ k_{form}/Mpc^{-1} $	& $ M_{PBH} $	& $ \delta _c $  &  $ f_{PBH} $ & $ f_{e-PBH} $\\
		\hline\rule{0pt}{10pt}
		$ 3.1548\times 10^{19} $ & $ 3.6975\times 10^{-27}M_{\odot} $	& 0.465 & $ 2.7181\times 10^{-1} $ & $ 4.0947\times 10^{-10} $\\
		\hline\rule{0pt}{10pt}
	$ 	7.9245\times 10^{13} $ & $ 5.8601\times 10^{-16}M_{\odot} $	& 0.465 & $ 9.0930\times 10^{-7} $ & $ 1.4448\times 10^{-15} $\\
		\hline\rule{0pt}{10pt}
		$ 3.1548\times 10^{7} $ & $ 3.6975\times 10^{-3}M_{\odot} $	& 0.465 & $ 4.9587\times 10^{-13} $ & $ 8.3559\times 10^{-22} $\\
		\hline
	\end{tabular}\\
	\caption{$f_{PBH}$ and $f_{e-PBH}$ for potential (\ref{multipotential}) and $\delta _c=0.465$.}
\label{Table4}
\end{table}

\section{scalar-induced gravitational waves}
\label{sec5}
Scalar-induced gravitational waves (SIGWs) are generated by linear scalar perturbations of the second-order \cite{Ananda:2006af,Baumann:2007zm}. After inflation, the universe should go through reheating epoch and will be dominated by radiation.
The energy density of the SIGWs in the radiation-dominated (RD) era can be estimated by \cite{Lu:2019sti,Cai:2018dig,Pi:2020otn,Chen:2021nio,Solbi:2021rse,Chen:2021nxo}
\begin{equation}\label{gwp}
	\Omega_{GW}(k,\tau_{c})=\frac{1}{6}\int_{0}^{\infty}dv\int_{\left|1-v\right|}^{1+v}du\left[\frac{4v^{2}
		-\left(1-u^{2}+v^{2}\right)^2}{4uv}\right]^{2}P_{S}(ku)P_{S}(kv)\overline{I_{RD}^{2}(u,v)}
\end{equation}
where the kernel function in the RD era takes the form \cite{Kohri:2018awv}
\begin{equation}\label{IRD}
	\overline{I_{RD}^{2}(u,v)}=\frac{9\left(u^{2}+v^{2}-3\right)^{2}}{32u^{6}v^{6}}\left\{ \left[-4uv+\left(u^{2}+v^{2}-3\right)\ln\left|\frac{3-(u+v)^{2}}{3-(u-v)^{2}}\right|\right]^{2}+\pi^{2}(u^{2}+v^{2}-3)^2\Theta(u+v-\sqrt{3})\right\}\,,
\end{equation}
where $\Theta$ denotes the Heaviside theta function.

\begin{figure}
	\centering
	\includegraphics[height=7.0cm,width=10.75cm]{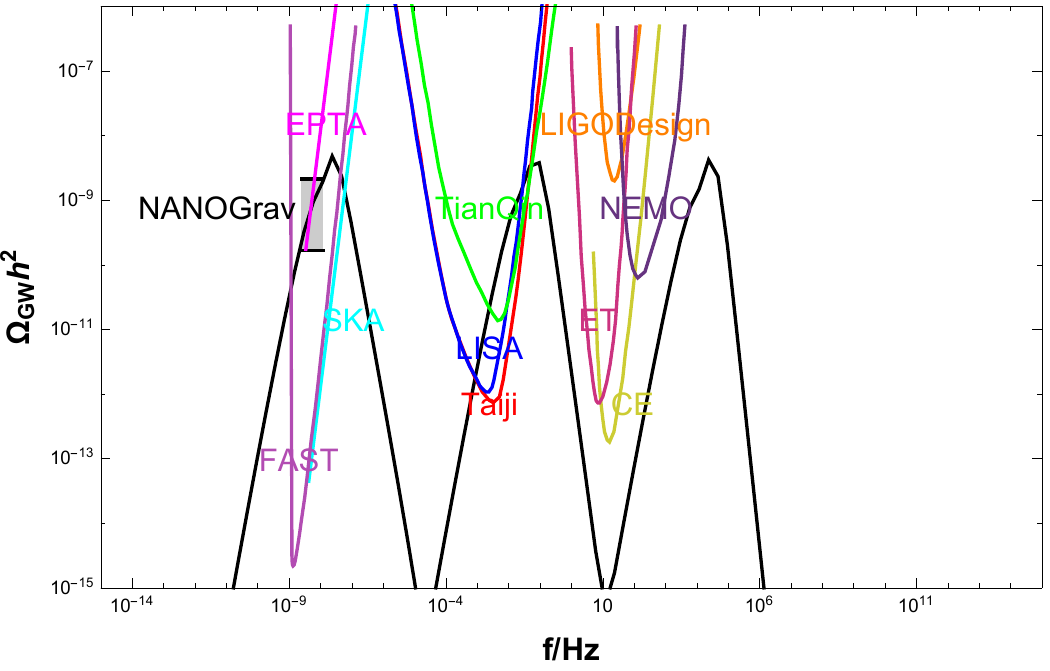}\\
	\caption{The energy spectra of scalar-induced GWs from multi-bumpy potential. The dashed region represents observational constrains by the NANOGrav data \cite{NANOGrav:2020bcs}. The colored lines are designed sensitivity curves of current and future gravitational wave detectors including FAST \cite{Nan:2011um}, SKA \cite{Janssen:2014dka}, EPTA \cite{Desvignes:2016yex}, TianQin \cite{TianQin:2015yph}, LISA \cite{LISA:2017pwj}, Taiji \cite{Hu:2017mde}, NEMO \cite{Ackley:2020atn},Einstein Telescope(ET) \cite{Hild:2010id},Cosmic Explorer(CE) \cite{LIGOScientific:2016wof}, LIGO Design \cite{LIGOScientific:2019vic}.  }\label{GWfig}
\end{figure}

The current energy density of the induced GWs can be related to their value in the RD era \cite{Inomata:2018epa}
\begin{equation}\label{gwc}
	\Omega_{GW}(k)h^{2}=0.83\left(\frac{g_{c}}{10.75}\right)^{-1/3}\Omega_{r,0}h^{2}\Omega_{GW}(k,\tau_{c})\,,
\end{equation}
where $\Omega_{r,0}h^{2}\approx4.2\times10^{-5}$ is the current energy density parameter of
radiation and $g_{c}\approx106.75$ is the effective degrees of freedom of the total radiation energy density at time $\tau_{c}$ \cite{Husdal:2016haj}. We convert to frequency $f$  from the comoving wavenumber $k$ via the relationship $k=2\pi f$, namely
\begin{equation}\label{dd}
	f=1.546\times10^{-15}\left(\frac{k}{\mathrm{Mpc}^{-1}}\right)\mathrm{Hz}\,,
\end{equation}
thus, we can plot the energy spectra of the SIGWs against its frequency $f$ in  Fig. \ref{GWfig}. We consider the SIGWs sourced from the multi-bumpy potential. Fig. \ref{GWfig} presents that three peaks of SIGWs that be generated correspondingly by the three bumps of inflation potential.  Interestingly, the curve of first peak at $f\sim [10^{-9},10^{-8}]$ Hz lies in the allowed region of the observational constraints by the NANOGrav result \cite{NANOGrav:2020bcs}.

\section{Conclusion}
\label{sec6}
In this study, we investigated the formation of PBHs from an inflation model with bumpy features on its potential. Although the case of a potential with one bump has been discussed in the literature, we discussed the extended case of multiple bumps, which can generate PBHs at different mass ranges. This will enlarge the detectability of PBHs from the coming development of astronomical observations. Specifically, we considered the power-law potential as the basic potential and add one or several bumps of Gaussian type, which makes the inflation roll from the slow-roll stage to the USR-like stage. With the potential, we constructed the power spectrum with single or multiple peaks in small scales, while keeping the large scale power spectrum consistent with CMB data. We numerically calculated the abundances of PBHs (fraction to dark matter) at the mass range given by the solo-bumpy potential, as well as three mass ranges given by the multi-bumpy potential. Owing to these peaks, PBHs can be formed at different mass ranges, including the asteroid mass range ($10^{-16}-10^{-14}M_\odot$), planet mass range ($10^{-6}-10^{-3}M_\odot$), and solar mass range (around $1M_\odot$), some of which can reach significant abundance. We also demonstrated that the abundance has a negative correlation with the threshold energy density $\delta_c$. We confronted the results to the current observational data, and all the abundances were determined to be consistent with the data constraints. Although we take three bumps for simplicity, in principle from such kind of potential, abundances at other mass ranges are also expected.   

Moreover, PBHs formed in the early universe can also be formed by ellipsoidal collapse. Therefore, in addition to considering the formation of spherical collapsed PBHs, we also considered the formation of PBHs under ellipsoidal collapse, owing to the correction brought by the ellipticity of the latter; hence, according to the Press-Schecther formalism, the abundance will be lower. We obtained the abundances of PBHs in the two types of collapse at all the mass ranges discussed above, and determined the numerical results consistent with the analytical analysis.

Considering the generation of scalar-induced gravitational waves, we have drawn the SIGWs generated by three bumps of inflation potential. Interestingly, the curve of the first peak lies within the area allowed by the NANOGrav observation constraint. 

Before ending, we briefly discuss the evaporation of the formed PBHs.  According to Hawking's theory of black hole radiation, these black holes will inevitably emit virtual particles and continuously reduce their mass until they vanish, which is known as Hawking evaporation \cite{Hawking:1974rv}. It has been pointed out that, considering the age of the universe, PBHs with initial mass less than $10^{15}\text{g}$ $(\sim 10^{-18}\text{M}_\odot)$ has been completely evaporated today. Therefore, for the multi-bumpy inflation model in our analysis, the PBHs standing for the leftmost peak in Fig. \ref{FIG6} will actually be vanish, and this cannot explain the dark matter today.  However, it does not imply that they are not important at all. Actually, they may still have a significant impact on the early universe, such as the process of Big Bang Nucleosynthesis, reheating, baryogenesis, etc \cite{Hawking:1974rv, Carr:1976zz, Baumann:2007yr, DeLuca:2021oer}. With the further development of observational techniques, we may be able to detect the traces left by the PBHs of this type, to find more evidence of their existence. Meanwhile, for other mass ranges, the PBHs are hardly evaporated to date, and can therefore act as dark matter. More details about the influence of the PBHs in our model on these other aspects of our universe, although very interesting, are postponed to an upcoming paper.

\begin{acknowledgements}
T. Q. thanks Yun-Gui Gong and Shi Pi for useful discussions. This work was supported by the National Natural Science Foundation of China under Grants No.~11653002 and No.~11875141.  J.S. was partially supported by the Fundamental Research Funds for the Central Universities (Innovation Funded Projects) under Grants No.~2020CXZZ105.
\end{acknowledgements}

\bibliographystyle{apsrev4-1}
\bibliography{PBH20210709}

\begin{thebibliography}{138}%
\makeatletter
\providecommand \@ifxundefined [1]{%
 \@ifx{#1\undefined}
}%
\providecommand \@ifnum [1]{%
 \ifnum #1\expandafter \@firstoftwo
 \else \expandafter \@secondoftwo
 \fi
}%
\providecommand \@ifx [1]{%
 \ifx #1\expandafter \@firstoftwo
 \else \expandafter \@secondoftwo
 \fi
}%
\providecommand \natexlab [1]{#1}%
\providecommand \enquote  [1]{``#1''}%
\providecommand \bibnamefont  [1]{#1}%
\providecommand \bibfnamefont [1]{#1}%
\providecommand \citenamefont [1]{#1}%
\providecommand \href@noop [0]{\@secondoftwo}%
\providecommand \href [0]{\begingroup \@sanitize@url \@href}%
\providecommand \@href[1]{\@@startlink{#1}\@@href}%
\providecommand \@@href[1]{\endgroup#1\@@endlink}%
\providecommand \@sanitize@url [0]{\catcode `\\12\catcode `\$12\catcode
  `\&12\catcode `\#12\catcode `\^12\catcode `\_12\catcode `\%12\relax}%
\providecommand \@@startlink[1]{}%
\providecommand \@@endlink[0]{}%
\providecommand \url  [0]{\begingroup\@sanitize@url \@url }%
\providecommand \@url [1]{\endgroup\@href {#1}{\urlprefix }}%
\providecommand \urlprefix  [0]{URL }%
\providecommand \Eprint [0]{\href }%
\providecommand \doibase [0]{http://dx.doi.org/}%
\providecommand \selectlanguage [0]{\@gobble}%
\providecommand \bibinfo  [0]{\@secondoftwo}%
\providecommand \bibfield  [0]{\@secondoftwo}%
\providecommand \translation [1]{[#1]}%
\providecommand \BibitemOpen [0]{}%
\providecommand \bibitemStop [0]{}%
\providecommand \bibitemNoStop [0]{.\EOS\space}%
\providecommand \EOS [0]{\spacefactor3000\relax}%
\providecommand \BibitemShut  [1]{\csname bibitem#1\endcsname}%
\let\auto@bib@innerbib\@empty
\bibitem [{\citenamefont {Abbott}\ \emph {et~al.}(2016)\citenamefont {Abbott}
  \emph {et~al.}}]{Abbott:2016blz}%
  \BibitemOpen
  \bibfield  {author} {\bibinfo {author} {\bibfnamefont {B.~P.}\ \bibnamefont
  {Abbott}} \emph {et~al.} (\bibinfo {collaboration} {LIGO Scientific,
  Virgo}),\ }\href {\doibase 10.1103/PhysRevLett.116.061102} {\bibfield
  {journal} {\bibinfo  {journal} {Phys. Rev. Lett.}\ }\textbf {\bibinfo
  {volume} {116}},\ \bibinfo {pages} {061102} (\bibinfo {year} {2016})},\
  \Eprint {http://arxiv.org/abs/1602.03837} {arXiv:1602.03837 [gr-qc]}
  \BibitemShut {NoStop}%
\bibitem [{\citenamefont {Abbott}\ \emph
  {et~al.}(2017{\natexlab{a}})\citenamefont {Abbott} \emph
  {et~al.}}]{TheLIGOScientific:2017qsa}%
  \BibitemOpen
  \bibfield  {author} {\bibinfo {author} {\bibfnamefont {B.~P.}\ \bibnamefont
  {Abbott}} \emph {et~al.} (\bibinfo {collaboration} {LIGO Scientific,
  Virgo}),\ }\href {\doibase 10.1103/PhysRevLett.119.161101} {\bibfield
  {journal} {\bibinfo  {journal} {Phys. Rev. Lett.}\ }\textbf {\bibinfo
  {volume} {119}},\ \bibinfo {pages} {161101} (\bibinfo {year}
  {2017}{\natexlab{a}})},\ \Eprint {http://arxiv.org/abs/1710.05832}
  {arXiv:1710.05832 [gr-qc]} \BibitemShut {NoStop}%
\bibitem [{\citenamefont {Akiyama}\ \emph {et~al.}(2019)\citenamefont {Akiyama}
  \emph {et~al.}}]{Akiyama:2019cqa}%
  \BibitemOpen
  \bibfield  {author} {\bibinfo {author} {\bibfnamefont {K.}~\bibnamefont
  {Akiyama}} \emph {et~al.} (\bibinfo {collaboration} {Event Horizon
  Telescope}),\ }\href {\doibase 10.3847/2041-8213/ab0ec7} {\bibfield
  {journal} {\bibinfo  {journal} {Astrophys. J. Lett.}\ }\textbf {\bibinfo
  {volume} {875}},\ \bibinfo {pages} {L1} (\bibinfo {year} {2019})},\ \Eprint
  {http://arxiv.org/abs/1906.11238} {arXiv:1906.11238 [astro-ph.GA]}
  \BibitemShut {NoStop}%
\bibitem [{\citenamefont {Hawking}(1971)}]{Hawking:1971ei}%
  \BibitemOpen
  \bibfield  {author} {\bibinfo {author} {\bibfnamefont {S.}~\bibnamefont
  {Hawking}},\ }\href@noop {} {\bibfield  {journal} {\bibinfo  {journal} {Mon.
  Not. Roy. Astron. Soc.}\ }\textbf {\bibinfo {volume} {152}},\ \bibinfo
  {pages} {75} (\bibinfo {year} {1971})}\BibitemShut {NoStop}%
\bibitem [{\citenamefont {Carr}\ and\ \citenamefont
  {Hawking}(1974)}]{Carr:1974nx}%
  \BibitemOpen
  \bibfield  {author} {\bibinfo {author} {\bibfnamefont {B.~J.}\ \bibnamefont
  {Carr}}\ and\ \bibinfo {author} {\bibfnamefont {S.~W.}\ \bibnamefont
  {Hawking}},\ }\href@noop {} {\bibfield  {journal} {\bibinfo  {journal} {Mon.
  Not. Roy. Astron. Soc.}\ }\textbf {\bibinfo {volume} {168}},\ \bibinfo
  {pages} {399} (\bibinfo {year} {1974})}\BibitemShut {NoStop}%
\bibitem [{\citenamefont {Khlopov}(2010)}]{Khlopov_2010}%
  \BibitemOpen
  \bibfield  {author} {\bibinfo {author} {\bibfnamefont {M.~Y.}\ \bibnamefont
  {Khlopov}},\ }\href {\doibase 10.1088/1674-4527/10/6/001} {\bibfield
  {journal} {\bibinfo  {journal} {Research in Astronomy and Astrophysics}\
  }\textbf {\bibinfo {volume} {10}},\ \bibinfo {pages} {495} (\bibinfo {year}
  {2010})}\BibitemShut {NoStop}%
\bibitem [{\citenamefont {Guth}(1981)}]{Guth:1980zm}%
  \BibitemOpen
  \bibfield  {author} {\bibinfo {author} {\bibfnamefont {A.~H.}\ \bibnamefont
  {Guth}},\ }\href {\doibase 10.1103/PhysRevD.23.347} {\bibfield  {journal}
  {\bibinfo  {journal} {Phys. Rev. D}\ }\textbf {\bibinfo {volume} {23}},\
  \bibinfo {pages} {347} (\bibinfo {year} {1981})}\BibitemShut {NoStop}%
\bibitem [{\citenamefont {Linde}(1982)}]{Linde:1981mu}%
  \BibitemOpen
  \bibfield  {author} {\bibinfo {author} {\bibfnamefont {A.~D.}\ \bibnamefont
  {Linde}},\ }\href {\doibase 10.1016/0370-2693(82)91219-9} {\bibfield
  {journal} {\bibinfo  {journal} {Phys. Lett. B}\ }\textbf {\bibinfo {volume}
  {108}},\ \bibinfo {pages} {389} (\bibinfo {year} {1982})}\BibitemShut
  {NoStop}%
\bibitem [{\citenamefont {Starobinsky}(1980)}]{Starobinsky:1980te}%
  \BibitemOpen
  \bibfield  {author} {\bibinfo {author} {\bibfnamefont {A.~A.}\ \bibnamefont
  {Starobinsky}},\ }\href {\doibase 10.1016/0370-2693(80)90670-X} {\bibfield
  {journal} {\bibinfo  {journal} {Phys. Lett. B}\ }\textbf {\bibinfo {volume}
  {91}},\ \bibinfo {pages} {99} (\bibinfo {year} {1980})}\BibitemShut {NoStop}%
\bibitem [{\citenamefont {Riotto}(2003)}]{Riotto:2002yw}%
  \BibitemOpen
  \bibfield  {author} {\bibinfo {author} {\bibfnamefont {A.}~\bibnamefont
  {Riotto}},\ }\href@noop {} {\bibfield  {journal} {\bibinfo  {journal} {ICTP
  Lect. Notes Ser.}\ }\textbf {\bibinfo {volume} {14}},\ \bibinfo {pages} {317}
  (\bibinfo {year} {2003})},\ \Eprint {http://arxiv.org/abs/hep-ph/0210162}
  {arXiv:hep-ph/0210162} \BibitemShut {NoStop}%
\bibitem [{\citenamefont {Cai}\ \emph {et~al.}(2020)\citenamefont {Cai},
  \citenamefont {Guo}, \citenamefont {Liu}, \citenamefont {Liu},\ and\
  \citenamefont {Yang}}]{Cai:2019bmk}%
  \BibitemOpen
  \bibfield  {author} {\bibinfo {author} {\bibfnamefont {R.-G.}\ \bibnamefont
  {Cai}}, \bibinfo {author} {\bibfnamefont {Z.-K.}\ \bibnamefont {Guo}},
  \bibinfo {author} {\bibfnamefont {J.}~\bibnamefont {Liu}}, \bibinfo {author}
  {\bibfnamefont {L.}~\bibnamefont {Liu}}, \ and\ \bibinfo {author}
  {\bibfnamefont {X.-Y.}\ \bibnamefont {Yang}},\ }\href {\doibase
  10.1088/1475-7516/2020/06/013} {\bibfield  {journal} {\bibinfo  {journal}
  {JCAP}\ }\textbf {\bibinfo {volume} {06}},\ \bibinfo {pages} {013} (\bibinfo
  {year} {2020})},\ \Eprint {http://arxiv.org/abs/1912.10437} {arXiv:1912.10437
  [astro-ph.CO]} \BibitemShut {NoStop}%
\bibitem [{\citenamefont {Garcia-Bellido}\ \emph {et~al.}(1996)\citenamefont
  {Garcia-Bellido}, \citenamefont {Linde},\ and\ \citenamefont
  {Wands}}]{GarciaBellido:1996qt}%
  \BibitemOpen
  \bibfield  {author} {\bibinfo {author} {\bibfnamefont {J.}~\bibnamefont
  {Garcia-Bellido}}, \bibinfo {author} {\bibfnamefont {A.~D.}\ \bibnamefont
  {Linde}}, \ and\ \bibinfo {author} {\bibfnamefont {D.}~\bibnamefont
  {Wands}},\ }\href {\doibase 10.1103/PhysRevD.54.6040} {\bibfield  {journal}
  {\bibinfo  {journal} {Phys. Rev. D}\ }\textbf {\bibinfo {volume} {54}},\
  \bibinfo {pages} {6040} (\bibinfo {year} {1996})},\ \Eprint
  {http://arxiv.org/abs/astro-ph/9605094} {arXiv:astro-ph/9605094} \BibitemShut
  {NoStop}%
\bibitem [{\citenamefont {Drees}\ and\ \citenamefont
  {Erfani}(2012)}]{Drees:2011yz}%
  \BibitemOpen
  \bibfield  {author} {\bibinfo {author} {\bibfnamefont {M.}~\bibnamefont
  {Drees}}\ and\ \bibinfo {author} {\bibfnamefont {E.}~\bibnamefont {Erfani}},\
  }\href {\doibase 10.1088/1475-7516/2012/01/035} {\bibfield  {journal}
  {\bibinfo  {journal} {JCAP}\ }\textbf {\bibinfo {volume} {01}},\ \bibinfo
  {pages} {035} (\bibinfo {year} {2012})},\ \Eprint
  {http://arxiv.org/abs/1110.6052} {arXiv:1110.6052 [astro-ph.CO]} \BibitemShut
  {NoStop}%
\bibitem [{\citenamefont {Garcia-Bellido}\ and\ \citenamefont
  {Ruiz~Morales}(2017)}]{Garcia-Bellido:2017mdw}%
  \BibitemOpen
  \bibfield  {author} {\bibinfo {author} {\bibfnamefont {J.}~\bibnamefont
  {Garcia-Bellido}}\ and\ \bibinfo {author} {\bibfnamefont {E.}~\bibnamefont
  {Ruiz~Morales}},\ }\href {\doibase 10.1016/j.dark.2017.09.007} {\bibfield
  {journal} {\bibinfo  {journal} {Phys. Dark Univ.}\ }\textbf {\bibinfo
  {volume} {18}},\ \bibinfo {pages} {47} (\bibinfo {year} {2017})},\ \Eprint
  {http://arxiv.org/abs/1702.03901} {arXiv:1702.03901 [astro-ph.CO]}
  \BibitemShut {NoStop}%
\bibitem [{\citenamefont {Domcke}\ \emph {et~al.}(2017)\citenamefont {Domcke},
  \citenamefont {Muia}, \citenamefont {Pieroni},\ and\ \citenamefont
  {Witkowski}}]{Domcke:2017fix}%
  \BibitemOpen
  \bibfield  {author} {\bibinfo {author} {\bibfnamefont {V.}~\bibnamefont
  {Domcke}}, \bibinfo {author} {\bibfnamefont {F.}~\bibnamefont {Muia}},
  \bibinfo {author} {\bibfnamefont {M.}~\bibnamefont {Pieroni}}, \ and\
  \bibinfo {author} {\bibfnamefont {L.~T.}\ \bibnamefont {Witkowski}},\ }\href
  {\doibase 10.1088/1475-7516/2017/07/048} {\bibfield  {journal} {\bibinfo
  {journal} {JCAP}\ }\textbf {\bibinfo {volume} {07}},\ \bibinfo {pages} {048}
  (\bibinfo {year} {2017})},\ \Eprint {http://arxiv.org/abs/1704.03464}
  {arXiv:1704.03464 [astro-ph.CO]} \BibitemShut {NoStop}%
\bibitem [{\citenamefont {Ballesteros}\ and\ \citenamefont
  {Taoso}(2018)}]{Ballesteros:2017fsr}%
  \BibitemOpen
  \bibfield  {author} {\bibinfo {author} {\bibfnamefont {G.}~\bibnamefont
  {Ballesteros}}\ and\ \bibinfo {author} {\bibfnamefont {M.}~\bibnamefont
  {Taoso}},\ }\href {\doibase 10.1103/PhysRevD.97.023501} {\bibfield  {journal}
  {\bibinfo  {journal} {Phys. Rev. D}\ }\textbf {\bibinfo {volume} {97}},\
  \bibinfo {pages} {023501} (\bibinfo {year} {2018})},\ \Eprint
  {http://arxiv.org/abs/1709.05565} {arXiv:1709.05565 [hep-ph]} \BibitemShut
  {NoStop}%
\bibitem [{\citenamefont {Kannike}\ \emph {et~al.}(2017)\citenamefont
  {Kannike}, \citenamefont {Marzola}, \citenamefont {Raidal},\ and\
  \citenamefont {Veerm\"ae}}]{Kannike:2017bxn}%
  \BibitemOpen
  \bibfield  {author} {\bibinfo {author} {\bibfnamefont {K.}~\bibnamefont
  {Kannike}}, \bibinfo {author} {\bibfnamefont {L.}~\bibnamefont {Marzola}},
  \bibinfo {author} {\bibfnamefont {M.}~\bibnamefont {Raidal}}, \ and\ \bibinfo
  {author} {\bibfnamefont {H.}~\bibnamefont {Veerm\"ae}},\ }\href {\doibase
  10.1088/1475-7516/2017/09/020} {\bibfield  {journal} {\bibinfo  {journal}
  {JCAP}\ }\textbf {\bibinfo {volume} {09}},\ \bibinfo {pages} {020} (\bibinfo
  {year} {2017})},\ \Eprint {http://arxiv.org/abs/1705.06225} {arXiv:1705.06225
  [astro-ph.CO]} \BibitemShut {NoStop}%
\bibitem [{\citenamefont {Carr}\ \emph {et~al.}(2017)\citenamefont {Carr},
  \citenamefont {Tenkanen},\ and\ \citenamefont {Vaskonen}}]{Carr:2017edp}%
  \BibitemOpen
  \bibfield  {author} {\bibinfo {author} {\bibfnamefont {B.}~\bibnamefont
  {Carr}}, \bibinfo {author} {\bibfnamefont {T.}~\bibnamefont {Tenkanen}}, \
  and\ \bibinfo {author} {\bibfnamefont {V.}~\bibnamefont {Vaskonen}},\ }\href
  {\doibase 10.1103/PhysRevD.96.063507} {\bibfield  {journal} {\bibinfo
  {journal} {Phys. Rev. D}\ }\textbf {\bibinfo {volume} {96}},\ \bibinfo
  {pages} {063507} (\bibinfo {year} {2017})},\ \Eprint
  {http://arxiv.org/abs/1706.03746} {arXiv:1706.03746 [astro-ph.CO]}
  \BibitemShut {NoStop}%
\bibitem [{\citenamefont {Germani}\ and\ \citenamefont
  {Prokopec}(2017)}]{Germani:2017bcs}%
  \BibitemOpen
  \bibfield  {author} {\bibinfo {author} {\bibfnamefont {C.}~\bibnamefont
  {Germani}}\ and\ \bibinfo {author} {\bibfnamefont {T.}~\bibnamefont
  {Prokopec}},\ }\href {\doibase 10.1016/j.dark.2017.09.001} {\bibfield
  {journal} {\bibinfo  {journal} {Phys. Dark Univ.}\ }\textbf {\bibinfo
  {volume} {18}},\ \bibinfo {pages} {6} (\bibinfo {year} {2017})},\ \Eprint
  {http://arxiv.org/abs/1706.04226} {arXiv:1706.04226 [astro-ph.CO]}
  \BibitemShut {NoStop}%
\bibitem [{\citenamefont {Motohashi}\ and\ \citenamefont
  {Hu}(2017)}]{Motohashi:2017kbs}%
  \BibitemOpen
  \bibfield  {author} {\bibinfo {author} {\bibfnamefont {H.}~\bibnamefont
  {Motohashi}}\ and\ \bibinfo {author} {\bibfnamefont {W.}~\bibnamefont {Hu}},\
  }\href {\doibase 10.1103/PhysRevD.96.063503} {\bibfield  {journal} {\bibinfo
  {journal} {Phys. Rev. D}\ }\textbf {\bibinfo {volume} {96}},\ \bibinfo
  {pages} {063503} (\bibinfo {year} {2017})},\ \Eprint
  {http://arxiv.org/abs/1706.06784} {arXiv:1706.06784 [astro-ph.CO]}
  \BibitemShut {NoStop}%
\bibitem [{\citenamefont {Di}\ and\ \citenamefont {Gong}(2018)}]{Gong:2017qlj}%
  \BibitemOpen
  \bibfield  {author} {\bibinfo {author} {\bibfnamefont {H.}~\bibnamefont
  {Di}}\ and\ \bibinfo {author} {\bibfnamefont {Y.}~\bibnamefont {Gong}},\
  }\href {\doibase 10.1088/1475-7516/2018/07/007} {\bibfield  {journal}
  {\bibinfo  {journal} {JCAP}\ }\textbf {\bibinfo {volume} {07}},\ \bibinfo
  {pages} {007} (\bibinfo {year} {2018})},\ \Eprint
  {http://arxiv.org/abs/1707.09578} {arXiv:1707.09578 [astro-ph.CO]}
  \BibitemShut {NoStop}%
\bibitem [{\citenamefont {Pi}\ \emph {et~al.}(2018)\citenamefont {Pi},
  \citenamefont {Zhang}, \citenamefont {Huang},\ and\ \citenamefont
  {Sasaki}}]{Pi:2017gih}%
  \BibitemOpen
  \bibfield  {author} {\bibinfo {author} {\bibfnamefont {S.}~\bibnamefont
  {Pi}}, \bibinfo {author} {\bibfnamefont {Y.-l.}\ \bibnamefont {Zhang}},
  \bibinfo {author} {\bibfnamefont {Q.-G.}\ \bibnamefont {Huang}}, \ and\
  \bibinfo {author} {\bibfnamefont {M.}~\bibnamefont {Sasaki}},\ }\href
  {\doibase 10.1088/1475-7516/2018/05/042} {\bibfield  {journal} {\bibinfo
  {journal} {JCAP}\ }\textbf {\bibinfo {volume} {05}},\ \bibinfo {pages} {042}
  (\bibinfo {year} {2018})},\ \Eprint {http://arxiv.org/abs/1712.09896}
  {arXiv:1712.09896 [astro-ph.CO]} \BibitemShut {NoStop}%
\bibitem [{\citenamefont {\"Ozsoy}\ \emph {et~al.}(2018)\citenamefont
  {\"Ozsoy}, \citenamefont {Parameswaran}, \citenamefont {Tasinato},\ and\
  \citenamefont {Zavala}}]{Ozsoy:2018flq}%
  \BibitemOpen
  \bibfield  {author} {\bibinfo {author} {\bibfnamefont {O.}~\bibnamefont
  {\"Ozsoy}}, \bibinfo {author} {\bibfnamefont {S.}~\bibnamefont
  {Parameswaran}}, \bibinfo {author} {\bibfnamefont {G.}~\bibnamefont
  {Tasinato}}, \ and\ \bibinfo {author} {\bibfnamefont {I.}~\bibnamefont
  {Zavala}},\ }\href {\doibase 10.1088/1475-7516/2018/07/005} {\bibfield
  {journal} {\bibinfo  {journal} {JCAP}\ }\textbf {\bibinfo {volume} {07}},\
  \bibinfo {pages} {005} (\bibinfo {year} {2018})},\ \Eprint
  {http://arxiv.org/abs/1803.07626} {arXiv:1803.07626 [hep-th]} \BibitemShut
  {NoStop}%
\bibitem [{\citenamefont {Biagetti}\ \emph {et~al.}(2018)\citenamefont
  {Biagetti}, \citenamefont {Franciolini}, \citenamefont {Kehagias},\ and\
  \citenamefont {Riotto}}]{Biagetti:2018pjj}%
  \BibitemOpen
  \bibfield  {author} {\bibinfo {author} {\bibfnamefont {M.}~\bibnamefont
  {Biagetti}}, \bibinfo {author} {\bibfnamefont {G.}~\bibnamefont
  {Franciolini}}, \bibinfo {author} {\bibfnamefont {A.}~\bibnamefont
  {Kehagias}}, \ and\ \bibinfo {author} {\bibfnamefont {A.}~\bibnamefont
  {Riotto}},\ }\href {\doibase 10.1088/1475-7516/2018/07/032} {\bibfield
  {journal} {\bibinfo  {journal} {JCAP}\ }\textbf {\bibinfo {volume} {07}},\
  \bibinfo {pages} {032} (\bibinfo {year} {2018})},\ \Eprint
  {http://arxiv.org/abs/1804.07124} {arXiv:1804.07124 [astro-ph.CO]}
  \BibitemShut {NoStop}%
\bibitem [{\citenamefont {Cai}\ \emph {et~al.}(2018)\citenamefont {Cai},
  \citenamefont {Tong}, \citenamefont {Wang},\ and\ \citenamefont
  {Yan}}]{Cai:2018tuh}%
  \BibitemOpen
  \bibfield  {author} {\bibinfo {author} {\bibfnamefont {Y.-F.}\ \bibnamefont
  {Cai}}, \bibinfo {author} {\bibfnamefont {X.}~\bibnamefont {Tong}}, \bibinfo
  {author} {\bibfnamefont {D.-G.}\ \bibnamefont {Wang}}, \ and\ \bibinfo
  {author} {\bibfnamefont {S.-F.}\ \bibnamefont {Yan}},\ }\href {\doibase
  10.1103/PhysRevLett.121.081306} {\bibfield  {journal} {\bibinfo  {journal}
  {Phys. Rev. Lett.}\ }\textbf {\bibinfo {volume} {121}},\ \bibinfo {pages}
  {081306} (\bibinfo {year} {2018})},\ \Eprint
  {http://arxiv.org/abs/1805.03639} {arXiv:1805.03639 [astro-ph.CO]}
  \BibitemShut {NoStop}%
\bibitem [{\citenamefont {Gao}\ and\ \citenamefont {Guo}(2018)}]{Gao:2018pvq}%
  \BibitemOpen
  \bibfield  {author} {\bibinfo {author} {\bibfnamefont {T.-J.}\ \bibnamefont
  {Gao}}\ and\ \bibinfo {author} {\bibfnamefont {Z.-K.}\ \bibnamefont {Guo}},\
  }\href {\doibase 10.1103/PhysRevD.98.063526} {\bibfield  {journal} {\bibinfo
  {journal} {Phys. Rev. D}\ }\textbf {\bibinfo {volume} {98}},\ \bibinfo
  {pages} {063526} (\bibinfo {year} {2018})},\ \Eprint
  {http://arxiv.org/abs/1806.09320} {arXiv:1806.09320 [hep-ph]} \BibitemShut
  {NoStop}%
\bibitem [{\citenamefont {Cai}\ \emph {et~al.}(2019)\citenamefont {Cai},
  \citenamefont {Pi},\ and\ \citenamefont {Sasaki}}]{Cai:2018dig}%
  \BibitemOpen
  \bibfield  {author} {\bibinfo {author} {\bibfnamefont {R.-g.}\ \bibnamefont
  {Cai}}, \bibinfo {author} {\bibfnamefont {S.}~\bibnamefont {Pi}}, \ and\
  \bibinfo {author} {\bibfnamefont {M.}~\bibnamefont {Sasaki}},\ }\href
  {\doibase 10.1103/PhysRevLett.122.201101} {\bibfield  {journal} {\bibinfo
  {journal} {Phys. Rev. Lett.}\ }\textbf {\bibinfo {volume} {122}},\ \bibinfo
  {pages} {201101} (\bibinfo {year} {2019})},\ \Eprint
  {http://arxiv.org/abs/1810.11000} {arXiv:1810.11000 [astro-ph.CO]}
  \BibitemShut {NoStop}%
\bibitem [{\citenamefont {Ballesteros}\ \emph {et~al.}(2019)\citenamefont
  {Ballesteros}, \citenamefont {Beltran~Jimenez},\ and\ \citenamefont
  {Pieroni}}]{Ballesteros:2018wlw}%
  \BibitemOpen
  \bibfield  {author} {\bibinfo {author} {\bibfnamefont {G.}~\bibnamefont
  {Ballesteros}}, \bibinfo {author} {\bibfnamefont {J.}~\bibnamefont
  {Beltran~Jimenez}}, \ and\ \bibinfo {author} {\bibfnamefont {M.}~\bibnamefont
  {Pieroni}},\ }\href {\doibase 10.1088/1475-7516/2019/06/016} {\bibfield
  {journal} {\bibinfo  {journal} {JCAP}\ }\textbf {\bibinfo {volume} {06}},\
  \bibinfo {pages} {016} (\bibinfo {year} {2019})},\ \Eprint
  {http://arxiv.org/abs/1811.03065} {arXiv:1811.03065 [astro-ph.CO]}
  \BibitemShut {NoStop}%
\bibitem [{\citenamefont {Byrnes}\ \emph {et~al.}(2019)\citenamefont {Byrnes},
  \citenamefont {Cole},\ and\ \citenamefont {Patil}}]{Byrnes:2018txb}%
  \BibitemOpen
  \bibfield  {author} {\bibinfo {author} {\bibfnamefont {C.~T.}\ \bibnamefont
  {Byrnes}}, \bibinfo {author} {\bibfnamefont {P.~S.}\ \bibnamefont {Cole}}, \
  and\ \bibinfo {author} {\bibfnamefont {S.~P.}\ \bibnamefont {Patil}},\ }\href
  {\doibase 10.1088/1475-7516/2019/06/028} {\bibfield  {journal} {\bibinfo
  {journal} {JCAP}\ }\textbf {\bibinfo {volume} {06}},\ \bibinfo {pages} {028}
  (\bibinfo {year} {2019})},\ \Eprint {http://arxiv.org/abs/1811.11158}
  {arXiv:1811.11158 [astro-ph.CO]} \BibitemShut {NoStop}%
\bibitem [{\citenamefont {Dalianis}(2019)}]{Dalianis:2018ymb}%
  \BibitemOpen
  \bibfield  {author} {\bibinfo {author} {\bibfnamefont {I.}~\bibnamefont
  {Dalianis}},\ }\href {\doibase 10.1088/1475-7516/2019/08/032} {\bibfield
  {journal} {\bibinfo  {journal} {JCAP}\ }\textbf {\bibinfo {volume} {08}},\
  \bibinfo {pages} {032} (\bibinfo {year} {2019})},\ \Eprint
  {http://arxiv.org/abs/1812.09807} {arXiv:1812.09807 [astro-ph.CO]}
  \BibitemShut {NoStop}%
\bibitem [{\citenamefont {Pi}\ \emph {et~al.}(2019)\citenamefont {Pi},
  \citenamefont {Sasaki},\ and\ \citenamefont {Zhang}}]{Pi:2019ihn}%
  \BibitemOpen
  \bibfield  {author} {\bibinfo {author} {\bibfnamefont {S.}~\bibnamefont
  {Pi}}, \bibinfo {author} {\bibfnamefont {M.}~\bibnamefont {Sasaki}}, \ and\
  \bibinfo {author} {\bibfnamefont {Y.-l.}\ \bibnamefont {Zhang}},\ }\href
  {\doibase 10.1088/1475-7516/2019/06/049} {\bibfield  {journal} {\bibinfo
  {journal} {JCAP}\ }\textbf {\bibinfo {volume} {06}},\ \bibinfo {pages} {049}
  (\bibinfo {year} {2019})},\ \Eprint {http://arxiv.org/abs/1904.06304}
  {arXiv:1904.06304 [gr-qc]} \BibitemShut {NoStop}%
\bibitem [{\citenamefont {Vallejo-Pe\~na}\ and\ \citenamefont
  {Romano}(2019)}]{Vallejo-Pena:2019lfo}%
  \BibitemOpen
  \bibfield  {author} {\bibinfo {author} {\bibfnamefont {S.~A.}\ \bibnamefont
  {Vallejo-Pe\~na}}\ and\ \bibinfo {author} {\bibfnamefont {A.~E.}\
  \bibnamefont {Romano}},\ }\href {\doibase 10.1088/1475-7516/2019/11/015}
  {\bibfield  {journal} {\bibinfo  {journal} {JCAP}\ }\textbf {\bibinfo
  {volume} {11}},\ \bibinfo {pages} {015} (\bibinfo {year} {2019})},\ \Eprint
  {http://arxiv.org/abs/1904.07503} {arXiv:1904.07503 [astro-ph.CO]}
  \BibitemShut {NoStop}%
\bibitem [{\citenamefont {Dalianis}\ and\ \citenamefont
  {Tringas}(2019)}]{Dalianis:2019asr}%
  \BibitemOpen
  \bibfield  {author} {\bibinfo {author} {\bibfnamefont {I.}~\bibnamefont
  {Dalianis}}\ and\ \bibinfo {author} {\bibfnamefont {G.}~\bibnamefont
  {Tringas}},\ }\href {\doibase 10.1103/PhysRevD.100.083512} {\bibfield
  {journal} {\bibinfo  {journal} {Phys. Rev. D}\ }\textbf {\bibinfo {volume}
  {100}},\ \bibinfo {pages} {083512} (\bibinfo {year} {2019})},\ \Eprint
  {http://arxiv.org/abs/1905.01741} {arXiv:1905.01741 [astro-ph.CO]}
  \BibitemShut {NoStop}%
\bibitem [{\citenamefont {Bhaumik}\ and\ \citenamefont
  {Jain}(2020)}]{Bhaumik:2019tvl}%
  \BibitemOpen
  \bibfield  {author} {\bibinfo {author} {\bibfnamefont {N.}~\bibnamefont
  {Bhaumik}}\ and\ \bibinfo {author} {\bibfnamefont {R.~K.}\ \bibnamefont
  {Jain}},\ }\href {\doibase 10.1088/1475-7516/2020/01/037} {\bibfield
  {journal} {\bibinfo  {journal} {JCAP}\ }\textbf {\bibinfo {volume} {01}},\
  \bibinfo {pages} {037} (\bibinfo {year} {2020})},\ \Eprint
  {http://arxiv.org/abs/1907.04125} {arXiv:1907.04125 [astro-ph.CO]}
  \BibitemShut {NoStop}%
\bibitem [{\citenamefont {Fu}\ \emph {et~al.}(2019)\citenamefont {Fu},
  \citenamefont {Wu},\ and\ \citenamefont {Yu}}]{Fu:2019ttf}%
  \BibitemOpen
  \bibfield  {author} {\bibinfo {author} {\bibfnamefont {C.}~\bibnamefont
  {Fu}}, \bibinfo {author} {\bibfnamefont {P.}~\bibnamefont {Wu}}, \ and\
  \bibinfo {author} {\bibfnamefont {H.}~\bibnamefont {Yu}},\ }\href {\doibase
  10.1103/PhysRevD.100.063532} {\bibfield  {journal} {\bibinfo  {journal}
  {Phys. Rev. D}\ }\textbf {\bibinfo {volume} {100}},\ \bibinfo {pages}
  {063532} (\bibinfo {year} {2019})},\ \Eprint
  {http://arxiv.org/abs/1907.05042} {arXiv:1907.05042 [astro-ph.CO]}
  \BibitemShut {NoStop}%
\bibitem [{\citenamefont {Xu}\ \emph {et~al.}(2020)\citenamefont {Xu},
  \citenamefont {Liu}, \citenamefont {Gao},\ and\ \citenamefont
  {Guo}}]{Xu:2019bdp}%
  \BibitemOpen
  \bibfield  {author} {\bibinfo {author} {\bibfnamefont {W.-T.}\ \bibnamefont
  {Xu}}, \bibinfo {author} {\bibfnamefont {J.}~\bibnamefont {Liu}}, \bibinfo
  {author} {\bibfnamefont {T.-J.}\ \bibnamefont {Gao}}, \ and\ \bibinfo
  {author} {\bibfnamefont {Z.-K.}\ \bibnamefont {Guo}},\ }\href {\doibase
  10.1103/PhysRevD.101.023505} {\bibfield  {journal} {\bibinfo  {journal}
  {Phys. Rev. D}\ }\textbf {\bibinfo {volume} {101}},\ \bibinfo {pages}
  {023505} (\bibinfo {year} {2020})},\ \Eprint
  {http://arxiv.org/abs/1907.05213} {arXiv:1907.05213 [astro-ph.CO]}
  \BibitemShut {NoStop}%
\bibitem [{\citenamefont {Liu}\ \emph {et~al.}(2020)\citenamefont {Liu},
  \citenamefont {Guo},\ and\ \citenamefont {Cai}}]{Liu:2019lul}%
  \BibitemOpen
  \bibfield  {author} {\bibinfo {author} {\bibfnamefont {J.}~\bibnamefont
  {Liu}}, \bibinfo {author} {\bibfnamefont {Z.-K.}\ \bibnamefont {Guo}}, \ and\
  \bibinfo {author} {\bibfnamefont {R.-G.}\ \bibnamefont {Cai}},\ }\href
  {\doibase 10.1103/PhysRevD.101.023513} {\bibfield  {journal} {\bibinfo
  {journal} {Phys. Rev. D}\ }\textbf {\bibinfo {volume} {101}},\ \bibinfo
  {pages} {023513} (\bibinfo {year} {2020})},\ \Eprint
  {http://arxiv.org/abs/1908.02662} {arXiv:1908.02662 [astro-ph.CO]}
  \BibitemShut {NoStop}%
\bibitem [{\citenamefont {Chen}\ and\ \citenamefont
  {Cai}(2019)}]{Chen:2019zza}%
  \BibitemOpen
  \bibfield  {author} {\bibinfo {author} {\bibfnamefont {C.}~\bibnamefont
  {Chen}}\ and\ \bibinfo {author} {\bibfnamefont {Y.-F.}\ \bibnamefont {Cai}},\
  }\href {\doibase 10.1088/1475-7516/2019/10/068} {\bibfield  {journal}
  {\bibinfo  {journal} {JCAP}\ }\textbf {\bibinfo {volume} {10}},\ \bibinfo
  {pages} {068} (\bibinfo {year} {2019})},\ \Eprint
  {http://arxiv.org/abs/1908.03942} {arXiv:1908.03942 [astro-ph.CO]}
  \BibitemShut {NoStop}%
\bibitem [{\citenamefont {Arya}(2020)}]{Arya:2019wck}%
  \BibitemOpen
  \bibfield  {author} {\bibinfo {author} {\bibfnamefont {R.}~\bibnamefont
  {Arya}},\ }\href {\doibase 10.1088/1475-7516/2020/09/042} {\bibfield
  {journal} {\bibinfo  {journal} {JCAP}\ }\textbf {\bibinfo {volume} {09}},\
  \bibinfo {pages} {042} (\bibinfo {year} {2020})},\ \Eprint
  {http://arxiv.org/abs/1910.05238} {arXiv:1910.05238 [astro-ph.CO]}
  \BibitemShut {NoStop}%
\bibitem [{\citenamefont {Mahbub}(2020{\natexlab{a}})}]{Mahbub:2019uhl}%
  \BibitemOpen
  \bibfield  {author} {\bibinfo {author} {\bibfnamefont {R.}~\bibnamefont
  {Mahbub}},\ }\href {\doibase 10.1103/PhysRevD.101.023533} {\bibfield
  {journal} {\bibinfo  {journal} {Phys. Rev. D}\ }\textbf {\bibinfo {volume}
  {101}},\ \bibinfo {pages} {023533} (\bibinfo {year} {2020}{\natexlab{a}})},\
  \Eprint {http://arxiv.org/abs/1910.10602} {arXiv:1910.10602 [astro-ph.CO]}
  \BibitemShut {NoStop}%
\bibitem [{\citenamefont {Mishra}\ and\ \citenamefont
  {Sahni}(2020)}]{Mishra:2019pzq}%
  \BibitemOpen
  \bibfield  {author} {\bibinfo {author} {\bibfnamefont {S.~S.}\ \bibnamefont
  {Mishra}}\ and\ \bibinfo {author} {\bibfnamefont {V.}~\bibnamefont {Sahni}},\
  }\href {\doibase 10.1088/1475-7516/2020/04/007} {\bibfield  {journal}
  {\bibinfo  {journal} {JCAP}\ }\textbf {\bibinfo {volume} {04}},\ \bibinfo
  {pages} {007} (\bibinfo {year} {2020})},\ \Eprint
  {http://arxiv.org/abs/1911.00057} {arXiv:1911.00057 [gr-qc]} \BibitemShut
  {NoStop}%
\bibitem [{\citenamefont {Lin}\ \emph {et~al.}(2020)\citenamefont {Lin},
  \citenamefont {Gao}, \citenamefont {Gong}, \citenamefont {Lu}, \citenamefont
  {Zhang},\ and\ \citenamefont {Zhang}}]{Lin:2020goi}%
  \BibitemOpen
  \bibfield  {author} {\bibinfo {author} {\bibfnamefont {J.}~\bibnamefont
  {Lin}}, \bibinfo {author} {\bibfnamefont {Q.}~\bibnamefont {Gao}}, \bibinfo
  {author} {\bibfnamefont {Y.}~\bibnamefont {Gong}}, \bibinfo {author}
  {\bibfnamefont {Y.}~\bibnamefont {Lu}}, \bibinfo {author} {\bibfnamefont
  {C.}~\bibnamefont {Zhang}}, \ and\ \bibinfo {author} {\bibfnamefont
  {F.}~\bibnamefont {Zhang}},\ }\href {\doibase 10.1103/PhysRevD.101.103515}
  {\bibfield  {journal} {\bibinfo  {journal} {Phys. Rev. D}\ }\textbf {\bibinfo
  {volume} {101}},\ \bibinfo {pages} {103515} (\bibinfo {year} {2020})},\
  \Eprint {http://arxiv.org/abs/2001.05909} {arXiv:2001.05909 [gr-qc]}
  \BibitemShut {NoStop}%
\bibitem [{\citenamefont {Fu}\ \emph {et~al.}(2020)\citenamefont {Fu},
  \citenamefont {Wu},\ and\ \citenamefont {Yu}}]{Fu:2020lob}%
  \BibitemOpen
  \bibfield  {author} {\bibinfo {author} {\bibfnamefont {C.}~\bibnamefont
  {Fu}}, \bibinfo {author} {\bibfnamefont {P.}~\bibnamefont {Wu}}, \ and\
  \bibinfo {author} {\bibfnamefont {H.}~\bibnamefont {Yu}},\ }\href {\doibase
  10.1103/PhysRevD.102.043527} {\bibfield  {journal} {\bibinfo  {journal}
  {Phys. Rev. D}\ }\textbf {\bibinfo {volume} {102}},\ \bibinfo {pages}
  {043527} (\bibinfo {year} {2020})},\ \Eprint
  {http://arxiv.org/abs/2006.03768} {arXiv:2006.03768 [astro-ph.CO]}
  \BibitemShut {NoStop}%
\bibitem [{\citenamefont {Ballesteros}\ \emph {et~al.}(2020)\citenamefont
  {Ballesteros}, \citenamefont {Rey}, \citenamefont {Taoso},\ and\
  \citenamefont {Urbano}}]{Ballesteros:2020sre}%
  \BibitemOpen
  \bibfield  {author} {\bibinfo {author} {\bibfnamefont {G.}~\bibnamefont
  {Ballesteros}}, \bibinfo {author} {\bibfnamefont {J.}~\bibnamefont {Rey}},
  \bibinfo {author} {\bibfnamefont {M.}~\bibnamefont {Taoso}}, \ and\ \bibinfo
  {author} {\bibfnamefont {A.}~\bibnamefont {Urbano}},\ }\href {\doibase
  10.1088/1475-7516/2020/08/043} {\bibfield  {journal} {\bibinfo  {journal}
  {JCAP}\ }\textbf {\bibinfo {volume} {08}},\ \bibinfo {pages} {043} (\bibinfo
  {year} {2020})},\ \Eprint {http://arxiv.org/abs/2006.14597} {arXiv:2006.14597
  [astro-ph.CO]} \BibitemShut {NoStop}%
\bibitem [{\citenamefont {\"Ozsoy}\ and\ \citenamefont
  {Lalak}(2021)}]{Ozsoy:2020kat}%
  \BibitemOpen
  \bibfield  {author} {\bibinfo {author} {\bibfnamefont {O.}~\bibnamefont
  {\"Ozsoy}}\ and\ \bibinfo {author} {\bibfnamefont {Z.}~\bibnamefont
  {Lalak}},\ }\href {\doibase 10.1088/1475-7516/2021/01/040} {\bibfield
  {journal} {\bibinfo  {journal} {JCAP}\ }\textbf {\bibinfo {volume} {01}},\
  \bibinfo {pages} {040} (\bibinfo {year} {2021})},\ \Eprint
  {http://arxiv.org/abs/2008.07549} {arXiv:2008.07549 [astro-ph.CO]}
  \BibitemShut {NoStop}%
\bibitem [{\citenamefont {Anguelova}(2021)}]{Anguelova:2020nzl}%
  \BibitemOpen
  \bibfield  {author} {\bibinfo {author} {\bibfnamefont {L.}~\bibnamefont
  {Anguelova}},\ }\href {\doibase 10.1088/1475-7516/2021/06/004} {\bibfield
  {journal} {\bibinfo  {journal} {JCAP}\ }\textbf {\bibinfo {volume} {06}},\
  \bibinfo {pages} {004} (\bibinfo {year} {2021})},\ \Eprint
  {http://arxiv.org/abs/2012.03705} {arXiv:2012.03705 [hep-th]} \BibitemShut
  {NoStop}%
\bibitem [{\citenamefont {Solbi}\ and\ \citenamefont
  {Karami}(2021{\natexlab{a}})}]{Solbi:2021wbo}%
  \BibitemOpen
  \bibfield  {author} {\bibinfo {author} {\bibfnamefont {M.}~\bibnamefont
  {Solbi}}\ and\ \bibinfo {author} {\bibfnamefont {K.}~\bibnamefont {Karami}},\
  }\href@noop {} {\  (\bibinfo {year} {2021}{\natexlab{a}})},\ \Eprint
  {http://arxiv.org/abs/2102.05651} {arXiv:2102.05651 [astro-ph.CO]}
  \BibitemShut {NoStop}%
\bibitem [{\citenamefont {Choi}\ \emph {et~al.}(2021)\citenamefont {Choi},
  \citenamefont {Kang},\ and\ \citenamefont {Raveendran}}]{Choi:2021yxz}%
  \BibitemOpen
  \bibfield  {author} {\bibinfo {author} {\bibfnamefont {K.-Y.}\ \bibnamefont
  {Choi}}, \bibinfo {author} {\bibfnamefont {S.-b.}\ \bibnamefont {Kang}}, \
  and\ \bibinfo {author} {\bibfnamefont {R.~N.}\ \bibnamefont {Raveendran}},\
  }\href@noop {} {\  (\bibinfo {year} {2021})},\ \Eprint
  {http://arxiv.org/abs/2102.02461} {arXiv:2102.02461 [astro-ph.CO]}
  \BibitemShut {NoStop}%
\bibitem [{\citenamefont {Gao}(2021)}]{Gao:2021vxb}%
  \BibitemOpen
  \bibfield  {author} {\bibinfo {author} {\bibfnamefont {Q.}~\bibnamefont
  {Gao}},\ }\href@noop {} {\  (\bibinfo {year} {2021})},\ \Eprint
  {http://arxiv.org/abs/2102.07369} {arXiv:2102.07369 [gr-qc]} \BibitemShut
  {NoStop}%
\bibitem [{\citenamefont {Inomata}\ \emph {et~al.}(2021)\citenamefont
  {Inomata}, \citenamefont {Mcdonough},\ and\ \citenamefont
  {Hu}}]{Inomata:2021uqj}%
  \BibitemOpen
  \bibfield  {author} {\bibinfo {author} {\bibfnamefont {K.}~\bibnamefont
  {Inomata}}, \bibinfo {author} {\bibfnamefont {E.}~\bibnamefont {Mcdonough}},
  \ and\ \bibinfo {author} {\bibfnamefont {W.}~\bibnamefont {Hu}},\ }\href@noop
  {} {\  (\bibinfo {year} {2021})},\ \Eprint {http://arxiv.org/abs/2104.03972}
  {arXiv:2104.03972 [astro-ph.CO]} \BibitemShut {NoStop}%
\bibitem [{\citenamefont {Biagetti}\ \emph {et~al.}(2021)\citenamefont
  {Biagetti}, \citenamefont {De~Luca}, \citenamefont {Franciolini},
  \citenamefont {Kehagias},\ and\ \citenamefont {Riotto}}]{Biagetti:2021eep}%
  \BibitemOpen
  \bibfield  {author} {\bibinfo {author} {\bibfnamefont {M.}~\bibnamefont
  {Biagetti}}, \bibinfo {author} {\bibfnamefont {V.}~\bibnamefont {De~Luca}},
  \bibinfo {author} {\bibfnamefont {G.}~\bibnamefont {Franciolini}}, \bibinfo
  {author} {\bibfnamefont {A.}~\bibnamefont {Kehagias}}, \ and\ \bibinfo
  {author} {\bibfnamefont {A.}~\bibnamefont {Riotto}},\ }\href@noop {} {\
  (\bibinfo {year} {2021})},\ \Eprint {http://arxiv.org/abs/2105.07810}
  {arXiv:2105.07810 [astro-ph.CO]} \BibitemShut {NoStop}%
\bibitem [{\citenamefont {Drees}\ and\ \citenamefont
  {Xu}(2021)}]{Drees:2019xpp}%
  \BibitemOpen
  \bibfield  {author} {\bibinfo {author} {\bibfnamefont {M.}~\bibnamefont
  {Drees}}\ and\ \bibinfo {author} {\bibfnamefont {Y.}~\bibnamefont {Xu}},\
  }\href {\doibase 10.1140/epjc/s10052-021-08976-2} {\bibfield  {journal}
  {\bibinfo  {journal} {Eur. Phys. J. C}\ }\textbf {\bibinfo {volume} {81}},\
  \bibinfo {pages} {182} (\bibinfo {year} {2021})},\ \Eprint
  {http://arxiv.org/abs/1905.13581} {arXiv:1905.13581 [hep-ph]} \BibitemShut
  {NoStop}%
\bibitem [{\citenamefont {Carr}\ and\ \citenamefont
  {Kuhnel}(2019)}]{Carr:2018poi}%
  \BibitemOpen
  \bibfield  {author} {\bibinfo {author} {\bibfnamefont {B.}~\bibnamefont
  {Carr}}\ and\ \bibinfo {author} {\bibfnamefont {F.}~\bibnamefont {Kuhnel}},\
  }\href {\doibase 10.1103/PhysRevD.99.103535} {\bibfield  {journal} {\bibinfo
  {journal} {Phys. Rev. D}\ }\textbf {\bibinfo {volume} {99}},\ \bibinfo
  {pages} {103535} (\bibinfo {year} {2019})},\ \Eprint
  {http://arxiv.org/abs/1811.06532} {arXiv:1811.06532 [astro-ph.CO]}
  \BibitemShut {NoStop}%
\bibitem [{\citenamefont {Atal}\ \emph {et~al.}(2019)\citenamefont {Atal},
  \citenamefont {Garriga},\ and\ \citenamefont
  {Marcos-Caballero}}]{Atal:2019cdz}%
  \BibitemOpen
  \bibfield  {author} {\bibinfo {author} {\bibfnamefont {V.}~\bibnamefont
  {Atal}}, \bibinfo {author} {\bibfnamefont {J.}~\bibnamefont {Garriga}}, \
  and\ \bibinfo {author} {\bibfnamefont {A.}~\bibnamefont {Marcos-Caballero}},\
  }\href {\doibase 10.1088/1475-7516/2019/09/073} {\bibfield  {journal}
  {\bibinfo  {journal} {JCAP}\ }\textbf {\bibinfo {volume} {09}},\ \bibinfo
  {pages} {073} (\bibinfo {year} {2019})},\ \Eprint
  {http://arxiv.org/abs/1905.13202} {arXiv:1905.13202 [astro-ph.CO]}
  \BibitemShut {NoStop}%
\bibitem [{\citenamefont {Niikura}\ \emph
  {et~al.}(2019{\natexlab{a}})\citenamefont {Niikura} \emph
  {et~al.}}]{Niikura:2017zjd}%
  \BibitemOpen
  \bibfield  {author} {\bibinfo {author} {\bibfnamefont {H.}~\bibnamefont
  {Niikura}} \emph {et~al.},\ }\href {\doibase 10.1038/s41550-019-0723-1}
  {\bibfield  {journal} {\bibinfo  {journal} {Nature Astron.}\ }\textbf
  {\bibinfo {volume} {3}},\ \bibinfo {pages} {524} (\bibinfo {year}
  {2019}{\natexlab{a}})},\ \Eprint {http://arxiv.org/abs/1701.02151}
  {arXiv:1701.02151 [astro-ph.CO]} \BibitemShut {NoStop}%
\bibitem [{\citenamefont {Tisserand}\ \emph {et~al.}(2007)\citenamefont
  {Tisserand} \emph {et~al.}}]{Tisserand:2006zx}%
  \BibitemOpen
  \bibfield  {author} {\bibinfo {author} {\bibfnamefont {P.}~\bibnamefont
  {Tisserand}} \emph {et~al.} (\bibinfo {collaboration} {EROS-2}),\ }\href
  {\doibase 10.1051/0004-6361:20066017} {\bibfield  {journal} {\bibinfo
  {journal} {Astron. Astrophys.}\ }\textbf {\bibinfo {volume} {469}},\ \bibinfo
  {pages} {387} (\bibinfo {year} {2007})},\ \Eprint
  {http://arxiv.org/abs/astro-ph/0607207} {arXiv:astro-ph/0607207} \BibitemShut
  {NoStop}%
\bibitem [{\citenamefont {Niikura}\ \emph
  {et~al.}(2019{\natexlab{b}})\citenamefont {Niikura}, \citenamefont {Takada},
  \citenamefont {Yokoyama}, \citenamefont {Sumi},\ and\ \citenamefont
  {Masaki}}]{Niikura:2019kqi}%
  \BibitemOpen
  \bibfield  {author} {\bibinfo {author} {\bibfnamefont {H.}~\bibnamefont
  {Niikura}}, \bibinfo {author} {\bibfnamefont {M.}~\bibnamefont {Takada}},
  \bibinfo {author} {\bibfnamefont {S.}~\bibnamefont {Yokoyama}}, \bibinfo
  {author} {\bibfnamefont {T.}~\bibnamefont {Sumi}}, \ and\ \bibinfo {author}
  {\bibfnamefont {S.}~\bibnamefont {Masaki}},\ }\href {\doibase
  10.1103/PhysRevD.99.083503} {\bibfield  {journal} {\bibinfo  {journal} {Phys.
  Rev. D}\ }\textbf {\bibinfo {volume} {99}},\ \bibinfo {pages} {083503}
  (\bibinfo {year} {2019}{\natexlab{b}})},\ \Eprint
  {http://arxiv.org/abs/1901.07120} {arXiv:1901.07120 [astro-ph.CO]}
  \BibitemShut {NoStop}%
\bibitem [{\citenamefont {Serpico}\ \emph {et~al.}(2020)\citenamefont
  {Serpico}, \citenamefont {Poulin}, \citenamefont {Inman},\ and\ \citenamefont
  {Kohri}}]{Serpico:2020ehh}%
  \BibitemOpen
  \bibfield  {author} {\bibinfo {author} {\bibfnamefont {P.~D.}\ \bibnamefont
  {Serpico}}, \bibinfo {author} {\bibfnamefont {V.}~\bibnamefont {Poulin}},
  \bibinfo {author} {\bibfnamefont {D.}~\bibnamefont {Inman}}, \ and\ \bibinfo
  {author} {\bibfnamefont {K.}~\bibnamefont {Kohri}},\ }\href {\doibase
  10.1103/PhysRevResearch.2.023204} {\bibfield  {journal} {\bibinfo  {journal}
  {Phys. Rev. Res.}\ }\textbf {\bibinfo {volume} {2}},\ \bibinfo {pages}
  {023204} (\bibinfo {year} {2020})},\ \Eprint
  {http://arxiv.org/abs/2002.10771} {arXiv:2002.10771 [astro-ph.CO]}
  \BibitemShut {NoStop}%
\bibitem [{\citenamefont {Barnacka}\ \emph {et~al.}(2012)\citenamefont
  {Barnacka}, \citenamefont {Glicenstein},\ and\ \citenamefont
  {Moderski}}]{Barnacka:2012bm}%
  \BibitemOpen
  \bibfield  {author} {\bibinfo {author} {\bibfnamefont {A.}~\bibnamefont
  {Barnacka}}, \bibinfo {author} {\bibfnamefont {J.~F.}\ \bibnamefont
  {Glicenstein}}, \ and\ \bibinfo {author} {\bibfnamefont {R.}~\bibnamefont
  {Moderski}},\ }\href {\doibase 10.1103/PhysRevD.86.043001} {\bibfield
  {journal} {\bibinfo  {journal} {Phys. Rev. D}\ }\textbf {\bibinfo {volume}
  {86}},\ \bibinfo {pages} {043001} (\bibinfo {year} {2012})},\ \Eprint
  {http://arxiv.org/abs/1204.2056} {arXiv:1204.2056 [astro-ph.CO]} \BibitemShut
  {NoStop}%
\bibitem [{\citenamefont {Graham}\ \emph {et~al.}(2015)\citenamefont {Graham},
  \citenamefont {Rajendran},\ and\ \citenamefont {Varela}}]{Graham:2015apa}%
  \BibitemOpen
  \bibfield  {author} {\bibinfo {author} {\bibfnamefont {P.~W.}\ \bibnamefont
  {Graham}}, \bibinfo {author} {\bibfnamefont {S.}~\bibnamefont {Rajendran}}, \
  and\ \bibinfo {author} {\bibfnamefont {J.}~\bibnamefont {Varela}},\ }\href
  {\doibase 10.1103/PhysRevD.92.063007} {\bibfield  {journal} {\bibinfo
  {journal} {Phys. Rev. D}\ }\textbf {\bibinfo {volume} {92}},\ \bibinfo
  {pages} {063007} (\bibinfo {year} {2015})},\ \Eprint
  {http://arxiv.org/abs/1505.04444} {arXiv:1505.04444 [hep-ph]} \BibitemShut
  {NoStop}%
\bibitem [{\citenamefont {Katz}\ \emph {et~al.}(2018)\citenamefont {Katz},
  \citenamefont {Kopp}, \citenamefont {Sibiryakov},\ and\ \citenamefont
  {Xue}}]{Katz:2018zrn}%
  \BibitemOpen
  \bibfield  {author} {\bibinfo {author} {\bibfnamefont {A.}~\bibnamefont
  {Katz}}, \bibinfo {author} {\bibfnamefont {J.}~\bibnamefont {Kopp}}, \bibinfo
  {author} {\bibfnamefont {S.}~\bibnamefont {Sibiryakov}}, \ and\ \bibinfo
  {author} {\bibfnamefont {W.}~\bibnamefont {Xue}},\ }\href {\doibase
  10.1088/1475-7516/2018/12/005} {\bibfield  {journal} {\bibinfo  {journal}
  {JCAP}\ }\textbf {\bibinfo {volume} {12}},\ \bibinfo {pages} {005} (\bibinfo
  {year} {2018})},\ \Eprint {http://arxiv.org/abs/1807.11495} {arXiv:1807.11495
  [astro-ph.CO]} \BibitemShut {NoStop}%
\bibitem [{\citenamefont {Montero-Camacho}\ \emph {et~al.}(2019)\citenamefont
  {Montero-Camacho}, \citenamefont {Fang}, \citenamefont {Vasquez},
  \citenamefont {Silva},\ and\ \citenamefont
  {Hirata}}]{Montero-Camacho:2019jte}%
  \BibitemOpen
  \bibfield  {author} {\bibinfo {author} {\bibfnamefont {P.}~\bibnamefont
  {Montero-Camacho}}, \bibinfo {author} {\bibfnamefont {X.}~\bibnamefont
  {Fang}}, \bibinfo {author} {\bibfnamefont {G.}~\bibnamefont {Vasquez}},
  \bibinfo {author} {\bibfnamefont {M.}~\bibnamefont {Silva}}, \ and\ \bibinfo
  {author} {\bibfnamefont {C.~M.}\ \bibnamefont {Hirata}},\ }\href {\doibase
  10.1088/1475-7516/2019/08/031} {\bibfield  {journal} {\bibinfo  {journal}
  {JCAP}\ }\textbf {\bibinfo {volume} {08}},\ \bibinfo {pages} {031} (\bibinfo
  {year} {2019})},\ \Eprint {http://arxiv.org/abs/1906.05950} {arXiv:1906.05950
  [astro-ph.CO]} \BibitemShut {NoStop}%
\bibitem [{\citenamefont {Laha}(2019)}]{Laha:2019ssq}%
  \BibitemOpen
  \bibfield  {author} {\bibinfo {author} {\bibfnamefont {R.}~\bibnamefont
  {Laha}},\ }\href {\doibase 10.1103/PhysRevLett.123.251101} {\bibfield
  {journal} {\bibinfo  {journal} {Phys. Rev. Lett.}\ }\textbf {\bibinfo
  {volume} {123}},\ \bibinfo {pages} {251101} (\bibinfo {year} {2019})},\
  \Eprint {http://arxiv.org/abs/1906.09994} {arXiv:1906.09994 [astro-ph.HE]}
  \BibitemShut {NoStop}%
\bibitem [{\citenamefont {Dasgupta}\ \emph {et~al.}(2020)\citenamefont
  {Dasgupta}, \citenamefont {Laha},\ and\ \citenamefont
  {Ray}}]{Dasgupta:2019cae}%
  \BibitemOpen
  \bibfield  {author} {\bibinfo {author} {\bibfnamefont {B.}~\bibnamefont
  {Dasgupta}}, \bibinfo {author} {\bibfnamefont {R.}~\bibnamefont {Laha}}, \
  and\ \bibinfo {author} {\bibfnamefont {A.}~\bibnamefont {Ray}},\ }\href
  {\doibase 10.1103/PhysRevLett.125.101101} {\bibfield  {journal} {\bibinfo
  {journal} {Phys. Rev. Lett.}\ }\textbf {\bibinfo {volume} {125}},\ \bibinfo
  {pages} {101101} (\bibinfo {year} {2020})},\ \Eprint
  {http://arxiv.org/abs/1912.01014} {arXiv:1912.01014 [hep-ph]} \BibitemShut
  {NoStop}%
\bibitem [{\citenamefont {Laha}\ \emph {et~al.}(2020)\citenamefont {Laha},
  \citenamefont {Mu\~noz},\ and\ \citenamefont {Slatyer}}]{Laha:2020ivk}%
  \BibitemOpen
  \bibfield  {author} {\bibinfo {author} {\bibfnamefont {R.}~\bibnamefont
  {Laha}}, \bibinfo {author} {\bibfnamefont {J.~B.}\ \bibnamefont {Mu\~noz}}, \
  and\ \bibinfo {author} {\bibfnamefont {T.~R.}\ \bibnamefont {Slatyer}},\
  }\href {\doibase 10.1103/PhysRevD.101.123514} {\bibfield  {journal} {\bibinfo
   {journal} {Phys. Rev. D}\ }\textbf {\bibinfo {volume} {101}},\ \bibinfo
  {pages} {123514} (\bibinfo {year} {2020})},\ \Eprint
  {http://arxiv.org/abs/2004.00627} {arXiv:2004.00627 [astro-ph.CO]}
  \BibitemShut {NoStop}%
\bibitem [{\citenamefont {Cai}\ \emph {et~al.}(2021{\natexlab{a}})\citenamefont
  {Cai}, \citenamefont {Ding}, \citenamefont {Yang},\ and\ \citenamefont
  {Zhou}}]{Cai:2020fnq}%
  \BibitemOpen
  \bibfield  {author} {\bibinfo {author} {\bibfnamefont {R.-G.}\ \bibnamefont
  {Cai}}, \bibinfo {author} {\bibfnamefont {Y.-C.}\ \bibnamefont {Ding}},
  \bibinfo {author} {\bibfnamefont {X.-Y.}\ \bibnamefont {Yang}}, \ and\
  \bibinfo {author} {\bibfnamefont {Y.-F.}\ \bibnamefont {Zhou}},\ }\href
  {\doibase 10.1088/1475-7516/2021/03/057} {\bibfield  {journal} {\bibinfo
  {journal} {JCAP}\ }\textbf {\bibinfo {volume} {03}},\ \bibinfo {pages} {057}
  (\bibinfo {year} {2021}{\natexlab{a}})},\ \Eprint
  {http://arxiv.org/abs/2007.11804} {arXiv:2007.11804 [astro-ph.CO]}
  \BibitemShut {NoStop}%
\bibitem [{\citenamefont {Carr}\ \emph {et~al.}(2010)\citenamefont {Carr},
  \citenamefont {Kohri}, \citenamefont {Sendouda},\ and\ \citenamefont
  {Yokoyama}}]{Carr:2009jm}%
  \BibitemOpen
  \bibfield  {author} {\bibinfo {author} {\bibfnamefont {B.~J.}\ \bibnamefont
  {Carr}}, \bibinfo {author} {\bibfnamefont {K.}~\bibnamefont {Kohri}},
  \bibinfo {author} {\bibfnamefont {Y.}~\bibnamefont {Sendouda}}, \ and\
  \bibinfo {author} {\bibfnamefont {J.}~\bibnamefont {Yokoyama}},\ }\href
  {\doibase 10.1103/PhysRevD.81.104019} {\bibfield  {journal} {\bibinfo
  {journal} {Phys. Rev. D}\ }\textbf {\bibinfo {volume} {81}},\ \bibinfo
  {pages} {104019} (\bibinfo {year} {2010})},\ \Eprint
  {http://arxiv.org/abs/0912.5297} {arXiv:0912.5297 [astro-ph.CO]} \BibitemShut
  {NoStop}%
\bibitem [{\citenamefont {Capela}\ \emph {et~al.}(2013)\citenamefont {Capela},
  \citenamefont {Pshirkov},\ and\ \citenamefont {Tinyakov}}]{Capela:2013yf}%
  \BibitemOpen
  \bibfield  {author} {\bibinfo {author} {\bibfnamefont {F.}~\bibnamefont
  {Capela}}, \bibinfo {author} {\bibfnamefont {M.}~\bibnamefont {Pshirkov}}, \
  and\ \bibinfo {author} {\bibfnamefont {P.}~\bibnamefont {Tinyakov}},\ }\href
  {\doibase 10.1103/PhysRevD.87.123524} {\bibfield  {journal} {\bibinfo
  {journal} {Phys. Rev. D}\ }\textbf {\bibinfo {volume} {87}},\ \bibinfo
  {pages} {123524} (\bibinfo {year} {2013})},\ \Eprint
  {http://arxiv.org/abs/1301.4984} {arXiv:1301.4984 [astro-ph.CO]} \BibitemShut
  {NoStop}%
\bibitem [{\citenamefont {Chen}\ \emph {et~al.}(2020)\citenamefont {Chen},
  \citenamefont {Yuan},\ and\ \citenamefont {Huang}}]{Chen:2019xse}%
  \BibitemOpen
  \bibfield  {author} {\bibinfo {author} {\bibfnamefont {Z.-C.}\ \bibnamefont
  {Chen}}, \bibinfo {author} {\bibfnamefont {C.}~\bibnamefont {Yuan}}, \ and\
  \bibinfo {author} {\bibfnamefont {Q.-G.}\ \bibnamefont {Huang}},\ }\href
  {\doibase 10.1103/PhysRevLett.124.251101} {\bibfield  {journal} {\bibinfo
  {journal} {Phys. Rev. Lett.}\ }\textbf {\bibinfo {volume} {124}},\ \bibinfo
  {pages} {251101} (\bibinfo {year} {2020})},\ \Eprint
  {http://arxiv.org/abs/1910.12239} {arXiv:1910.12239 [astro-ph.CO]}
  \BibitemShut {NoStop}%
\bibitem [{\citenamefont {Wong}\ \emph {et~al.}(2021)\citenamefont {Wong},
  \citenamefont {Franciolini}, \citenamefont {De~Luca}, \citenamefont
  {Baibhav}, \citenamefont {Berti}, \citenamefont {Pani},\ and\ \citenamefont
  {Riotto}}]{Wong:2020yig}%
  \BibitemOpen
  \bibfield  {author} {\bibinfo {author} {\bibfnamefont {K.~W.~K.}\
  \bibnamefont {Wong}}, \bibinfo {author} {\bibfnamefont {G.}~\bibnamefont
  {Franciolini}}, \bibinfo {author} {\bibfnamefont {V.}~\bibnamefont
  {De~Luca}}, \bibinfo {author} {\bibfnamefont {V.}~\bibnamefont {Baibhav}},
  \bibinfo {author} {\bibfnamefont {E.}~\bibnamefont {Berti}}, \bibinfo
  {author} {\bibfnamefont {P.}~\bibnamefont {Pani}}, \ and\ \bibinfo {author}
  {\bibfnamefont {A.}~\bibnamefont {Riotto}},\ }\href {\doibase
  10.1103/PhysRevD.103.023026} {\bibfield  {journal} {\bibinfo  {journal}
  {Phys. Rev. D}\ }\textbf {\bibinfo {volume} {103}},\ \bibinfo {pages}
  {023026} (\bibinfo {year} {2021})},\ \Eprint
  {http://arxiv.org/abs/2011.01865} {arXiv:2011.01865 [gr-qc]} \BibitemShut
  {NoStop}%
\bibitem [{\citenamefont {Kavanagh}\ \emph {et~al.}(2018)\citenamefont
  {Kavanagh}, \citenamefont {Gaggero},\ and\ \citenamefont
  {Bertone}}]{Kavanagh:2018ggo}%
  \BibitemOpen
  \bibfield  {author} {\bibinfo {author} {\bibfnamefont {B.~J.}\ \bibnamefont
  {Kavanagh}}, \bibinfo {author} {\bibfnamefont {D.}~\bibnamefont {Gaggero}}, \
  and\ \bibinfo {author} {\bibfnamefont {G.}~\bibnamefont {Bertone}},\ }\href
  {\doibase 10.1103/PhysRevD.98.023536} {\bibfield  {journal} {\bibinfo
  {journal} {Phys. Rev. D}\ }\textbf {\bibinfo {volume} {98}},\ \bibinfo
  {pages} {023536} (\bibinfo {year} {2018})},\ \Eprint
  {http://arxiv.org/abs/1805.09034} {arXiv:1805.09034 [astro-ph.CO]}
  \BibitemShut {NoStop}%
\bibitem [{\citenamefont {Lu}\ and\ \citenamefont {Wu}(2019)}]{Lu:2019ktw}%
  \BibitemOpen
  \bibfield  {author} {\bibinfo {author} {\bibfnamefont {B.-Q.}\ \bibnamefont
  {Lu}}\ and\ \bibinfo {author} {\bibfnamefont {Y.-L.}\ \bibnamefont {Wu}},\
  }\href {\doibase 10.1103/PhysRevD.99.123023} {\bibfield  {journal} {\bibinfo
  {journal} {Phys. Rev. D}\ }\textbf {\bibinfo {volume} {99}},\ \bibinfo
  {pages} {123023} (\bibinfo {year} {2019})},\ \Eprint
  {http://arxiv.org/abs/1906.10463} {arXiv:1906.10463 [astro-ph.HE]}
  \BibitemShut {NoStop}%
\bibitem [{\citenamefont {Jung}\ and\ \citenamefont
  {Shin}(2019)}]{Jung:2017flg}%
  \BibitemOpen
  \bibfield  {author} {\bibinfo {author} {\bibfnamefont {S.}~\bibnamefont
  {Jung}}\ and\ \bibinfo {author} {\bibfnamefont {C.~S.}\ \bibnamefont
  {Shin}},\ }\href {\doibase 10.1103/PhysRevLett.122.041103} {\bibfield
  {journal} {\bibinfo  {journal} {Phys. Rev. Lett.}\ }\textbf {\bibinfo
  {volume} {122}},\ \bibinfo {pages} {041103} (\bibinfo {year} {2019})},\
  \Eprint {http://arxiv.org/abs/1712.01396} {arXiv:1712.01396 [astro-ph.CO]}
  \BibitemShut {NoStop}%
\bibitem [{\citenamefont {Ray}\ \emph {et~al.}(2021)\citenamefont {Ray},
  \citenamefont {Laha}, \citenamefont {Mu\~noz},\ and\ \citenamefont
  {Caputo}}]{Ray:2021mxu}%
  \BibitemOpen
  \bibfield  {author} {\bibinfo {author} {\bibfnamefont {A.}~\bibnamefont
  {Ray}}, \bibinfo {author} {\bibfnamefont {R.}~\bibnamefont {Laha}}, \bibinfo
  {author} {\bibfnamefont {J.~B.}\ \bibnamefont {Mu\~noz}}, \ and\ \bibinfo
  {author} {\bibfnamefont {R.}~\bibnamefont {Caputo}},\ }\href@noop {} {\
  (\bibinfo {year} {2021})},\ \Eprint {http://arxiv.org/abs/2102.06714}
  {arXiv:2102.06714 [astro-ph.CO]} \BibitemShut {NoStop}%
\bibitem [{\citenamefont {Acharya}\ and\ \citenamefont
  {Khatri}(2020)}]{Acharya:2020jbv}%
  \BibitemOpen
  \bibfield  {author} {\bibinfo {author} {\bibfnamefont {S.~K.}\ \bibnamefont
  {Acharya}}\ and\ \bibinfo {author} {\bibfnamefont {R.}~\bibnamefont
  {Khatri}},\ }\href {\doibase 10.1088/1475-7516/2020/06/018} {\bibfield
  {journal} {\bibinfo  {journal} {JCAP}\ }\textbf {\bibinfo {volume} {06}},\
  \bibinfo {pages} {018} (\bibinfo {year} {2020})},\ \Eprint
  {http://arxiv.org/abs/2002.00898} {arXiv:2002.00898 [astro-ph.CO]}
  \BibitemShut {NoStop}%
\bibitem [{\citenamefont {Carr}\ \emph {et~al.}(2020)\citenamefont {Carr},
  \citenamefont {Kohri}, \citenamefont {Sendouda},\ and\ \citenamefont
  {Yokoyama}}]{Carr:2020gox}%
  \BibitemOpen
  \bibfield  {author} {\bibinfo {author} {\bibfnamefont {B.}~\bibnamefont
  {Carr}}, \bibinfo {author} {\bibfnamefont {K.}~\bibnamefont {Kohri}},
  \bibinfo {author} {\bibfnamefont {Y.}~\bibnamefont {Sendouda}}, \ and\
  \bibinfo {author} {\bibfnamefont {J.}~\bibnamefont {Yokoyama}},\ }\href@noop
  {} {\  (\bibinfo {year} {2020})},\ \Eprint {http://arxiv.org/abs/2002.12778}
  {arXiv:2002.12778 [astro-ph.CO]} \BibitemShut {NoStop}%
\bibitem [{\citenamefont {Allahverdi}\ \emph {et~al.}(2020)\citenamefont
  {Allahverdi} \emph {et~al.}}]{Allahverdi:2020bys}%
  \BibitemOpen
  \bibfield  {author} {\bibinfo {author} {\bibfnamefont {R.}~\bibnamefont
  {Allahverdi}} \emph {et~al.},\ }\href {\doibase 10.21105/astro.2006.16182} {\
   (\bibinfo {year} {2020}),\ 10.21105/astro.2006.16182},\ \Eprint
  {http://arxiv.org/abs/2006.16182} {arXiv:2006.16182 [astro-ph.CO]}
  \BibitemShut {NoStop}%
\bibitem [{\citenamefont {Cai}\ \emph {et~al.}(2021{\natexlab{b}})\citenamefont
  {Cai}, \citenamefont {Chen}, \citenamefont {Ding},\ and\ \citenamefont
  {Wang}}]{Cai:2021zxo}%
  \BibitemOpen
  \bibfield  {author} {\bibinfo {author} {\bibfnamefont {Y.-F.}\ \bibnamefont
  {Cai}}, \bibinfo {author} {\bibfnamefont {C.}~\bibnamefont {Chen}}, \bibinfo
  {author} {\bibfnamefont {Q.}~\bibnamefont {Ding}}, \ and\ \bibinfo {author}
  {\bibfnamefont {Y.}~\bibnamefont {Wang}},\ }\href@noop {} {\  (\bibinfo
  {year} {2021}{\natexlab{b}})},\ \Eprint {http://arxiv.org/abs/2105.11481}
  {arXiv:2105.11481 [astro-ph.CO]} \BibitemShut {NoStop}%
\bibitem [{\citenamefont {Tada}\ and\ \citenamefont
  {Yokoyama}(2019)}]{Tada:2019amh}%
  \BibitemOpen
  \bibfield  {author} {\bibinfo {author} {\bibfnamefont {Y.}~\bibnamefont
  {Tada}}\ and\ \bibinfo {author} {\bibfnamefont {S.}~\bibnamefont
  {Yokoyama}},\ }\href {\doibase 10.1103/PhysRevD.100.023537} {\bibfield
  {journal} {\bibinfo  {journal} {Phys. Rev. D}\ }\textbf {\bibinfo {volume}
  {100}},\ \bibinfo {pages} {023537} (\bibinfo {year} {2019})},\ \Eprint
  {http://arxiv.org/abs/1904.10298} {arXiv:1904.10298 [astro-ph.CO]}
  \BibitemShut {NoStop}%
\bibitem [{\citenamefont {Braglia}\ \emph {et~al.}(2020)\citenamefont
  {Braglia}, \citenamefont {Hazra}, \citenamefont {Finelli}, \citenamefont
  {Smoot}, \citenamefont {Sriramkumar},\ and\ \citenamefont
  {Starobinsky}}]{Braglia:2020eai}%
  \BibitemOpen
  \bibfield  {author} {\bibinfo {author} {\bibfnamefont {M.}~\bibnamefont
  {Braglia}}, \bibinfo {author} {\bibfnamefont {D.~K.}\ \bibnamefont {Hazra}},
  \bibinfo {author} {\bibfnamefont {F.}~\bibnamefont {Finelli}}, \bibinfo
  {author} {\bibfnamefont {G.~F.}\ \bibnamefont {Smoot}}, \bibinfo {author}
  {\bibfnamefont {L.}~\bibnamefont {Sriramkumar}}, \ and\ \bibinfo {author}
  {\bibfnamefont {A.~A.}\ \bibnamefont {Starobinsky}},\ }\href {\doibase
  10.1088/1475-7516/2020/08/001} {\bibfield  {journal} {\bibinfo  {journal}
  {JCAP}\ }\textbf {\bibinfo {volume} {08}},\ \bibinfo {pages} {001} (\bibinfo
  {year} {2020})},\ \Eprint {http://arxiv.org/abs/2005.02895} {arXiv:2005.02895
  [astro-ph.CO]} \BibitemShut {NoStop}%
\bibitem [{\citenamefont {Ananda}\ \emph {et~al.}(2007)\citenamefont {Ananda},
  \citenamefont {Clarkson},\ and\ \citenamefont {Wands}}]{Ananda:2006af}%
  \BibitemOpen
  \bibfield  {author} {\bibinfo {author} {\bibfnamefont {K.~N.}\ \bibnamefont
  {Ananda}}, \bibinfo {author} {\bibfnamefont {C.}~\bibnamefont {Clarkson}}, \
  and\ \bibinfo {author} {\bibfnamefont {D.}~\bibnamefont {Wands}},\ }\href
  {\doibase 10.1103/PhysRevD.75.123518} {\bibfield  {journal} {\bibinfo
  {journal} {Phys. Rev. D}\ }\textbf {\bibinfo {volume} {75}},\ \bibinfo
  {pages} {123518} (\bibinfo {year} {2007})},\ \Eprint
  {http://arxiv.org/abs/gr-qc/0612013} {arXiv:gr-qc/0612013} \BibitemShut
  {NoStop}%
\bibitem [{\citenamefont {Baumann}\ \emph
  {et~al.}(2007{\natexlab{a}})\citenamefont {Baumann}, \citenamefont
  {Steinhardt}, \citenamefont {Takahashi},\ and\ \citenamefont
  {Ichiki}}]{Baumann:2007zm}%
  \BibitemOpen
  \bibfield  {author} {\bibinfo {author} {\bibfnamefont {D.}~\bibnamefont
  {Baumann}}, \bibinfo {author} {\bibfnamefont {P.~J.}\ \bibnamefont
  {Steinhardt}}, \bibinfo {author} {\bibfnamefont {K.}~\bibnamefont
  {Takahashi}}, \ and\ \bibinfo {author} {\bibfnamefont {K.}~\bibnamefont
  {Ichiki}},\ }\href {\doibase 10.1103/PhysRevD.76.084019} {\bibfield
  {journal} {\bibinfo  {journal} {Phys. Rev. D}\ }\textbf {\bibinfo {volume}
  {76}},\ \bibinfo {pages} {084019} (\bibinfo {year} {2007}{\natexlab{a}})},\
  \Eprint {http://arxiv.org/abs/hep-th/0703290} {arXiv:hep-th/0703290}
  \BibitemShut {NoStop}%
\bibitem [{\citenamefont {Baumann}(2009)}]{Baumann:2009ds}%
  \BibitemOpen
  \bibfield  {author} {\bibinfo {author} {\bibfnamefont {D.}~\bibnamefont
  {Baumann}},\ }in\ \href {\doibase 10.1142/9789814327183_0010} {\emph
  {\bibinfo {booktitle} {{Theoretical Advanced Study Institute in Elementary
  Particle Physics}: {Physics of the Large and the Small}}}}\ (\bibinfo {year}
  {2009})\ \Eprint {http://arxiv.org/abs/0907.5424} {arXiv:0907.5424 [hep-th]}
  \BibitemShut {NoStop}%
\bibitem [{\citenamefont {Arnowitt}\ \emph {et~al.}(2008)\citenamefont
  {Arnowitt}, \citenamefont {Deser},\ and\ \citenamefont
  {Misner}}]{Arnowitt:1962hi}%
  \BibitemOpen
  \bibfield  {author} {\bibinfo {author} {\bibfnamefont {R.~L.}\ \bibnamefont
  {Arnowitt}}, \bibinfo {author} {\bibfnamefont {S.}~\bibnamefont {Deser}}, \
  and\ \bibinfo {author} {\bibfnamefont {C.~W.}\ \bibnamefont {Misner}},\
  }\href {\doibase 10.1007/s10714-008-0661-1} {\bibfield  {journal} {\bibinfo
  {journal} {Gen. Rel. Grav.}\ }\textbf {\bibinfo {volume} {40}},\ \bibinfo
  {pages} {1997} (\bibinfo {year} {2008})},\ \Eprint
  {http://arxiv.org/abs/gr-qc/0405109} {arXiv:gr-qc/0405109} \BibitemShut
  {NoStop}%
\bibitem [{\citenamefont {Maldacena}(2003)}]{Maldacena:2002vr}%
  \BibitemOpen
  \bibfield  {author} {\bibinfo {author} {\bibfnamefont {J.~M.}\ \bibnamefont
  {Maldacena}},\ }\href {\doibase 10.1088/1126-6708/2003/05/013} {\bibfield
  {journal} {\bibinfo  {journal} {JHEP}\ }\textbf {\bibinfo {volume} {05}},\
  \bibinfo {pages} {013} (\bibinfo {year} {2003})},\ \Eprint
  {http://arxiv.org/abs/astro-ph/0210603} {arXiv:astro-ph/0210603} \BibitemShut
  {NoStop}%
\bibitem [{\citenamefont {Chen}\ \emph {et~al.}(2007)\citenamefont {Chen},
  \citenamefont {Huang}, \citenamefont {Kachru},\ and\ \citenamefont
  {Shiu}}]{Chen:2006nt}%
  \BibitemOpen
  \bibfield  {author} {\bibinfo {author} {\bibfnamefont {X.}~\bibnamefont
  {Chen}}, \bibinfo {author} {\bibfnamefont {M.-x.}\ \bibnamefont {Huang}},
  \bibinfo {author} {\bibfnamefont {S.}~\bibnamefont {Kachru}}, \ and\ \bibinfo
  {author} {\bibfnamefont {G.}~\bibnamefont {Shiu}},\ }\href {\doibase
  10.1088/1475-7516/2007/01/002} {\bibfield  {journal} {\bibinfo  {journal}
  {JCAP}\ }\textbf {\bibinfo {volume} {01}},\ \bibinfo {pages} {002} (\bibinfo
  {year} {2007})},\ \Eprint {http://arxiv.org/abs/hep-th/0605045}
  {arXiv:hep-th/0605045} \BibitemShut {NoStop}%
\bibitem [{\citenamefont {Sasaki}(1986)}]{Sasaki:1986hm}%
  \BibitemOpen
  \bibfield  {author} {\bibinfo {author} {\bibfnamefont {M.}~\bibnamefont
  {Sasaki}},\ }\href {\doibase 10.1143/PTP.76.1036} {\bibfield  {journal}
  {\bibinfo  {journal} {Prog. Theor. Phys.}\ }\textbf {\bibinfo {volume}
  {76}},\ \bibinfo {pages} {1036} (\bibinfo {year} {1986})}\BibitemShut
  {NoStop}%
\bibitem [{\citenamefont {Mukhanov}(1988)}]{Mukhanov:1988jd}%
  \BibitemOpen
  \bibfield  {author} {\bibinfo {author} {\bibfnamefont {V.~F.}\ \bibnamefont
  {Mukhanov}},\ }\href@noop {} {\bibfield  {journal} {\bibinfo  {journal} {Sov.
  Phys. JETP}\ }\textbf {\bibinfo {volume} {67}},\ \bibinfo {pages} {1297}
  (\bibinfo {year} {1988})}\BibitemShut {NoStop}%
\bibitem [{\citenamefont {Bunch}\ and\ \citenamefont
  {Davies}(1978)}]{Bunch:1978yq}%
  \BibitemOpen
  \bibfield  {author} {\bibinfo {author} {\bibfnamefont {T.~S.}\ \bibnamefont
  {Bunch}}\ and\ \bibinfo {author} {\bibfnamefont {P.~C.~W.}\ \bibnamefont
  {Davies}},\ }\href {\doibase 10.1098/rspa.1978.0060} {\bibfield  {journal}
  {\bibinfo  {journal} {Proc. Roy. Soc. Lond. A}\ }\textbf {\bibinfo {volume}
  {360}},\ \bibinfo {pages} {117} (\bibinfo {year} {1978})}\BibitemShut
  {NoStop}%
\bibitem [{\citenamefont {Akrami}\ \emph {et~al.}(2020)\citenamefont {Akrami}
  \emph {et~al.}}]{Akrami:2018odb}%
  \BibitemOpen
  \bibfield  {author} {\bibinfo {author} {\bibfnamefont {Y.}~\bibnamefont
  {Akrami}} \emph {et~al.} (\bibinfo {collaboration} {Planck}),\ }\href
  {\doibase 10.1051/0004-6361/201833887} {\bibfield  {journal} {\bibinfo
  {journal} {Astron. Astrophys.}\ }\textbf {\bibinfo {volume} {641}},\ \bibinfo
  {pages} {A10} (\bibinfo {year} {2020})},\ \Eprint
  {http://arxiv.org/abs/1807.06211} {arXiv:1807.06211 [astro-ph.CO]}
  \BibitemShut {NoStop}%
\bibitem [{\citenamefont {Kinney}(2005)}]{Kinney:2005vj}%
  \BibitemOpen
  \bibfield  {author} {\bibinfo {author} {\bibfnamefont {W.~H.}\ \bibnamefont
  {Kinney}},\ }\href {\doibase 10.1103/PhysRevD.72.023515} {\bibfield
  {journal} {\bibinfo  {journal} {Phys. Rev. D}\ }\textbf {\bibinfo {volume}
  {72}},\ \bibinfo {pages} {023515} (\bibinfo {year} {2005})},\ \Eprint
  {http://arxiv.org/abs/gr-qc/0503017} {arXiv:gr-qc/0503017} \BibitemShut
  {NoStop}%
\bibitem [{\citenamefont {Namjoo}\ \emph {et~al.}(2013)\citenamefont {Namjoo},
  \citenamefont {Firouzjahi},\ and\ \citenamefont {Sasaki}}]{Namjoo:2012aa}%
  \BibitemOpen
  \bibfield  {author} {\bibinfo {author} {\bibfnamefont {M.~H.}\ \bibnamefont
  {Namjoo}}, \bibinfo {author} {\bibfnamefont {H.}~\bibnamefont {Firouzjahi}},
  \ and\ \bibinfo {author} {\bibfnamefont {M.}~\bibnamefont {Sasaki}},\ }\href
  {\doibase 10.1209/0295-5075/101/39001} {\bibfield  {journal} {\bibinfo
  {journal} {EPL}\ }\textbf {\bibinfo {volume} {101}},\ \bibinfo {pages}
  {39001} (\bibinfo {year} {2013})},\ \Eprint {http://arxiv.org/abs/1210.3692}
  {arXiv:1210.3692 [astro-ph.CO]} \BibitemShut {NoStop}%
\bibitem [{\citenamefont {Martin}\ \emph {et~al.}(2013)\citenamefont {Martin},
  \citenamefont {Motohashi},\ and\ \citenamefont {Suyama}}]{Martin:2012pe}%
  \BibitemOpen
  \bibfield  {author} {\bibinfo {author} {\bibfnamefont {J.}~\bibnamefont
  {Martin}}, \bibinfo {author} {\bibfnamefont {H.}~\bibnamefont {Motohashi}}, \
  and\ \bibinfo {author} {\bibfnamefont {T.}~\bibnamefont {Suyama}},\ }\href
  {\doibase 10.1103/PhysRevD.87.023514} {\bibfield  {journal} {\bibinfo
  {journal} {Phys. Rev. D}\ }\textbf {\bibinfo {volume} {87}},\ \bibinfo
  {pages} {023514} (\bibinfo {year} {2013})},\ \Eprint
  {http://arxiv.org/abs/1211.0083} {arXiv:1211.0083 [astro-ph.CO]} \BibitemShut
  {NoStop}%
\bibitem [{\citenamefont {Ragavendra}(2021)}]{Ragavendra:2021qdu}%
  \BibitemOpen
  \bibfield  {author} {\bibinfo {author} {\bibfnamefont {H.~V.}\ \bibnamefont
  {Ragavendra}},\ }\href@noop {} {\  (\bibinfo {year} {2021})},\ \Eprint
  {http://arxiv.org/abs/2108.04193} {arXiv:2108.04193 [astro-ph.CO]}
  \BibitemShut {NoStop}%
\bibitem [{\citenamefont {Sasaki}\ \emph {et~al.}(2018)\citenamefont {Sasaki},
  \citenamefont {Suyama}, \citenamefont {Tanaka},\ and\ \citenamefont
  {Yokoyama}}]{Sasaki:2018dmp}%
  \BibitemOpen
  \bibfield  {author} {\bibinfo {author} {\bibfnamefont {M.}~\bibnamefont
  {Sasaki}}, \bibinfo {author} {\bibfnamefont {T.}~\bibnamefont {Suyama}},
  \bibinfo {author} {\bibfnamefont {T.}~\bibnamefont {Tanaka}}, \ and\ \bibinfo
  {author} {\bibfnamefont {S.}~\bibnamefont {Yokoyama}},\ }\href {\doibase
  10.1088/1361-6382/aaa7b4} {\bibfield  {journal} {\bibinfo  {journal} {Class.
  Quant. Grav.}\ }\textbf {\bibinfo {volume} {35}},\ \bibinfo {pages} {063001}
  (\bibinfo {year} {2018})},\ \Eprint {http://arxiv.org/abs/1801.05235}
  {arXiv:1801.05235 [astro-ph.CO]} \BibitemShut {NoStop}%
\bibitem [{\citenamefont {Green}\ and\ \citenamefont
  {Liddle}(1997)}]{Green:1997sz}%
  \BibitemOpen
  \bibfield  {author} {\bibinfo {author} {\bibfnamefont {A.~M.}\ \bibnamefont
  {Green}}\ and\ \bibinfo {author} {\bibfnamefont {A.~R.}\ \bibnamefont
  {Liddle}},\ }\href {\doibase 10.1103/PhysRevD.56.6166} {\bibfield  {journal}
  {\bibinfo  {journal} {Phys. Rev. D}\ }\textbf {\bibinfo {volume} {56}},\
  \bibinfo {pages} {6166} (\bibinfo {year} {1997})},\ \Eprint
  {http://arxiv.org/abs/astro-ph/9704251} {arXiv:astro-ph/9704251} \BibitemShut
  {NoStop}%
\bibitem [{\citenamefont {Kolb}\ and\ \citenamefont
  {Turner}(1990)}]{Kolb:1990vq}%
  \BibitemOpen
  \bibfield  {author} {\bibinfo {author} {\bibfnamefont {E.~W.}\ \bibnamefont
  {Kolb}}\ and\ \bibinfo {author} {\bibfnamefont {M.~S.}\ \bibnamefont
  {Turner}},\ }\href@noop {} {\emph {\bibinfo {title} {{The Early
  Universe}}}},\ Vol.~\bibinfo {volume} {69}\ (\bibinfo {year}
  {1990})\BibitemShut {NoStop}%
\bibitem [{\citenamefont {Inomata}\ \emph
  {et~al.}(2017{\natexlab{a}})\citenamefont {Inomata}, \citenamefont
  {Kawasaki}, \citenamefont {Mukaida}, \citenamefont {Tada},\ and\
  \citenamefont {Yanagida}}]{Inomata:2017okj}%
  \BibitemOpen
  \bibfield  {author} {\bibinfo {author} {\bibfnamefont {K.}~\bibnamefont
  {Inomata}}, \bibinfo {author} {\bibfnamefont {M.}~\bibnamefont {Kawasaki}},
  \bibinfo {author} {\bibfnamefont {K.}~\bibnamefont {Mukaida}}, \bibinfo
  {author} {\bibfnamefont {Y.}~\bibnamefont {Tada}}, \ and\ \bibinfo {author}
  {\bibfnamefont {T.~T.}\ \bibnamefont {Yanagida}},\ }\href {\doibase
  10.1103/PhysRevD.96.043504} {\bibfield  {journal} {\bibinfo  {journal} {Phys.
  Rev. D}\ }\textbf {\bibinfo {volume} {96}},\ \bibinfo {pages} {043504}
  (\bibinfo {year} {2017}{\natexlab{a}})},\ \Eprint
  {http://arxiv.org/abs/1701.02544} {arXiv:1701.02544 [astro-ph.CO]}
  \BibitemShut {NoStop}%
\bibitem [{\citenamefont {Press}\ and\ \citenamefont
  {Schechter}(1974)}]{Press:1973iz}%
  \BibitemOpen
  \bibfield  {author} {\bibinfo {author} {\bibfnamefont {W.~H.}\ \bibnamefont
  {Press}}\ and\ \bibinfo {author} {\bibfnamefont {P.}~\bibnamefont
  {Schechter}},\ }\href {\doibase 10.1086/152650} {\bibfield  {journal}
  {\bibinfo  {journal} {Astrophys. J.}\ }\textbf {\bibinfo {volume} {187}},\
  \bibinfo {pages} {425} (\bibinfo {year} {1974})}\BibitemShut {NoStop}%
\bibitem [{\citenamefont {Sureda}\ \emph {et~al.}(2020)\citenamefont {Sureda},
  \citenamefont {Magana}, \citenamefont {Araya},\ and\ \citenamefont
  {Padilla}}]{Sureda:2020vgi}%
  \BibitemOpen
  \bibfield  {author} {\bibinfo {author} {\bibfnamefont {J.}~\bibnamefont
  {Sureda}}, \bibinfo {author} {\bibfnamefont {J.}~\bibnamefont {Magana}},
  \bibinfo {author} {\bibfnamefont {I.~J.}\ \bibnamefont {Araya}}, \ and\
  \bibinfo {author} {\bibfnamefont {N.~D.}\ \bibnamefont {Padilla}},\
  }\href@noop {} {\  (\bibinfo {year} {2020})},\ \Eprint
  {http://arxiv.org/abs/2008.09683} {arXiv:2008.09683 [astro-ph.CO]}
  \BibitemShut {NoStop}%
\bibitem [{\citenamefont {Wu}(2020)}]{Wu:2020ilx}%
  \BibitemOpen
  \bibfield  {author} {\bibinfo {author} {\bibfnamefont {Y.-P.}\ \bibnamefont
  {Wu}},\ }\href {\doibase 10.1016/j.dark.2020.100654} {\bibfield  {journal}
  {\bibinfo  {journal} {Phys. Dark Univ.}\ }\textbf {\bibinfo {volume} {30}},\
  \bibinfo {pages} {100654} (\bibinfo {year} {2020})},\ \Eprint
  {http://arxiv.org/abs/2005.00441} {arXiv:2005.00441 [astro-ph.CO]}
  \BibitemShut {NoStop}%
\bibitem [{\citenamefont {Mahbub}(2020{\natexlab{b}})}]{Mahbub:2020row}%
  \BibitemOpen
  \bibfield  {author} {\bibinfo {author} {\bibfnamefont {R.}~\bibnamefont
  {Mahbub}},\ }\href {\doibase 10.1103/PhysRevD.102.023538} {\bibfield
  {journal} {\bibinfo  {journal} {Phys. Rev. D}\ }\textbf {\bibinfo {volume}
  {102}},\ \bibinfo {pages} {023538} (\bibinfo {year} {2020}{\natexlab{b}})},\
  \Eprint {http://arxiv.org/abs/2005.03618} {arXiv:2005.03618 [astro-ph.CO]}
  \BibitemShut {NoStop}%
\bibitem [{\citenamefont {Inomata}\ \emph
  {et~al.}(2017{\natexlab{b}})\citenamefont {Inomata}, \citenamefont
  {Kawasaki}, \citenamefont {Mukaida}, \citenamefont {Tada},\ and\
  \citenamefont {Yanagida}}]{Inomata:2017uaw}%
  \BibitemOpen
  \bibfield  {author} {\bibinfo {author} {\bibfnamefont {K.}~\bibnamefont
  {Inomata}}, \bibinfo {author} {\bibfnamefont {M.}~\bibnamefont {Kawasaki}},
  \bibinfo {author} {\bibfnamefont {K.}~\bibnamefont {Mukaida}}, \bibinfo
  {author} {\bibfnamefont {Y.}~\bibnamefont {Tada}}, \ and\ \bibinfo {author}
  {\bibfnamefont {T.~T.}\ \bibnamefont {Yanagida}},\ }\href {\doibase
  10.1103/PhysRevD.96.123527} {\bibfield  {journal} {\bibinfo  {journal} {Phys.
  Rev. D}\ }\textbf {\bibinfo {volume} {96}},\ \bibinfo {pages} {123527}
  (\bibinfo {year} {2017}{\natexlab{b}})},\ \Eprint
  {http://arxiv.org/abs/1709.07865} {arXiv:1709.07865 [astro-ph.CO]}
  \BibitemShut {NoStop}%
\bibitem [{\citenamefont {Carr}(1975)}]{Carr:1975qj}%
  \BibitemOpen
  \bibfield  {author} {\bibinfo {author} {\bibfnamefont {B.~J.}\ \bibnamefont
  {Carr}},\ }\href {\doibase 10.1086/153853} {\bibfield  {journal} {\bibinfo
  {journal} {Astrophys. J.}\ }\textbf {\bibinfo {volume} {201}},\ \bibinfo
  {pages} {1} (\bibinfo {year} {1975})}\BibitemShut {NoStop}%
\bibitem [{\citenamefont {Blais}\ \emph {et~al.}(2003)\citenamefont {Blais},
  \citenamefont {Bringmann}, \citenamefont {Kiefer},\ and\ \citenamefont
  {Polarski}}]{Blais:2002gw}%
  \BibitemOpen
  \bibfield  {author} {\bibinfo {author} {\bibfnamefont {D.}~\bibnamefont
  {Blais}}, \bibinfo {author} {\bibfnamefont {T.}~\bibnamefont {Bringmann}},
  \bibinfo {author} {\bibfnamefont {C.}~\bibnamefont {Kiefer}}, \ and\ \bibinfo
  {author} {\bibfnamefont {D.}~\bibnamefont {Polarski}},\ }\href {\doibase
  10.1103/PhysRevD.67.024024} {\bibfield  {journal} {\bibinfo  {journal} {Phys.
  Rev. D}\ }\textbf {\bibinfo {volume} {67}},\ \bibinfo {pages} {024024}
  (\bibinfo {year} {2003})},\ \Eprint {http://arxiv.org/abs/astro-ph/0206262}
  {arXiv:astro-ph/0206262} \BibitemShut {NoStop}%
\bibitem [{\citenamefont {Tokeshi}\ \emph {et~al.}(2020)\citenamefont
  {Tokeshi}, \citenamefont {Inomata},\ and\ \citenamefont
  {Yokoyama}}]{Tokeshi:2020tjq}%
  \BibitemOpen
  \bibfield  {author} {\bibinfo {author} {\bibfnamefont {K.}~\bibnamefont
  {Tokeshi}}, \bibinfo {author} {\bibfnamefont {K.}~\bibnamefont {Inomata}}, \
  and\ \bibinfo {author} {\bibfnamefont {J.}~\bibnamefont {Yokoyama}},\ }\href
  {\doibase 10.1088/1475-7516/2020/12/038} {\bibfield  {journal} {\bibinfo
  {journal} {JCAP}\ }\textbf {\bibinfo {volume} {12}},\ \bibinfo {pages} {038}
  (\bibinfo {year} {2020})},\ \Eprint {http://arxiv.org/abs/2005.07153}
  {arXiv:2005.07153 [astro-ph.CO]} \BibitemShut {NoStop}%
\bibitem [{\citenamefont {Young}(2019)}]{Young:2019osy}%
  \BibitemOpen
  \bibfield  {author} {\bibinfo {author} {\bibfnamefont {S.}~\bibnamefont
  {Young}},\ }\href {\doibase 10.1142/S0218271820300025} {\bibfield  {journal}
  {\bibinfo  {journal} {Int. J. Mod. Phys. D}\ }\textbf {\bibinfo {volume}
  {29}},\ \bibinfo {pages} {2030002} (\bibinfo {year} {2019})},\ \Eprint
  {http://arxiv.org/abs/1905.01230} {arXiv:1905.01230 [astro-ph.CO]}
  \BibitemShut {NoStop}%
\bibitem [{\citenamefont {Musco}\ \emph {et~al.}(2020)\citenamefont {Musco},
  \citenamefont {De~Luca}, \citenamefont {Franciolini},\ and\ \citenamefont
  {Riotto}}]{Musco:2020jjb}%
  \BibitemOpen
  \bibfield  {author} {\bibinfo {author} {\bibfnamefont {I.}~\bibnamefont
  {Musco}}, \bibinfo {author} {\bibfnamefont {V.}~\bibnamefont {De~Luca}},
  \bibinfo {author} {\bibfnamefont {G.}~\bibnamefont {Franciolini}}, \ and\
  \bibinfo {author} {\bibfnamefont {A.}~\bibnamefont {Riotto}},\ }\href@noop {}
  {\  (\bibinfo {year} {2020})},\ \Eprint {http://arxiv.org/abs/2011.03014}
  {arXiv:2011.03014 [astro-ph.CO]} \BibitemShut {NoStop}%
\bibitem [{\citenamefont {Sato-Polito}\ \emph {et~al.}(2019)\citenamefont
  {Sato-Polito}, \citenamefont {Kovetz},\ and\ \citenamefont
  {Kamionkowski}}]{Sato-Polito:2019hws}%
  \BibitemOpen
  \bibfield  {author} {\bibinfo {author} {\bibfnamefont {G.}~\bibnamefont
  {Sato-Polito}}, \bibinfo {author} {\bibfnamefont {E.~D.}\ \bibnamefont
  {Kovetz}}, \ and\ \bibinfo {author} {\bibfnamefont {M.}~\bibnamefont
  {Kamionkowski}},\ }\href {\doibase 10.1103/PhysRevD.100.063521} {\bibfield
  {journal} {\bibinfo  {journal} {Phys. Rev. D}\ }\textbf {\bibinfo {volume}
  {100}},\ \bibinfo {pages} {063521} (\bibinfo {year} {2019})},\ \Eprint
  {http://arxiv.org/abs/1904.10971} {arXiv:1904.10971 [astro-ph.CO]}
  \BibitemShut {NoStop}%
\bibitem [{\citenamefont {Siraj}\ and\ \citenamefont
  {Loeb}(2021)}]{Siraj:2021arp}%
  \BibitemOpen
  \bibfield  {author} {\bibinfo {author} {\bibfnamefont {A.}~\bibnamefont
  {Siraj}}\ and\ \bibinfo {author} {\bibfnamefont {A.}~\bibnamefont {Loeb}},\
  }\href@noop {} {\  (\bibinfo {year} {2021})},\ \Eprint
  {http://arxiv.org/abs/2104.07672} {arXiv:2104.07672 [astro-ph.CO]}
  \BibitemShut {NoStop}%
\bibitem [{\citenamefont {Kavanagh}(2019)}]{bradley_j_kavanagh_2019_3538999}%
  \BibitemOpen
  \bibfield  {author} {\bibinfo {author} {\bibfnamefont {B.~J.}\ \bibnamefont
  {Kavanagh}},\ }\href {\doibase 10.5281/zenodo.3538999} {\enquote {\bibinfo
  {title} {bradkav/pbhbounds: Release version},}\ } (\bibinfo {year}
  {2019})\BibitemShut {NoStop}%
\bibitem [{\citenamefont {Arbey}\ \emph {et~al.}(2020)\citenamefont {Arbey},
  \citenamefont {Auffinger},\ and\ \citenamefont {Silk}}]{Arbey:2020yzj}%
  \BibitemOpen
  \bibfield  {author} {\bibinfo {author} {\bibfnamefont {A.}~\bibnamefont
  {Arbey}}, \bibinfo {author} {\bibfnamefont {J.}~\bibnamefont {Auffinger}}, \
  and\ \bibinfo {author} {\bibfnamefont {J.}~\bibnamefont {Silk}},\ }in\
  \href@noop {} {\emph {\bibinfo {booktitle} {{40th International Conference on
  High Energy Physics}}}}\ (\bibinfo {year} {2020})\ \Eprint
  {http://arxiv.org/abs/2012.14767} {arXiv:2012.14767 [astro-ph.CO]}
  \BibitemShut {NoStop}%
\bibitem [{\citenamefont {Mirbabayi}\ \emph {et~al.}(2020)\citenamefont
  {Mirbabayi}, \citenamefont {Gruzinov},\ and\ \citenamefont
  {Nore\~na}}]{Mirbabayi:2019uph}%
  \BibitemOpen
  \bibfield  {author} {\bibinfo {author} {\bibfnamefont {M.}~\bibnamefont
  {Mirbabayi}}, \bibinfo {author} {\bibfnamefont {A.}~\bibnamefont {Gruzinov}},
  \ and\ \bibinfo {author} {\bibfnamefont {J.}~\bibnamefont {Nore\~na}},\
  }\href {\doibase 10.1088/1475-7516/2020/03/017} {\bibfield  {journal}
  {\bibinfo  {journal} {JCAP}\ }\textbf {\bibinfo {volume} {03}},\ \bibinfo
  {pages} {017} (\bibinfo {year} {2020})},\ \Eprint
  {http://arxiv.org/abs/1901.05963} {arXiv:1901.05963 [astro-ph.CO]}
  \BibitemShut {NoStop}%
\bibitem [{\citenamefont {K\"uhnel}\ and\ \citenamefont
  {Sandstad}(2016)}]{Kuhnel:2016exn}%
  \BibitemOpen
  \bibfield  {author} {\bibinfo {author} {\bibfnamefont {F.}~\bibnamefont
  {K\"uhnel}}\ and\ \bibinfo {author} {\bibfnamefont {M.}~\bibnamefont
  {Sandstad}},\ }\href {\doibase 10.1103/PhysRevD.94.063514} {\bibfield
  {journal} {\bibinfo  {journal} {Phys. Rev. D}\ }\textbf {\bibinfo {volume}
  {94}},\ \bibinfo {pages} {063514} (\bibinfo {year} {2016})},\ \Eprint
  {http://arxiv.org/abs/1602.04815} {arXiv:1602.04815 [astro-ph.CO]}
  \BibitemShut {NoStop}%
\bibitem [{\citenamefont {Sheth}\ \emph {et~al.}(2001)\citenamefont {Sheth},
  \citenamefont {Mo},\ and\ \citenamefont {Tormen}}]{Sheth:1999su}%
  \BibitemOpen
  \bibfield  {author} {\bibinfo {author} {\bibfnamefont {R.~K.}\ \bibnamefont
  {Sheth}}, \bibinfo {author} {\bibfnamefont {H.~J.}\ \bibnamefont {Mo}}, \
  and\ \bibinfo {author} {\bibfnamefont {G.}~\bibnamefont {Tormen}},\ }\href
  {\doibase 10.1046/j.1365-8711.2001.04006.x} {\bibfield  {journal} {\bibinfo
  {journal} {Mon. Not. Roy. Astron. Soc.}\ }\textbf {\bibinfo {volume} {323}},\
  \bibinfo {pages} {1} (\bibinfo {year} {2001})},\ \Eprint
  {http://arxiv.org/abs/astro-ph/9907024} {arXiv:astro-ph/9907024} \BibitemShut
  {NoStop}%
\bibitem [{\citenamefont {Lu}\ \emph {et~al.}(2019)\citenamefont {Lu},
  \citenamefont {Gong}, \citenamefont {Yi},\ and\ \citenamefont
  {Zhang}}]{Lu:2019sti}%
  \BibitemOpen
  \bibfield  {author} {\bibinfo {author} {\bibfnamefont {Y.}~\bibnamefont
  {Lu}}, \bibinfo {author} {\bibfnamefont {Y.}~\bibnamefont {Gong}}, \bibinfo
  {author} {\bibfnamefont {Z.}~\bibnamefont {Yi}}, \ and\ \bibinfo {author}
  {\bibfnamefont {F.}~\bibnamefont {Zhang}},\ }\href {\doibase
  10.1088/1475-7516/2019/12/031} {\bibfield  {journal} {\bibinfo  {journal}
  {JCAP}\ }\textbf {\bibinfo {volume} {12}},\ \bibinfo {pages} {031} (\bibinfo
  {year} {2019})},\ \Eprint {http://arxiv.org/abs/1907.11896} {arXiv:1907.11896
  [gr-qc]} \BibitemShut {NoStop}%
\bibitem [{\citenamefont {Pi}\ and\ \citenamefont {Sasaki}(2020)}]{Pi:2020otn}%
  \BibitemOpen
  \bibfield  {author} {\bibinfo {author} {\bibfnamefont {S.}~\bibnamefont
  {Pi}}\ and\ \bibinfo {author} {\bibfnamefont {M.}~\bibnamefont {Sasaki}},\
  }\href {\doibase 10.1088/1475-7516/2020/09/037} {\bibfield  {journal}
  {\bibinfo  {journal} {JCAP}\ }\textbf {\bibinfo {volume} {09}},\ \bibinfo
  {pages} {037} (\bibinfo {year} {2020})},\ \Eprint
  {http://arxiv.org/abs/2005.12306} {arXiv:2005.12306 [gr-qc]} \BibitemShut
  {NoStop}%
\bibitem [{\citenamefont {Chen}\ \emph
  {et~al.}(2021{\natexlab{a}})\citenamefont {Chen}, \citenamefont {Koh},\ and\
  \citenamefont {Tumurtushaa}}]{Chen:2021nio}%
  \BibitemOpen
  \bibfield  {author} {\bibinfo {author} {\bibfnamefont {P.}~\bibnamefont
  {Chen}}, \bibinfo {author} {\bibfnamefont {S.}~\bibnamefont {Koh}}, \ and\
  \bibinfo {author} {\bibfnamefont {G.}~\bibnamefont {Tumurtushaa}},\
  }\href@noop {} {\  (\bibinfo {year} {2021}{\natexlab{a}})},\ \Eprint
  {http://arxiv.org/abs/2107.08638} {arXiv:2107.08638 [gr-qc]} \BibitemShut
  {NoStop}%
\bibitem [{\citenamefont {Solbi}\ and\ \citenamefont
  {Karami}(2021{\natexlab{b}})}]{Solbi:2021rse}%
  \BibitemOpen
  \bibfield  {author} {\bibinfo {author} {\bibfnamefont {M.}~\bibnamefont
  {Solbi}}\ and\ \bibinfo {author} {\bibfnamefont {K.}~\bibnamefont {Karami}},\
  }\href@noop {} {\  (\bibinfo {year} {2021}{\natexlab{b}})},\ \Eprint
  {http://arxiv.org/abs/2106.02863} {arXiv:2106.02863 [astro-ph.CO]}
  \BibitemShut {NoStop}%
\bibitem [{\citenamefont {Chen}\ \emph
  {et~al.}(2021{\natexlab{b}})\citenamefont {Chen}, \citenamefont {Yuan},\ and\
  \citenamefont {Huang}}]{Chen:2021nxo}%
  \BibitemOpen
  \bibfield  {author} {\bibinfo {author} {\bibfnamefont {Z.-C.}\ \bibnamefont
  {Chen}}, \bibinfo {author} {\bibfnamefont {C.}~\bibnamefont {Yuan}}, \ and\
  \bibinfo {author} {\bibfnamefont {Q.-G.}\ \bibnamefont {Huang}},\ }\href@noop
  {} {\  (\bibinfo {year} {2021}{\natexlab{b}})},\ \Eprint
  {http://arxiv.org/abs/2108.11740} {arXiv:2108.11740 [astro-ph.CO]}
  \BibitemShut {NoStop}%
\bibitem [{\citenamefont {Kohri}\ and\ \citenamefont
  {Terada}(2018)}]{Kohri:2018awv}%
  \BibitemOpen
  \bibfield  {author} {\bibinfo {author} {\bibfnamefont {K.}~\bibnamefont
  {Kohri}}\ and\ \bibinfo {author} {\bibfnamefont {T.}~\bibnamefont {Terada}},\
  }\href {\doibase 10.1103/PhysRevD.97.123532} {\bibfield  {journal} {\bibinfo
  {journal} {Phys. Rev. D}\ }\textbf {\bibinfo {volume} {97}},\ \bibinfo
  {pages} {123532} (\bibinfo {year} {2018})},\ \Eprint
  {http://arxiv.org/abs/1804.08577} {arXiv:1804.08577 [gr-qc]} \BibitemShut
  {NoStop}%
\bibitem [{\citenamefont {Inomata}\ and\ \citenamefont
  {Nakama}(2019)}]{Inomata:2018epa}%
  \BibitemOpen
  \bibfield  {author} {\bibinfo {author} {\bibfnamefont {K.}~\bibnamefont
  {Inomata}}\ and\ \bibinfo {author} {\bibfnamefont {T.}~\bibnamefont
  {Nakama}},\ }\href {\doibase 10.1103/PhysRevD.99.043511} {\bibfield
  {journal} {\bibinfo  {journal} {Phys. Rev. D}\ }\textbf {\bibinfo {volume}
  {99}},\ \bibinfo {pages} {043511} (\bibinfo {year} {2019})},\ \Eprint
  {http://arxiv.org/abs/1812.00674} {arXiv:1812.00674 [astro-ph.CO]}
  \BibitemShut {NoStop}%
\bibitem [{\citenamefont {Husdal}(2016)}]{Husdal:2016haj}%
  \BibitemOpen
  \bibfield  {author} {\bibinfo {author} {\bibfnamefont {L.}~\bibnamefont
  {Husdal}},\ }\href {\doibase 10.3390/galaxies4040078} {\bibfield  {journal}
  {\bibinfo  {journal} {Galaxies}\ }\textbf {\bibinfo {volume} {4}},\ \bibinfo
  {pages} {78} (\bibinfo {year} {2016})},\ \Eprint
  {http://arxiv.org/abs/1609.04979} {arXiv:1609.04979 [astro-ph.CO]}
  \BibitemShut {NoStop}%
\bibitem [{\citenamefont {Arzoumanian}\ \emph {et~al.}(2020)\citenamefont
  {Arzoumanian} \emph {et~al.}}]{NANOGrav:2020bcs}%
  \BibitemOpen
  \bibfield  {author} {\bibinfo {author} {\bibfnamefont {Z.}~\bibnamefont
  {Arzoumanian}} \emph {et~al.} (\bibinfo {collaboration} {NANOGrav}),\ }\href
  {\doibase 10.3847/2041-8213/abd401} {\bibfield  {journal} {\bibinfo
  {journal} {Astrophys. J. Lett.}\ }\textbf {\bibinfo {volume} {905}},\
  \bibinfo {pages} {L34} (\bibinfo {year} {2020})},\ \Eprint
  {http://arxiv.org/abs/2009.04496} {arXiv:2009.04496 [astro-ph.HE]}
  \BibitemShut {NoStop}%
\bibitem [{\citenamefont {Nan}\ \emph {et~al.}(2011)\citenamefont {Nan},
  \citenamefont {Li}, \citenamefont {Jin}, \citenamefont {Wang}, \citenamefont
  {Zhu}, \citenamefont {Zhu}, \citenamefont {Zhang}, \citenamefont {Yue},\ and\
  \citenamefont {Qian}}]{Nan:2011um}%
  \BibitemOpen
  \bibfield  {author} {\bibinfo {author} {\bibfnamefont {R.}~\bibnamefont
  {Nan}}, \bibinfo {author} {\bibfnamefont {D.}~\bibnamefont {Li}}, \bibinfo
  {author} {\bibfnamefont {C.}~\bibnamefont {Jin}}, \bibinfo {author}
  {\bibfnamefont {Q.}~\bibnamefont {Wang}}, \bibinfo {author} {\bibfnamefont
  {L.}~\bibnamefont {Zhu}}, \bibinfo {author} {\bibfnamefont {W.}~\bibnamefont
  {Zhu}}, \bibinfo {author} {\bibfnamefont {H.}~\bibnamefont {Zhang}}, \bibinfo
  {author} {\bibfnamefont {Y.}~\bibnamefont {Yue}}, \ and\ \bibinfo {author}
  {\bibfnamefont {L.}~\bibnamefont {Qian}},\ }\href {\doibase
  10.1142/S0218271811019335} {\bibfield  {journal} {\bibinfo  {journal} {Int.
  J. Mod. Phys. D}\ }\textbf {\bibinfo {volume} {20}},\ \bibinfo {pages} {989}
  (\bibinfo {year} {2011})},\ \Eprint {http://arxiv.org/abs/1105.3794}
  {arXiv:1105.3794 [astro-ph.IM]} \BibitemShut {NoStop}%
\bibitem [{\citenamefont {Janssen}\ \emph {et~al.}(2015)\citenamefont {Janssen}
  \emph {et~al.}}]{Janssen:2014dka}%
  \BibitemOpen
  \bibfield  {author} {\bibinfo {author} {\bibfnamefont {G.}~\bibnamefont
  {Janssen}} \emph {et~al.},\ }\href {\doibase 10.22323/1.215.0037} {\bibfield
  {journal} {\bibinfo  {journal} {PoS}\ }\textbf {\bibinfo {volume}
  {AASKA14}},\ \bibinfo {pages} {037} (\bibinfo {year} {2015})},\ \Eprint
  {http://arxiv.org/abs/1501.00127} {arXiv:1501.00127 [astro-ph.IM]}
  \BibitemShut {NoStop}%
\bibitem [{\citenamefont {Desvignes}\ \emph {et~al.}(2016)\citenamefont
  {Desvignes} \emph {et~al.}}]{Desvignes:2016yex}%
  \BibitemOpen
  \bibfield  {author} {\bibinfo {author} {\bibfnamefont {G.}~\bibnamefont
  {Desvignes}} \emph {et~al.},\ }\href {\doibase 10.1093/mnras/stw483}
  {\bibfield  {journal} {\bibinfo  {journal} {Mon. Not. Roy. Astron. Soc.}\
  }\textbf {\bibinfo {volume} {458}},\ \bibinfo {pages} {3341} (\bibinfo {year}
  {2016})},\ \Eprint {http://arxiv.org/abs/1602.08511} {arXiv:1602.08511
  [astro-ph.HE]} \BibitemShut {NoStop}%
\bibitem [{\citenamefont {Luo}\ \emph {et~al.}(2016)\citenamefont {Luo} \emph
  {et~al.}}]{TianQin:2015yph}%
  \BibitemOpen
  \bibfield  {author} {\bibinfo {author} {\bibfnamefont {J.}~\bibnamefont
  {Luo}} \emph {et~al.} (\bibinfo {collaboration} {TianQin}),\ }\href {\doibase
  10.1088/0264-9381/33/3/035010} {\bibfield  {journal} {\bibinfo  {journal}
  {Class. Quant. Grav.}\ }\textbf {\bibinfo {volume} {33}},\ \bibinfo {pages}
  {035010} (\bibinfo {year} {2016})},\ \Eprint
  {http://arxiv.org/abs/1512.02076} {arXiv:1512.02076 [astro-ph.IM]}
  \BibitemShut {NoStop}%
\bibitem [{\citenamefont {Amaro-Seoane}\ \emph {et~al.}(2017)\citenamefont
  {Amaro-Seoane} \emph {et~al.}}]{LISA:2017pwj}%
  \BibitemOpen
  \bibfield  {author} {\bibinfo {author} {\bibfnamefont {P.}~\bibnamefont
  {Amaro-Seoane}} \emph {et~al.} (\bibinfo {collaboration} {LISA}),\
  }\href@noop {} {\  (\bibinfo {year} {2017})},\ \Eprint
  {http://arxiv.org/abs/1702.00786} {arXiv:1702.00786 [astro-ph.IM]}
  \BibitemShut {NoStop}%
\bibitem [{\citenamefont {Hu}\ and\ \citenamefont {Wu}(2017)}]{Hu:2017mde}%
  \BibitemOpen
  \bibfield  {author} {\bibinfo {author} {\bibfnamefont {W.-R.}\ \bibnamefont
  {Hu}}\ and\ \bibinfo {author} {\bibfnamefont {Y.-L.}\ \bibnamefont {Wu}},\
  }\href {\doibase 10.1093/nsr/nwx116} {\bibfield  {journal} {\bibinfo
  {journal} {Natl. Sci. Rev.}\ }\textbf {\bibinfo {volume} {4}},\ \bibinfo
  {pages} {685} (\bibinfo {year} {2017})}\BibitemShut {NoStop}%
\bibitem [{\citenamefont {Ackley}\ \emph {et~al.}(2020)\citenamefont {Ackley}
  \emph {et~al.}}]{Ackley:2020atn}%
  \BibitemOpen
  \bibfield  {author} {\bibinfo {author} {\bibfnamefont {K.}~\bibnamefont
  {Ackley}} \emph {et~al.},\ }\href {\doibase 10.1017/pasa.2020.39} {\bibfield
  {journal} {\bibinfo  {journal} {Publ. Astron. Soc. Austral.}\ }\textbf
  {\bibinfo {volume} {37}},\ \bibinfo {pages} {e047} (\bibinfo {year}
  {2020})},\ \Eprint {http://arxiv.org/abs/2007.03128} {arXiv:2007.03128
  [astro-ph.HE]} \BibitemShut {NoStop}%
\bibitem [{\citenamefont {Hild}\ \emph {et~al.}(2011)\citenamefont {Hild} \emph
  {et~al.}}]{Hild:2010id}%
  \BibitemOpen
  \bibfield  {author} {\bibinfo {author} {\bibfnamefont {S.}~\bibnamefont
  {Hild}} \emph {et~al.},\ }\href {\doibase 10.1088/0264-9381/28/9/094013}
  {\bibfield  {journal} {\bibinfo  {journal} {Class. Quant. Grav.}\ }\textbf
  {\bibinfo {volume} {28}},\ \bibinfo {pages} {094013} (\bibinfo {year}
  {2011})},\ \Eprint {http://arxiv.org/abs/1012.0908} {arXiv:1012.0908 [gr-qc]}
  \BibitemShut {NoStop}%
\bibitem [{\citenamefont {Abbott}\ \emph
  {et~al.}(2017{\natexlab{b}})\citenamefont {Abbott} \emph
  {et~al.}}]{LIGOScientific:2016wof}%
  \BibitemOpen
  \bibfield  {author} {\bibinfo {author} {\bibfnamefont {B.~P.}\ \bibnamefont
  {Abbott}} \emph {et~al.} (\bibinfo {collaboration} {LIGO Scientific}),\
  }\href {\doibase 10.1088/1361-6382/aa51f4} {\bibfield  {journal} {\bibinfo
  {journal} {Class. Quant. Grav.}\ }\textbf {\bibinfo {volume} {34}},\ \bibinfo
  {pages} {044001} (\bibinfo {year} {2017}{\natexlab{b}})},\ \Eprint
  {http://arxiv.org/abs/1607.08697} {arXiv:1607.08697 [astro-ph.IM]}
  \BibitemShut {NoStop}%
\bibitem [{\citenamefont {Abbott}\ \emph {et~al.}(2019)\citenamefont {Abbott}
  \emph {et~al.}}]{LIGOScientific:2019vic}%
  \BibitemOpen
  \bibfield  {author} {\bibinfo {author} {\bibfnamefont {B.~P.}\ \bibnamefont
  {Abbott}} \emph {et~al.} (\bibinfo {collaboration} {LIGO Scientific,
  Virgo}),\ }\href {\doibase 10.1103/PhysRevD.100.061101} {\bibfield  {journal}
  {\bibinfo  {journal} {Phys. Rev. D}\ }\textbf {\bibinfo {volume} {100}},\
  \bibinfo {pages} {061101} (\bibinfo {year} {2019})},\ \Eprint
  {http://arxiv.org/abs/1903.02886} {arXiv:1903.02886 [gr-qc]} \BibitemShut
  {NoStop}%
\bibitem [{\citenamefont {Hawking}(1974)}]{Hawking:1974rv}%
  \BibitemOpen
  \bibfield  {author} {\bibinfo {author} {\bibfnamefont {S.~W.}\ \bibnamefont
  {Hawking}},\ }\href {\doibase 10.1038/248030a0} {\bibfield  {journal}
  {\bibinfo  {journal} {Nature}\ }\textbf {\bibinfo {volume} {248}},\ \bibinfo
  {pages} {30} (\bibinfo {year} {1974})}\BibitemShut {NoStop}%
\bibitem [{\citenamefont {Carr}(1976)}]{Carr:1976zz}%
  \BibitemOpen
  \bibfield  {author} {\bibinfo {author} {\bibfnamefont {B.~J.}\ \bibnamefont
  {Carr}},\ }\href {\doibase 10.1086/154351} {\bibfield  {journal} {\bibinfo
  {journal} {Astrophys. J.}\ }\textbf {\bibinfo {volume} {206}},\ \bibinfo
  {pages} {8} (\bibinfo {year} {1976})}\BibitemShut {NoStop}%
\bibitem [{\citenamefont {Baumann}\ \emph
  {et~al.}(2007{\natexlab{b}})\citenamefont {Baumann}, \citenamefont
  {Steinhardt},\ and\ \citenamefont {Turok}}]{Baumann:2007yr}%
  \BibitemOpen
  \bibfield  {author} {\bibinfo {author} {\bibfnamefont {D.}~\bibnamefont
  {Baumann}}, \bibinfo {author} {\bibfnamefont {P.~J.}\ \bibnamefont
  {Steinhardt}}, \ and\ \bibinfo {author} {\bibfnamefont {N.}~\bibnamefont
  {Turok}},\ }\href@noop {} {\  (\bibinfo {year} {2007}{\natexlab{b}})},\
  \Eprint {http://arxiv.org/abs/hep-th/0703250} {arXiv:hep-th/0703250}
  \BibitemShut {NoStop}%
\bibitem [{\citenamefont {De~Luca}\ \emph {et~al.}(2021)\citenamefont
  {De~Luca}, \citenamefont {Franciolini}, \citenamefont {Kehagias},\ and\
  \citenamefont {Riotto}}]{DeLuca:2021oer}%
  \BibitemOpen
  \bibfield  {author} {\bibinfo {author} {\bibfnamefont {V.}~\bibnamefont
  {De~Luca}}, \bibinfo {author} {\bibfnamefont {G.}~\bibnamefont
  {Franciolini}}, \bibinfo {author} {\bibfnamefont {A.}~\bibnamefont
  {Kehagias}}, \ and\ \bibinfo {author} {\bibfnamefont {A.}~\bibnamefont
  {Riotto}},\ }\href {\doibase 10.1016/j.physletb.2021.136454} {\bibfield
  {journal} {\bibinfo  {journal} {Phys. Lett. B}\ }\textbf {\bibinfo {volume}
  {819}},\ \bibinfo {pages} {136454} (\bibinfo {year} {2021})},\ \Eprint
  {http://arxiv.org/abs/2102.07408} {arXiv:2102.07408 [astro-ph.CO]}
  \BibitemShut {NoStop}%
\end{thebibliography}%
\end{document}